\documentclass[
  paper    = a4,
  BCOR     = 15mm,
  DIV      = 12,
  twoside=on,
  fontsize = 12pt,
  fleqn,
  toc      = bibnumbered,
  toc      = listofnumbered,
  numbers  = noendperiod,
  headings = normal,
  listof   = leveldown,
  version  = 3.03
]{scrbook}

\usepackage[T1]{fontenc}
\usepackage[utf8]{inputenc}

\DeclareUnicodeCharacter{039B}{$\Lambda$}
\DeclareUnicodeCharacter{0127}{$\hbar$}

\usepackage[usenames,dvipsnames]{color}
\usepackage[
    colorlinks=false,
    citecolor=MidnightBlue,
    linkcolor=Maroon,
    urlcolor=OliveGreen,
]{hyperref}

\usepackage{ifthen}
\usepackage{amsmath}
\usepackage{float}
\usepackage{graphicx}
\usepackage[all]{nowidow}
\usepackage{siunitx}
\usepackage{listings}
\usepackage{setspace}
\usepackage{pdflscape}
\usepackage{rotating}
\usepackage{metalogo}
\usepackage{ragged2e}

\usepackage{multirow}
\usepackage{longtable,ltcaption,array}
\newlength{\DUtablewidth} 
\usepackage{tabularx}

\newcommand{\slstinputlisting}[1]{
    \section{\protect\detokenize{#1}}
    \lstinputlisting[caption=\protect\detokenize{#1}]{external/#1}
}

\newcommand{\labeledslstinputlisting}[2]{
    \section{\protect\detokenize{#1}}
    \label{#2}
    \lstinputlisting[caption=\protect\detokenize{#1}]{external/#1}
}

\newcommand{\labeledslstinputlistingc}[2]{
    \section{\protect\detokenize{#1}}
    \label{#2}
    \lstinputlisting[caption=\protect\detokenize{#1},language=C]{external/#1}
}

\newcommand{\labeledslstinputlistingsh}[2]{
    \section{\protect\detokenize{#1}}
    \label{#2}
    \lstinputlisting[caption=\protect\detokenize{#1},language=bash]{external/#1}
}

\newcommand*\mean[1]{\bar{#1}}
\newenvironment{DUlegend}{}{}

\usepackage[ngerman,english]{babel}
\usepackage{csquotes}

\usepackage[natbib=true,
            style=authoryear,
            backend=bibtex
           ]{biblatex}
\usepackage{listings}
\usepackage{setspace}
\usepackage[version=3]{mhchem}

\newcommand{\hefour}{\ce{^{4}He}}
\newcommand{\hethree}{\ce{^{3}He}}
\newcommand{\hp}{\ce{H+}}
\newcommand{\htwo}{\ce{H2}}
\newcommand{\co}{\ce{CO}}
\newcommand{\h}{\ce{H}}

\newcommand*{\sw}{\textbf}

\providecommand*{\DUdocumentsubtitle}[1]{{\large #1}}
\providecommand*{\DUfootnotemark}[3]{%
  \raisebox{1em}{\hypertarget{#1}{}}%
  \hyperlink{#2}{\textsuperscript{#3}}%
}
\providecommand{\DUfootnotetext}[4]{%
  \begingroup%
  \renewcommand{\thefootnote}{%
    \protect\raisebox{1em}{\protect\hypertarget{#1}{}}%
    \protect\hyperlink{#2}{#3}}%
  \footnotetext{#4}%
  \endgroup%
}

\providecommand*{\DUrole}[2]{%
  \ifcsname DUrole#1\endcsname%
    \csname DUrole#1\endcsname{#2}%
  \else
    \ifcsname docutilsrole#1\endcsname%
      \csname docutilsrole#1\endcsname{#2}%
    \else%
      #2%
    \fi%
  \fi%
}

\ifthenelse{\isundefined{\DUlegend}}{
  \newenvironment{DUlegend}{\small}{}
}{}

\ifthenelse{\isundefined{\hypersetup}}{
  \usepackage[colorlinks=true,linkcolor=blue,urlcolor=blue]{hyperref}
  \urlstyle{same} 
}{}
\hypersetup{
  pdftitle={A supernova feedback implementation for the astrophysical simulation software Arepo},
}

\usepackage{etoolbox}

\makeatletter
\AtBeginDocument{%
  \pretocmd{\@citex}{\def\_{_}\def\:{:}}{}{}%
}
\makeatother

\frontmatter
\KOMAoptions{cleardoublepage=empty}

\title{\phantomsection%
  A supernova feedback implementation for the astrophysical simulation software Arepo%
  \label{a-supernova-feedback-implementation-for-the-astrophysical-simulation-software-arepo}%
  \\ 
  \DUdocumentsubtitle{Master’s thesis in physics by André-Patrick Bubel, 2015}%
  \label{master-s-thesis-in-physics-by-andre-patrick-bubel-2015}}
\author{}
\date{}

\begin{document}
\maketitle

\newcommand*{\docutilsroleref}{\ref}
\newcommand*{\docutilsrolelabel}{\label}
\newcommand*{\docutilsrolecitet}{\citet}
\newcommand*{\docutilsrolecitep}{\citep}
\newcommand*{\docutilsrolecitealp}{\citealp}
\newcommand*{\docutilsrolesw}{\sw}
\newcommand*{\docutilsrolefunction}{\textit}
\newcommand*{\docutilsrolefile}{\textit}

\thispagestyle{empty}

\noindent

This Master's thesis has been carried out by André-Patrick Bubel (born in
Hadamar on July 9th, 1987) at the Institute for Theoretical Astrophysics (ITA) in
Heidelberg under the supervision of Prof. Dr. Ralf S. Klessen, Dr. Simon C. O. Glover and Dr. Rowan J. Smith.

\vspace{\fill}

\noindent
\includegraphics[width=10pt]{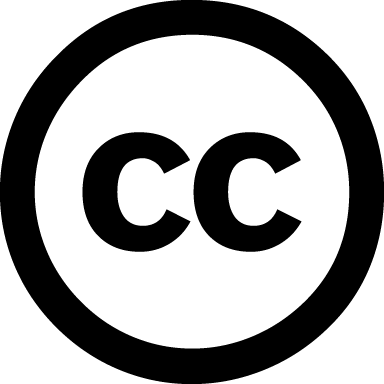}
\includegraphics[width=10pt]{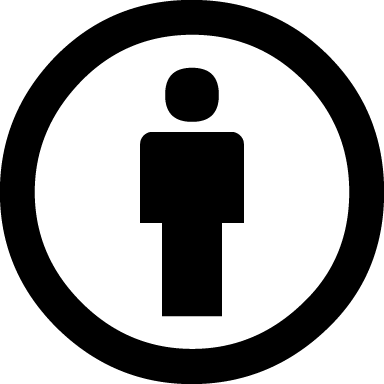}
\includegraphics[width=10pt]{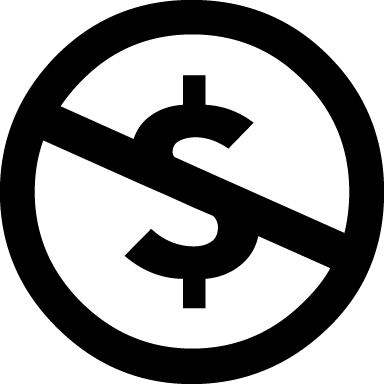}
\includegraphics[width=10pt]{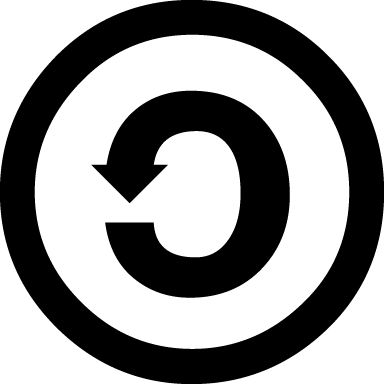}
\hspace{1pt}
\RaggedRight
This work by Andr\'{e}-Patrick Bubel is licensed under a \textit{Creative
Commons Attribution Non-Commercial ShareAlike 4.0 International
License}, with the exception of Appendix \ref{pint}, \ref{gpl} and
\ref{appendixpint}. For further information see
\url{http://creativecommons.org/licenses/by-nc-sa/4.0/}

\justifying
For Appendix \ref{appendixpint} the License under Appendix \ref{pint} applies.

The source code in Appendix \ref{appendixsource} is also licensed under the
GNU General Public License version 3.0. For the license text see Appendix
\ref{gpl}.

Original thesis submission date: August 01, 2015

Build date: \today

You can contact the author via email at \href{mailto:masterthesis@andre-bubel.de}{\texttt{masterthesis@andre-bubel.de}}

The sourcecode for generating this thesis and the software written for it will be available at \url{https://masterthesis.andre-bubel.de/}

\cleardoublepage
\thispagestyle{empty}

\begin{center}
In memory of my father, Wolfgang Bubel.
\end{center}

\cleardoublepage

\KOMAoptions{cleardoublepage=plain}

\chapter*{Abstract}

\section*{English}

Supernova (SN) explosions play an important role in the development of
galactic structures. The energy and momentum imparted on the interstellar
medium (ISM) in so-called "supernova feedback" drives turbulence, heats the
gas, enriches it with heavy elements, can lead to the formation of new stars
or even suppress star formation by disrupting stellar nurseries. In the
numerical simulation at the sub-galactic level, not including the energy and
momentum of supernovas in the physical description of the problem can also
lead to several problems that might partially be resolved by including a
description of supernovas.

In this thesis such an implementation is attempted
for the combined numerical hydrodynamics and N-body simulation software
\sw{Arepo} \citep{Springel2010} for the high density gas in the ISM only. This allows supernova driven turbulence in boxes of 400pc cubed to be studied. In a stochastic process a large amount of
thermal energy is imparted on a number of neighbouring cells, mimicking the
effect of a supernova explosions.

We test this approach by modelling the explosion of a single supernova in a
uniform density medium and comparing the evolution of the resulting
supernova remnant to the theoretically-predicted behaviour. We also run a
simulation with our feedback code and a fixed supernova rate derived from
the Kennicutt-Schmidt relation \citep{Kennicutt1998} for a duration of about
$\SI{20}{Myrs}$. We describe our method in detail in this text and discuss
the properties of our implementation.

\cleardoublepage

\section*{Deutsch}

\begin{otherlanguage}{ngerman}

Supernovas (SN) spielen eine wichtige Rolle in der Bildung von galaktischen
Strukturen. Energie und Impuls die vom sogenannten ``supernova feedback''
auf das interstellare Medium übertragen wird, erzeugt Turbulenz, heizt das
Gas, reichert es mit schweren Elementen an, führt zu der Bildung von neuen
Sternen oder gar zu der Unterdrückung von Sternentstehung, wenn deren
Entstehungsgebiete zerstört werden. Wenn solch ein Energie- und
Impulsübertrag durch Supernovas in numerischen Simulationen nicht
berücksichtigt wird, kann dies zu Problemen führen.

In dieser Arbeit wird eine Implementierung einer solchen Methode für die
Hydrodynamik- und N-Körpersimulationssoftware \sw{Arepo}
\citep{Springel2010} versucht. Durch einen stochastischen Prozess wird eine
große Menge thermischer Energie auf eine kleine Zahl benachbarter Zellen
aufgeprägt, welches den Einfluss einer Supernova imitiert.

Wir testen das generelle Verhalten zunächst von einzelnen Supernovas in
einem gleichförmigen Medium konstanter Dichte und vergleichen die
Entwicklung der Explosion mit deren theoretischem Verhalten. Außerdem lassen
wir eine Simulation mit unserer Implementation laufen, wobei die
Supernova-Rate von der Kennicutt-Schmidt-Relation \citep{Kennicutt1998}
abgeleitet ist. Diese Simulation umfasst ungefähr $\SI{20}{Myrs}$. Details
unserer Methode werden im folgenden Text diskutiert und Eigenschaften der
Implementation erläutert.

\end{otherlanguage}

\cleardoublepage

\chapter{Changelog%
  \label{changelog}%
}

This version was prepared for publication on \url{https://arxiv.org/}, and
published after submitting the original version for my master’s degree.

\section*{arXiv version 2}

Removed section 3.4 and most of the discussion in \DUrole{ref}{further-work}. I might
have worded my critique a bit too harshly, and a significant amount of issues
were already addressed since the original submission. These sections may need
rework and further objective underpinning.

I want to apologize to Prof. Springel and other Arepo coauthors if this
misrepresented the quality of their work.

Smaller changes:
\begin{itemize}

\item Removed unused lines from templates for Arepo config files

\item Added date of orginal submission to copyright section

\item Clarified applicability of the SN code in the abstract and conclusion

\end{itemize}

\section*{arXiv version 1}

Some
changes to the originally submitted version were necessary to accommodate for the arXiv publishing pipeline:
\begin{itemize}

\item Added reference for \DUrole{sw}{GNU parallel}

\item Reduced image size, to shrink PDF file size

\item Used \DUrole{sw}{pdflatex}, as \DUrole{latex}{\XeLaTeX} is not available with arXiv’s typesetting pipeline

\item Used default \DUrole{latex}{\LaTeX} fonts

\item \DUrole{file}{\_\_init\_\_.py} file removed from listings

\item Link added to source code

\item Improved formatting of license URL

\item Added Changelog section

\end{itemize}

\chapter{Acknowledgements%
  \label{acknowledgements}%
}

I foremost thank Dr. Simon C. O. Glover and Dr. Rowan J. Smith for mentoring, supervision and
feedback during writing of my thesis.

Thanks to Eduard Bopp, Jon Ramsey, Mai Sasaki, and the countless authors of
free software that I used in the preparation of this thesis.

Tobias Förtsch, Kai Salm and my brother Pascal Bubel gave useful feedback for
improving this text.

I thank Prof. Dr. Volker Springel for giving access to \DUrole{sw}{Arepo}.

The simulations in this thesis were carried out on the MilkyWay cluster, which is
supported by the DFG via SFB 881 (sub-project Z2) and SPP 1573.

Finally, without support of my friends and family my thesis and finishing my
undergraduate studies wouldn’t have been possible. Thank you all!

\chapter{Units%
  \label{units}%
}

Mainly SI units are used in this thesis. For handling unit conversions the
\DUrole{sw}{Python} package \DUrole{sw}{Pint} version 0.6 is used \DUrole{citep}{Grecco2015}.
Constant values are based mainly on \DUrole{citet}{Mohr2012}.

The mass of the isotope $\ce{^{1}H}$ - $\si{m_{\ce{H}}}$ - is not a
default unit provided by \DUrole{sw}{Pint}. We use the value of \DUrole{citet}{Emsley1995},
$\si{m_{\ce{H}}} = \SI{1.007825}{u}$.

The value of the solar mass ($\si{M_\odot} = \SI{1.989e30}{kg}$),
astronomical unit ($\si{au} = \SI{1.49598e11}{m}$) and parsec
($\si{pc} = \SI{3.085678e16}{m}$) is taken from the Arepo source code for
consistency. Constants in the \DUrole{sw}{Arepo} source code are those defined in the
file \DUrole{file}{allvars.c}.

\DUrole{sw}{Arepo} uses a user defined unit system, specified in the run-time
configuration file in cgs (“centimetre, grams, seconds”) units, internally as
well as in input and output files. The units used during the simulation step
therefore also needs to be know in the pre- and post-processing steps.

The default unit and constant definition file for \DUrole{sw}{Pint} is printed verbatim
in Appendix \DUrole{ref}{appendixpint}. The custom unit definition file
\DUrole{file}{units.txt} (see \DUrole{ref}{code:units-txt}) overwrites the default definitions
and adds custom units. The internal code units for \DUrole{sw}{Arepo} simulation runs
are also defined there.

Other units used in the text are
\begin{itemize}

\item The Unified Atomic Mass Unit: $\si{u} = \SI{1.660538782e-27}{kg}$ \DUrole{citep}{Mohr2012}

\item The erg, the energy unit of the CGS system: $\SI{1}{erg} = \SI{e-7}{J}$

\end{itemize}

\phantomsection\label{contents}
\pdfbookmark[2]{Contents}{contents}
\tableofcontents

\cleardoublepage

\mainmatter
\floatplacement{figure}{htbp}

\chapter{Introduction%
  \label{introduction}%
}

Feedback from supernovas (SN), the violent end of stars in energetic explosions,
plays an important part in driving the turbulence in the interstellar medium
(ISM) and pressurizing it \DUrole{citep}{Norman1996,MacLow2004,Joung2006,Gent2013}. It
also has a role in regulating the rate of star formation by on the one hand
disrupting stellar nurseries, where heavy, short lived stars explode relatively
soon after their birth, and on the other hand destabilizing molecular clouds
leading to their collapse and enriching the interstellar medium with gas and
heavy elements \DUrole{citep}{Ferriere2001,Hopkins2011}.

While N-body simulations of pure dark matter, without incorporating “baryonic”
physics historically have made important contributions in large scale structure
and galaxy formation, inconsistencies with observations, like the cusp-core
problem \DUrole{citep}{DeBlok2009}, hint at the limits of such simulations.  General
multi-purpose astrophysical hydrodynamic simulation software packages, like
\DUrole{sw}{FLASH} \DUrole{citep}{Fryxell2000}, \DUrole{sw}{ENZO} \DUrole{citep}{Oshea2005,Bryan2014},
\DUrole{sw}{GADGET} \DUrole{citep}{Springel2001,Springel2005b}, and its successor \DUrole{sw}{Arepo}
\DUrole{citep}{Springel2010}, among others, using various approaches, are now
state-of-the-art for large scale simulations, and have been applied to varying
scales, from star formation, planet formation, simulations of the interstellar
medium, galaxy formation and structure formation. Treatments of other physical
processes like chemistry in the ISM, radiative transfer and star formation and
feedback processes have been implemented in many of these packages. Supernova
feedback is needed for a realistic treatment of the ISM and galactic matter
cycle.

In this thesis a supernova feedback scheme by injection of thermal energy is
implemented for the multi-physics simulation software \DUrole{sw}{Arepo}.
Stochastically a fixed amount of thermal energy is added to a spherical region
in random locations in the simulation domain. The general approach is tested for
single supernova explosions, where the energy is applied in the initial
conditions. Problems in the volume estimates of the supernova region limits the
significance of these simulations. The actual implementation is tested in a
simulation of a periodic box with initially uniform density and a fixed
supernova rate.

\section{Stellar evolution and supernova explosions%
  \label{stellar-evolution-and-supernova-explosions}%
}

Stars form from clouds of gas that collapse gravitationally. The gas heats up
while being compressed, finally leading to the ignition of nuclear fusion in the
stellar core. The radiation pressure keeps the new formed star from further
collapse.

The main fusion process of lighter stars - the p-p chain - creates
\DUrole{latex}{\hefour} by first fusing two protons to form deuterium, which in turn
fuses to \DUrole{latex}{\hethree} and then into \DUrole{latex}{\hefour}. Heavier stars create
\DUrole{latex}{\hefour} in a catalytic process - the CNO-cycle - involving carbon,
nitrogen and oxygen \DUrole{citep}{Salaris2005}.

Although brightness and size of the star change during this burning phase, it
only changes massively when most of its hydrogen is used up. Then, without the
radiation pressure supporting the core, the star shrinks until the temperature
and density increase to the point at which helium fusion begins (Only the
lightest stars under $\SI{0.1}{\si{M_\odot}}$ might avoid this phase and
collapse to white dwarfs without igniting helium burning). After this point the
development of stars of different masses varies, though in general more heavy
elements are produced in cascading steps up to the production of iron. Just
heaviest stars reach the latest phases \DUrole{citep}{Heger2003}.

Heavier elements are not formed in this process, as the fusion of heavier
elements than \DUrole{latex}{\ce{^{56}Fe}} uses more energy than it produces. Depending
on the initial mass and metallicity (the fraction of heavier elements in
comparison to hydrogen and helium) of the star, stars end their life in
different ways. Most single stars, those with not enough mass to form neutron
stars, finally collapse to white dwarfs, with electron degeneracy pressure
stabilizing the remnant.  Heavier stars collapse to neutron stars or black
holes, or are destroyed in so-called pair-instability supernovas
\DUrole{citep}{Heger2003}.

Depending on the initial mass and metallicity, which determines the mass lost
during the stars lifetime in stellar winds, the star collapses directly to a
black hole without ejecting much mass, form black holes, after some mass is
ejected and falls back, or form neutron stars, remnants stabilized due to
neutron degeneracy pressure. Most of these processes release large amount of
energy, detectable as bright, temporary objects, so-called “supernovas”, often
outshining their host galaxies \DUrole{citep}{Heger2003}.

Another type of similarly violent events - Type Ia supernovas - are likely
produced when a white dwarf accretes so much matter from a companion star that
it collapses when the electron degeneracy pressure is overcome. Both processes
can produce more than $\SI{e44}{J}$ of kinetic energy, which is rapidly
converted to thermal energy by strong shocks in the expanding supernova ejecta.
However, in some cases, the energy input can be much less, for instance if the
star directly collapses to a black hole \DUrole{citep}{Heger2003}.

\section{Astrophysical simulations%
  \label{astrophysical-simulations}%
}

The typical approach of experimental physics is to disprove theories with
laboratory experiments, that are postulated in the context of theoretical
physics.  Real systems oftentimes can not be realistically replicated in a
laboratory setting, especially large scale planetary or astrophysical systems,
like the interior of stars, the structure of planetary systems and galaxies, or
even the universe as a whole. Then it is only possible to observe such systems
as they unfold in nature, and trying to explain them, again, with physical
theories.

Oftentimes simple theories are not able to produce results that are consistent
with reality and even simple systems, e.g. the general dynamics of more than
three gravitationally interacting particles, generally can not be solved
analytically.

The field of computational physics in some sense bridges the classical divide
between theoretical and experimental physics. It makes use of the possibility to
solve ever more complex systems numerically and complex interacting systems are
modelled with algorithms. These “simulations” can provide a “mock laboratory” to
study systems that are out of the scope of classical experiments.

As computer time is a limited resource, there is a trade-off between increasing
the accuracy of the numerical simulations (either by increasing the resolution
of discretization, or using better, but more computationally involved
algorithms), or adding additional physics to the simulation.

Galactic and cosmological simulations often limit themselves to purely
gravitationally interacting “particles”, which approximate, depending on the
scope of the simulation, whole galaxy clusters, down to individual stars, in so
called N-body simulations.

As gravitational forces are long reaching, the force of even far away masses
need to be considered in the calculation of the movement of the individual
particles. The computational cost of the naive approach, the direct summation of
gravitational forces of all other particles for each individual particle, scales
quadratically with the number of particles $N$, i.e. the computational
cost is of order $\mathcal O(N^2)$. Either solving the Poisson equation on
a grid, a so-called “particle-mesh” approach, which can be done efficiently in
Fourier space using the Fast Fourier Transform (FFT), or embedding the particles
in a tree structure (typically a quad-tree in 2D and an octree 3D in
simulations) and combining the masses of far away particles into a single
virtual one (known as the Barnes-Hut tree method \DUrole{citep}{Barnes1986}), can lower
the complexity down to $\mathcal O(N log N)$ on average, potentially
sacrificing accuracy and increasing the complexity of implementing the software.
Both methods can also be combined in a “TreePM” approach
\DUrole{citep}{Bagla2002,Springel2005b}.

Especially the simulation of dark matter, that largely dominates structure
formation, galaxy formation and the overall dynamics of galaxies, are well
suited for a pure N-body approach, as almost 84.5\% of the total matter content
of the universe is non-baryonic \DUrole{citep}{PlanckCollaboration2014}. One large
scale project was the Millennium simulation run \DUrole{citep}{Springel2005a}, a cold
dark matter N-body simulation, with cosmic expansion due to dark energy (ΛCDM),
using the software \DUrole{sw}{GADGET 2} \DUrole{citep}{Springel2005b} with $2160^3
\simeq 10^{10}$ particles.

But simulations only considering dark matter (e.g. of galaxy formation) show
several problems. One is the cusp-core problem, where the central bulge of
simulated CDM galaxies have a different radial density profile than real,
observed galaxies \DUrole{citep}{DeBlok2009}. Considering the baryonic content of
galaxies, and simulating interstellar gas, stellar feedback, e.g.  stellar
winds, supernovas, and radiative heating, seem to alleviate those problems to
some degree \DUrole{citep}{Valenzuela2007,DeBlok2009}.

On the other hand the study of the interstellar medium in itself is an
interesting field. A large fraction of the baryonic mass of galaxies is not
bound in stars, but spread throughout galaxies as molecular, atomic or ionized
gas (mostly hydrogen and helium, with varying degree of “metals”, all other
elements in the context of astrophysics \DUrole{citep}{SpitzerJr1978}), and small dust
grains.  Non-equilibrium hydrodynamics is needed to understand the behaviour of
this gas fraction. Complex heating and cooling processes, e.g. radiative heating
by stars and metal line cooling, add additional complexities to the numerical
descriptions of the ISM.

There are several general approaches to simulate hydrodynamical systems. The
most prominent are the Lagrangian smoothed-particle hydrodynamics (SPH) and
Eulerian, mesh-based schemes. SPH describes the gas as particles, with the gas
properties spread between them. A larger number of particles can be positioned,
where a high simulation resolution is desired (for instance at fluid boundaries,
in shocks and turbulent regions) e.g. by just advecting the particles with the
fluid flow. This high adaptability is desirable, as computer time is mostly
spend where it is needed.

In mesh based schemes the simulation domain is divided into cells, often regular
rectangles (in 2D) or cubes (in 3D), for a so-called Euclidean mesh. Several
approaches are available to solve the scale invariant Navier-Stokes equations
numerically on such meshes, with different trade-offs between computational
efforts and accuracy. To cover the high contrasts, for instance in density
between large sparse bubbles and collapsing high-density molecular clouds, a
high resolution mesh is needed. Although the computational complexity is not
quadratic, as in the case of gravity, because hydrodynamic forces are only short
reaching, the high resolution (which scales cubic in three dimensions when
dividing the domain in each Euclidean direction), makes it desirable, only to
spend computing time where needed. Adaptive mesh refinement (AMR) only increases
the resolution of the mesh, where it is deemed necessary, leaving it sparse
where little is happening. Different schemes exist to additionally simulate the
effect of magnetic fields, e.g. so-called magneto-hydrodynamics (MHD) solver.
Magnetic fields might be an important factor in driving the turbulence of the
ISM \DUrole{citep}{Norman1996,Joung2006,Gent2013}.

Additionally, instead of the Navier-Stokes equations, the discrete Boltzmann
equation can be solved, forming Lattice-Boltzmann methods (LBM) (though they
seem not to be used much in the context of astrophysics yet) (e.g.
\DUrole{citet}{Xu1997,Slyz1999}). Other kinds of physics are commonly added to both
schemes, e.g. treatment of star formation, radiative transfer and optic
shielding, swallowing of matter by black holes, chemistry, interstellar dust and
heating by cosmic rays.

SPH schemes can easily coupled to particle based N-body systems, and don’t lose
accuracy in regions of large bulk flow because of their Lagrangian formulation.
On the other hand, they often have problems resolving shocks (which are common
in astrophysics) as density gradients introduce spurious pressure forces
\DUrole{citep}{Agertz2007}. Newer modified schemes might avoid this
\DUrole{citep}{Hopkins2014}.

Mesh based schemes often have better accuracy in shocks and contact
discontinuities, but as they are generally not Galilean-invariant, large bulk
flows, relative to the fixed mesh, lead to lower accuracy
\DUrole{citep}{Springel2011}.

Common to all these numerical schemes is, that the time evolution is divided
into discrete “timesteps”. Depending on the approach, the timestep can be varied
globally, or only for a subset of the domain. A “timestep criterion” determines
the timestep needed to reach a certain accuracy and especially stability. Often
a “Courant–Friedrichs–Lewy condition” (CFL) is used for finite volume methods,
where the so-called Courant number
\begin{equation}
C = \frac{v \delta t}{\delta x} \leq C_{max}
\end{equation}
is limited to a certain value $C_{max}$, where $v$ is the velocity,
$\Delta t$ the timestep and $\Delta x$ is the spatial size of i.e.
of a cell in the domain. For explicit schemes, i.e. where fluid quantities at
time (t + dt) are derived from those at time t, $C$ must also be smaller
than one to ensure stability, i.e. it limits the movement of quantities to the
size of the discretization, so that e.g. no cells are “skipped”. Other limits
might be apply for stability, depending on the algorithm. The timestep criterion
can be either enforced locally for varying timesteps, or globally for a global
timestep.

\section{The astrophysical simulation software Arepo%
  \label{the-astrophysical-simulation-software-arepo}%
}

\DUrole{sw}{Arepo} \DUrole{citep}{Springel2011}, a multi-physics simulation software in
development by Prof.  Springel et. al., based on the SPH TreePM code \DUrole{sw}{GADGET
2} \DUrole{citep}{Springel2005b}, solves conservation laws of ideal hydrodynamics on an
unstructured, moving mesh, “defined by the Voronoi tessellation of a set of
discrete points”. A Voronoi tessellation or Voronoi decomposition, given a set
of mesh generating points (or “particles”), divides a domain into “cells”, with
each cell containing all points that are nearer to the corresponding mesh
generating point than any others (see Fig. \DUrole{ref}{fig:voronoi} for an example in
two dimensions). The cell boundaries in two dimensions are convex polygons, in
three dimension convex polyhedra (which are internally represented as a set of
triangles or tetrahedra). The tessellation can be done efficiently in a
distributed fashion. The hydrodynamic equations are solved on this mesh using an
approach “based on a second-order unsplit Godunov scheme”, i.e a finite-volume
method that solves a Riemann problem for each domain boundary, in this case with
an exact Riemann solver. When advecting the mesh generating points with the
fluid flow, an (almost, as still some mass is transfered between cells)
“Lagrangian formulation” is obtained, which mostly avoids the problems
associated with both SPH and Eulerian mesh-based schemes. In this way it also
adapts automatically the resolution of the mesh, as it depends on the local
density of particles. Optionally additional refinement and derefinement of the
mesh by merging and splitting of cells is supported, e.g. to limit the volume of
individual cells, or focus the resolution to certain regions
\DUrole{citep}{Springel2011}.

\begin{figure}
\noindent\makebox[\textwidth][c]{\includegraphics[scale=0.800000]{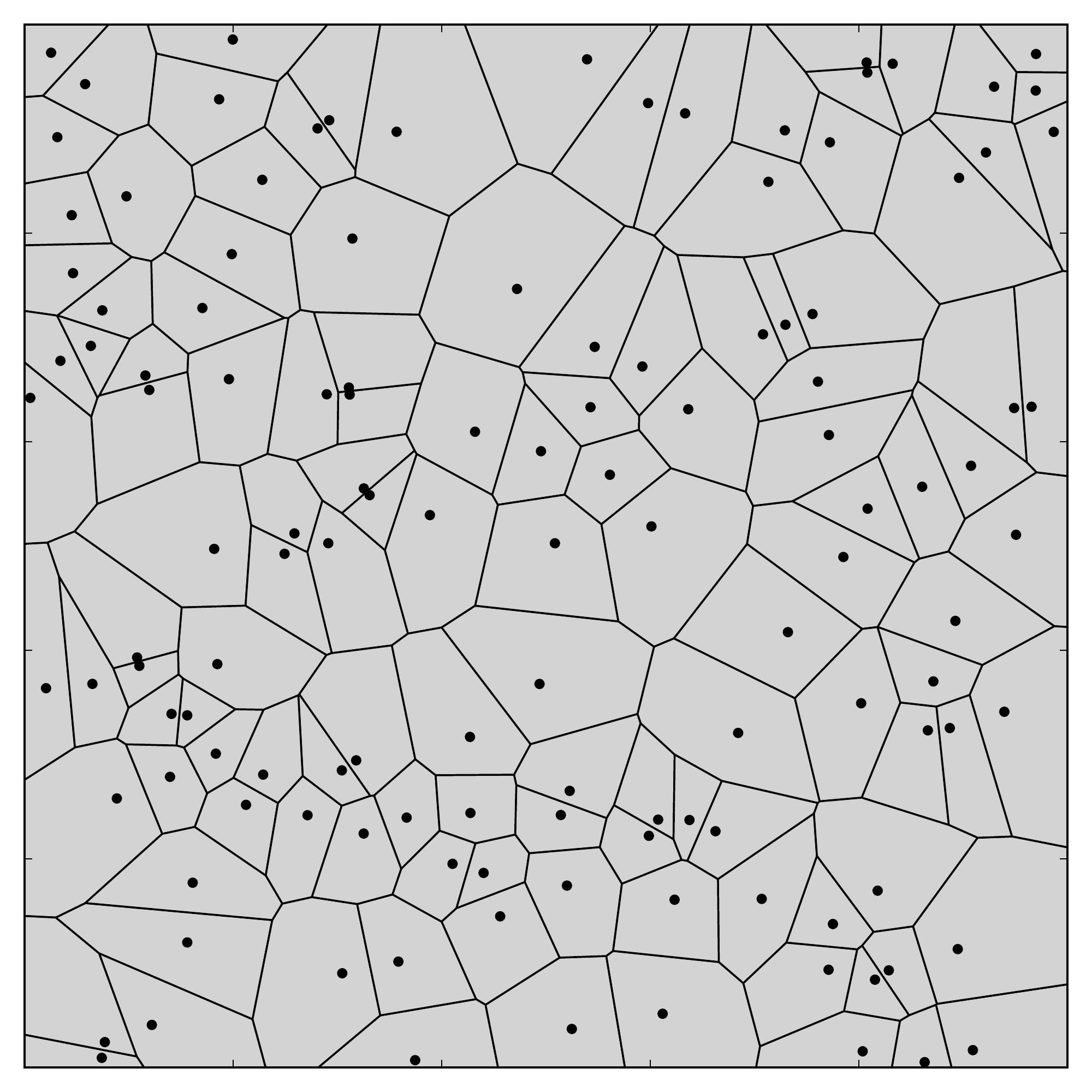}}
\caption{\DUrole{label}{fig:voronoi}
Voronoi tessellation of 128 randomly placed points in 2D. The dots are the
cell generating points of the cells they are contained in.}
\end{figure}

For gravity the TreePM implementation of newer versions of \DUrole{sw}{GADGET} is
retained, with the mesh generating points as another species of gravitationally
interacting particles. Self-gravity of gas can also be enabled \DUrole{citep}{Springel2011}.

\DUrole{sw}{Arepo} is implemented in C for distributed memory computers using \DUrole{sw}{MPI},
the “Message Passing Interface”, a standardized communication library
\DUrole{citep}{Springel2011}. At the time of writing the software source code is only
available on request, but should be made open source in the future.

\section{The Interstellar Medium (ISM)%
  \label{the-interstellar-medium-ism}%
}

The space between stars in our and other galaxies is not completely empty, but
filled with gas and dust. This “interstellar medium” (ISM) is not homogeneous.
The interstellar gas density and temperature varies over a large range - near
the Sun from as low as \DUrole{latex}{$\approx$ \SI{0.01}{u/cm^{3}}} and
\DUrole{latex}{$\approx$ \SI{1e6}{K}} in the “hot ionized medium” to \DUrole{latex}{$\approx$
\SIrange{e4}{e6}{u/cm^{3}}} and \DUrole{latex}{$\approx$ \SIrange{10}{20}{K}} in
the densest regions of molecular clouds. The ISM contributes \DUrole{latex}{$\approx$
\SIrange{10}{15}{\percent}} of the Galactic mass, and about half of it is
confined to discrete clouds \DUrole{citep}{Ferriere2001}.

The Galactic ISM is mainly composed out of hydrogen and helium, with about 1.5\%
in heavier elements near the sun, close to the composition of disk stars, though
often “metals” (elements beside Hydrogen and Helium) are “depleted” in
interstellar absorption line measurements. They are likely locked up in dust
grains \DUrole{citep}{Ferriere2001}.

The gas in the ISM can be found in different phases. A possible classification
is (with typical temperatures $T$ and number densities $n$ in the
solar neighbourhood) \DUrole{citep}{Ferriere2001}
\begin{itemize}

\item Molecular gas ($T = 10-20 \si{K}$, $n = 10^2 - 10^6
\si{cm^{-3}}$): Molecules, like \DUrole{latex}{\htwo} and \DUrole{latex}{\co}, or even mor
complex molecules, like the amino acid Glycine \DUrole{citep}{Kuan2003} and
polycyclic aromatic hydrocarbons \DUrole{citep}{Allamandola1989}, can be found in the
ISM. Most of the molecular gas is located in opaque clouds, largely shielded
from UV photons and cold enough to not be dissociated by collisions.

\item The Cold Neutral Medium (CNM) ($T = 50-100 \si{K}$, $n = 20 - 50
\si{cm^{-3}}$): Neutral atomic hydrogen can be detected by the emission and
absorption line of the 21 cm “forbidden” hyperfine transition of its ground
state, even with its very long lifetime of $10^7 \si{yrs}$. It can be
found in sheetlike clouds, but also surrounded by atomic gas or enveloping
atomic clouds. It often also can be found in thermal pressure equilibrium with
a warmer atomic phase, the WNM.

\item The Warm Neutral Medium (WNM) ($T = 60000-10000 \si{K}$, $n = 0.2
- 0.5 \si{cm^{-3}}$): Warmer hydrogen gas is mainly found spread out between
molecular and atomic clouds. It can also be detected via the 21 cm line.

\item The Warm Ionized Medium (WIM) ($T \sim 8000 \si{K}$, $n = 0.2 -
0.5 \si{cm^{-3}}$): Hot, young O and B stars ionize their surrounding hydrogen
gas, as they strongly emit in the UV spectrum. These ionized clouds are the
main contributors to ionized gas in the galactic plane. Additionally a thin
component outside of clouds and reaching above and below the galactic plane
can be observed, in parts explainable by channels driven into the ISM by
radiation and winds from O and B stars.

\item The Hot Ionized Medium (HIM) ($T \sim \SI{1e6}{K}$, $n \sim 0.0065
\si{cm^{-3}}$): Gas that is mainly shock heated (and pressurized) by supernova
explosions. They leave rarefied hot gas surrounded by a cold shell.
“Superbubbles” that are formed by interaction of clustered supernovas, likely
fill a significant volume of the ISM ($\sim \SI{20}{\percent}$ in the
solar neighbourhood). The hot gas is visible in the X-Ray spectrum and by
highly ionized metal absorption lines.

\end{itemize}

Heating of the ISM is provided by several mechanisms, e.g.
\begin{itemize}

\item Interaction of gas particles and dust grains with Cosmic Rays

\item Photoelectric heating due to dust

\item Photoionization, mainly due to UV photons

\item Chemical heating, i. e. in the formation of \DUrole{latex}{\htwo} gas on dust grains

\item Kinetic energy transfer of dust grains to the gas by collisions

\item Shock heating of gas in supernova explosion fronts

\end{itemize}

Radiative cooling is mostly provided by fine-structure line de-excitation, but
other line-cooling effects can be important, e.g. rotational lines of CO in
molecular clouds \DUrole{citep}{Klessen2014}.

Besides generating the hottest ISM phase \DUrole{citep}{McKee1977}, supernovas also
might be an important driver of turbulence in the interstellar gas
\DUrole{citep}{Norman1996,MacLow2004,Joung2006,Gent2013}.

Supernova feedback and stellar winds are also likely an important factor in
regulating star formation, with stellar winds of hot OB stars and SN explosions
of short lived stars removing gas from the star forming regions they are born
in. On the other hand stellar winds and supernovas make some of the gas that is
locked in stars again available for further star formation. Also
interaction of the shock front with otherwise stable clouds might trigger their
collapse \DUrole{citep}{Ferriere2001,Hopkins2011}.

\section{Supernova feedback%
  \label{supernova-feedback}%
}

\subsection{Supernova time evolution in a uniform ISM%
  \label{supernova-time-evolution-in-a-uniform-ism}%
}

This section mainly follows the formulation of \DUrole{citet}{Draine2010}.

A supernova ejects some fraction of its progenitor star’s mass
$M_{ejecta}$ into the surrounding interstellar mass with a typical energy
of $E_{inj} \approx \SI{1e51}{erg} = \SI{1e44}{J}$. The outer regions are
initially moving much faster than the local sound speed of the ambient medium.
This phase is called the free expansion phase. As the density of the ejecta is
much larger than the ambient medium, the forming shock can move ballistically
through it.  The region enclosed by the shock front is often called \emph{supernova
remnant} or \emph{SNR}.

As the ejecta thins and cools almost adiabatically while expanding, the pressure
soon drops close to that of the shocked ambient medium. A reverse shock is
driven back into the supernova remnant, shock-heating the rarefied gas. As it
reaches the centre of the SNR, the ejecta are heated and the pressure is much
higher than that of the ambient medium. The free expansion phase ends.  After
this, the following phases can be approximated by a point explosion.

Injecting a large amount of energy into a small volume creates a strong,
approximately spherical expanding shock wave. Assuming instantaneous injection
into a background medium of uniform density $\rho _b$, the shock undergoes
three phases.
\begin{itemize}

\item The blast-wave or Sedov-Taylor phase, where the shell expands adiabatically

\item The shell forming or radiative phase, where cooling of the shell becomes
important

\item The snowplow phase, where most of the thermal energy is spent but momentum is
conserved. The shell gains mass as it sweeps up the ambient medium

\end{itemize}

When the speed of the shock front drops to the sound speed of the ambient medium
it becomes a sound wave, with the shell now interacting freely with the ambient
medium.

\subsection{The Sedov-Taylor blast wave solution%
  \label{the-sedov-taylor-blast-wave-solution}%
}

A point explosion in a zero temperature uniform background medium of density
$\rho_0$, ignoring the background pressure, can be numerically solved.
This so-called “similarity solution”, as the structure of the blast wave only
depends on dimension-less functions, was found independently several times
during and after the second world war. It is known as the Sedov-Taylor
blast-wave solution, named after two of those authors
\DUrole{citep}{Sedov1946,Taylor1950a,Taylor1950}.

From simple dimensional analysis the temporal evolution of the shock front can
be found. If we write the radius $R(t)$ of the shock with the injection energy
$E$, the background density $\rho$ and time $t$ as
\begin{equation}
R_{ST}(t) = A \times E^\alpha \times \rho^\beta \times t^\gamma
\end{equation}
we can form a linear equation system for solving for $\alpha$,
$\beta$ and $\gamma$, by equating the power of time, length and mass
appearing in the equation above.

With the dimensions
\begin{align}
\si{[R]} = & \si{[length]} \\
\si{[E]} = & \si{[mass] x [length]^2 x [time]^{-2}} \\
\si{[\rho]} = & \si{[mass] x [length]^{-3}} \\
\si{[t]} = & \si{[time]}
\end{align}
we get
\begin{align}
\text{Length} & \text{:} & 1 = & 2 \alpha - 3 \beta \\
\text{Mass} & \text{:} &  0 = & \alpha + \beta \\
\text{Time} & \text{:} &  0 = & \gamma - 2 \alpha
\end{align}
The solution is $\alpha = \frac{1}{5}$, $\beta = -\frac{1}{5}$ and
$\gamma = \frac{2}{5}$. The dimensionless coefficient $A = 1.15167$
follows numerically from the blast-wave solution \DUrole{citep}{Draine2010}.

Therefore we get
\begin{equation}
\label{eq:sn-radius}
R_{ST}(t) = \SI{1.54e19}{cm} \times \left( \frac{E_{inj}}{10^{51} \si{erg}} \right)^{1/5} \times \left( \frac{\rho}{\si{u/cm^3}} \right)^{-1/5} \times \left( \frac{t}{10^3 \si{yrs}} \right)^{2/5}
\end{equation}
In a simulation without cooling, a point explosion will stay in the
Sedov-Taylor phase until the shock has cooled to the ambient temperature. If we
include cooling the gas behind the shock front can cool radiatively.

We take the time $t_{rad}$ (Eq. (39.20) in \DUrole{citet}{Draine2010})  and radius
$R_{rad}$ (Eq. (39.21) in \DUrole{citet}{Draine2010}) to estimate when the
remnant reaches the “radiative” phase, with the cooling function in Eq.
(39.15) of \DUrole{citet}{Draine2010} for solar-metallicity gas
\begin{align}
t_{rad} & = \SI{49.3e3}{yr}  \left( \frac{E_{inj}}{10^{51} \si{erg}} \right)^{0.22} \left( \frac{\rho_0}{\si{u/cm^3}} \right)^{-0.55}  \\
R_{rad} & = \SI{7.32e19}{cm} \left( \frac{E_{inj}}{10^{51} \si{erg}} \right)^{0.29} \left( \frac{\rho_0}{\si{u/cm^3}} \right)^{-0.42}
\end{align}
When the shell behind the shock front cools its pressure drops, and its
expansion stops. The pressure of the still hot inner region now takes over and
drives the expansion of the remnant, which expands adiabatically. An analytic
solution including the internal pressure is referred to as the
\emph{pressure-modified snowplow phase}.

For an adiabatic expansion of a monoatomic gas with the adiabatic index
$\gamma = \frac{5}{3}$, we have for the pressure P and volume V of
the gas
\begin{equation}
PV^\gamma = const.
\end{equation}
Because $V \propto R^3$ for the radius $R$, the pressure is $P
\propto R^{-5}$. Therefore the pressure inside the sphere at time $t_i >
t_{rad}$ is
\begin{equation}
P(t_i) = P(t_{rad}) \left( \frac{R_{rad}}{R(t_i)} \right)^5
\end{equation}
If we assume that the external pressure can be ignored, the force on the shell
is
\begin{equation}
F(t_i) = P(t_i) A(t_i) = P(t_i) 4 \pi R(t_i)^2 = 4 \pi P(t_{rad}) R_{rad}^5 R(t_i)^{-3}
\end{equation}
As the shell has swept up most of the gas inside the sphere, its mass at time
$t_i$ is
\begin{equation}
M(t_i) = \frac{4}{3} \pi \rho R(t_i)^3
\end{equation}
The change of the momentum of the shell $\frac{d}{dt}(M \times v)$ is
equal to the force acting on it.

As the momentum scales as $Mv \propto R^3 v$, we have
\begin{equation}
\frac{d}{dt}(M \times v) \propto R^{-3}
\end{equation}
With $v = \frac{dR}{dt} = R'$ and the Ansatz $v \propto R^\nu$ (or
$v = \alpha R^\nu)$, we get
\begin{align}
\frac{d}{dt} (R^3 v) & = (R^3 v)' \\
                     & = (R^3)' v + R^3 v' \\
                     & = 3 R^2 R' v + \alpha R^3 (R^\nu)' \\
                     & = 3 R^2 v^2 + \alpha \nu R^3 R^{\nu-1} R' \\
                     & = 3 \alpha R^{2+\nu} v + \alpha \nu R^{2+\nu} v \\
                     & \propto R^{2+\nu} v \\
                     & \propto R^{2+2\nu}
\end{align}
So we have
\begin{equation}
R^{2+2\nu} \propto R^{-3} \implies 2+2\nu = -3 \implies \nu = -\frac{5}{2}
\end{equation}
Therefore we get $v \propto R^{-5/2}$. Again, as $v =
\frac{dR}{dt}$, we find by integration that \DUrole{citep}{McKee1977}
\begin{equation}
R \propto t^{\frac{2}{7}}
\end{equation}
and from this
\begin{equation}
v \propto t^{-\frac{5}{7}}
\end{equation}
As the radius during the snowplow phase needs to connect continuously to the
Sedov-Taylor phase (e.g. see Fig. \DUrole{ref}{fig:radial-1u-blank}), we get
\begin{equation}
R_{SP}(t) = R_{ST}(t_{rad})\left( \frac{t}{t_{rad}} \right)^{\frac{2}{7}}
\end{equation}
\begin{figure}
\noindent\makebox[\textwidth][c]{\includegraphics[width=1.000\linewidth]{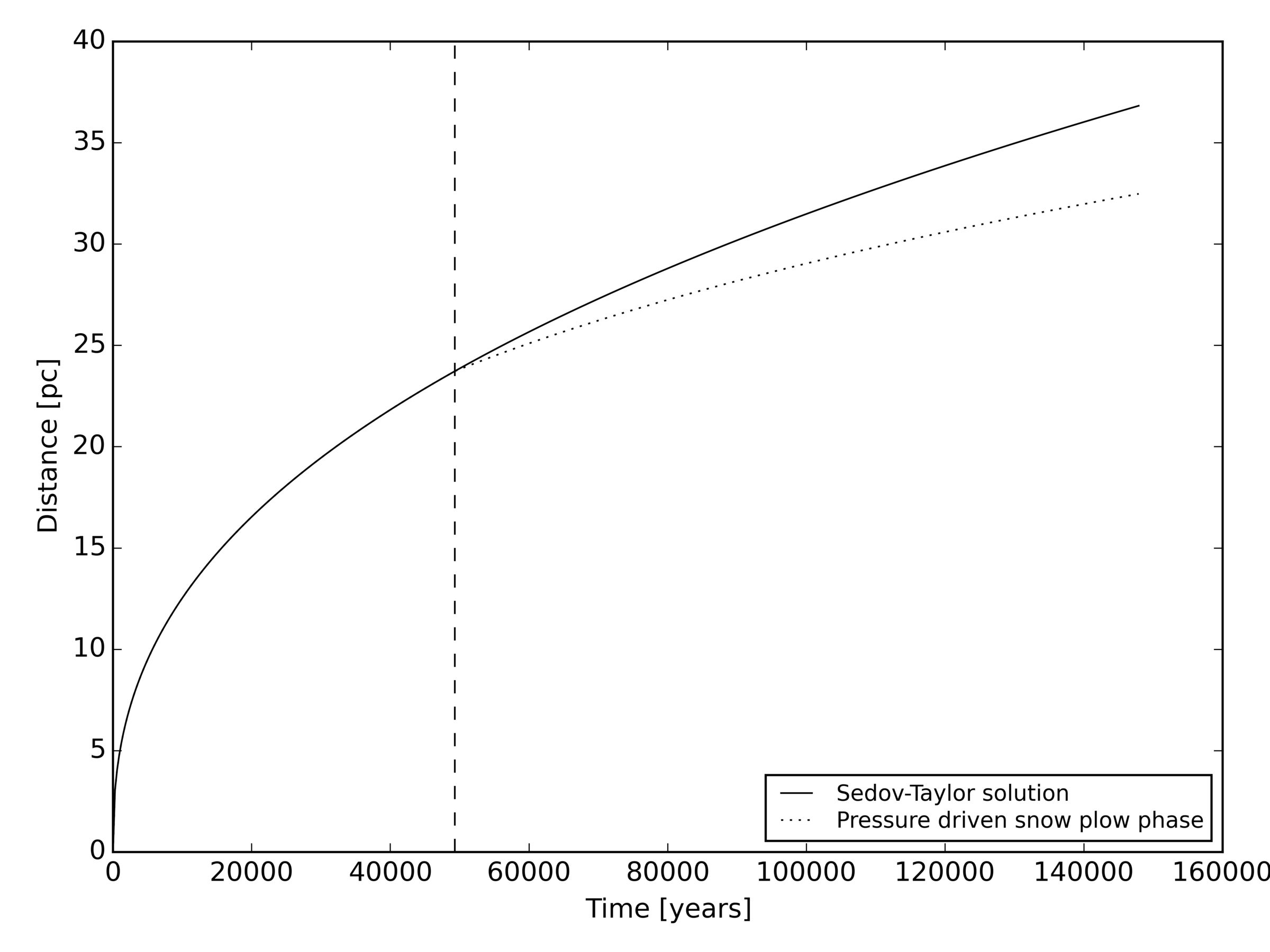}}
\caption{\DUrole{label}{fig:radial-1u-blank}
Radius of the supernova shell during the Sedov-Taylor and pressure driven
snowplow phase. The background density is $\SI{1}{u/cm^3}$ and
injection energy is $\SI{e51}{erg}$. The vertical dashed line marks the
beginning of the radiative phase, where the Sedov-Taylor phase ends.}
\end{figure}

When the internal pressure drops low enough it doesn’t contribute much to the
momentum of the shell anymore. Now, as the momentum is approximately conserved,
but the shell still sweeps up more gas, the mass increases $M \propto R^3$
and therefore $v \propto R^{-3}$.

Again we integrate and find
\begin{equation}
R \propto t^{\frac{1}{4}}
\end{equation}
and
\begin{equation}
v \propto t^{-\frac{3}{4}}
\end{equation}
Finally, when the velocity drops to the ambient sound speed, the shock turns
into a sound wave.

\section{Implementation of supernova feedback in astrophysical simulations%
  \label{implementation-of-supernova-feedback-in-astrophysical-simulations}%
}

One of the most common approaches to implement SN feedback in recent
astrophysical simulations is by injection of a certain amount of energy
(typically $\SI{e44}{J}$ for the most energetic
supernova classes) into the gas in a spherical region inside the simulation
domain, often with the gas equally spread over the SN particles or cells,
approximating the situation just before the Sedov-Taylor phase of a point
explosion. Less energetic SN types are often ignored. The size of the supernova
remnant is often chosen such, that a certain target mass is contained inside the
sphere (see below in section \DUrole{ref}{recent-works}).

SN events are often either placed randomly in space and time, or clustered, e.g.
in star-forming regions. This mimics the observation, of a fraction of
supernovas occurring in a seemingly uncorrelated fashion, and others occurring
close by, forming large superbubbles (see also \DUrole{ref}{recent-works}).

Supernova feedback treatment that only inject a fixed amount of thermal energy
have problems generating enough momentum in under-resolved regions. Consequently
the snowplow phase of the SNR has qualitatively different behaviour than found
in high resolution simulations (i.e. \DUrole{citet}{Simpson2014}). When resolution is
lacking, the injected energy is spread over a large amount of mass, leading to a
too low mean temperature of the gas in the designated SN cell(s). As the cooling
time of gas at $\sim \SI{e4}{K}$ is much lower than around $\sim
\SI{e6}{K}$, the gas in the SN remnant cools before enough of the thermal energy
is converted to kinetic energy, leading to “overcooling”. Adding even a small
amount of the supernova energy as kinetic energy seems to alleviate this problem
(\DUrole{citet}{Simpson2014,Martizzi2014,Gatto2014}), and seems more appropriate than
disabling cooling for a time after a supernova occurs.

\section{Recent works%
  \label{recent-works}%
}

In this section, we give a brief overview of some recent work implementing
supernova feedback in large-scale simulations of the ISM.

\DUrole{citet}{Joung2006, Joung2009} implement SN feedback in \DUrole{sw}{FLASH}, a
multi-scale, multi-physics adaptive mesh refinement hydrodynamic simulation
software (e.g.  see \DUrole{citet}{Fryxell2000,Dubey2008,Dubey2013}) for studying the “Turbulent structure of a stratified
supernova-driven interstellar medium”. The mass in a spherical region containing
$\SI{60}{M_\odot}$ is smoothed out and thermal energy is added. The
clustering of SN is simulated by accounting for superbubbles. Each superbubble
subregion has multiple SN occurring at the same location over its lifetime. The
simulation include diffuse heating and cooling dependent on the local
metallicity, but not thermal conduction, self-gravity and explicit treatment of
star formation. The star forming rate is estimated by a “modified Jeans
criterion”.

\DUrole{citet}{Hill2012} use the supernova feedback model of \DUrole{citet}{Joung2006,
Joung2009} but with a “stable, positivity-preserving scheme for ideal MHD”.
Effects of self-gravity, thermal conductivity, cosmic rays and the UV field are
not included. They find that “the qualitative structure of the ISM is similar in
magnetized and unmagnetized models”.

\DUrole{citet}{Simpson2014} explore the importance of adding kinetic energy in a SN
feedback scheme. Simulations are created with the AMR code \DUrole{sw}{Enzo}
\DUrole{citep}{Bryan2014}. Momentum is applied to the SNR gas cells via a 3D, 3x3x3
cell virtual cloud stencil; mass, metallicity and thermal energy, as they are
scalar quantities, via a single cube stencil. After the centre of the SN is
chosen, mass, metallicity and energy is applied according to the volume overlap
with the stencil. The amount of momentum added to each cell, corresponding to
the desired fraction of the SN energy in kinetic energy, is more complicated,
but can be solved analytically, so that the total momentum added vanishes.
Simulations of varying resolution and different amount of kinetic energy are run
to test the scheme in uniform density media. The feedback scheme is also used
“to model the formation of a low-mass dwarf system in a cosmological setting”.
Even a low amount of kinetic energy leads to different behaviour of the system
compared to simulation runs with just thermal energy injection.

\DUrole{citet}{Gatto2014} look at the effect of different kinds of supernova driving on
the ISM and different initial densities. The simulations are also created with
\DUrole{sw}{FLASH}. For regions that can resolve the Sedov-Taylor phase energy is
injected purely as thermal energy. In unresolved regions the cells are heated to
$\SI{e4}{K}$ and an amount of momentum is added, calculated from the end
of the Sedov-Taylor phase from the mean density of the injection region. The
supernova rate is derived from the Kennicutt-Schmidt relation
\DUrole{citep}{Kennicutt1998} for the initial density, and simulations varying the rate
by a factor of two are compared. SN are either placed randomly, at the global
density maximum, or with a fixed ratio of supernovas at the peak and
random places.  The effect of varying this fraction is also explored.

\DUrole{citet}{Walch2014} implement feedback from supernovas for the SILCC project
(SImulating the Life-Cycle of molecular Clouds) which “aims at a more self
consistent understanding of the interstellar medium (ISM) on small scales and
its link to galaxy evolution” \DUrole{citep}{Walch2014}. The physical model includes
“an external galactic potential, self-gravity, radiative heating and cooling
coupled to, a chemical network to follow the formation of \DUrole{latex}{\htwo} and
\DUrole{latex}{\co}, diffuse heating and its attenuation by dust shielding, and
feedback from supernova explosions of massive stars”. The method is similar to
\DUrole{citet}{Gatto2014}, and abundances of \DUrole{latex}{\htwo} is reduced “assuming that
all \DUrole{latex}{\htwo} is dissociated by the SN”, increasing the \DUrole{latex}{\hp}
abundances accordingly. Different schemes for choosing the locations of the
supernovas in space and time are studied, finding that “Only random or clustered
(in space and time) models with self-gravity (…) are in agreement with
observations”.

\DUrole{citet}{Hopkins2014} include stellar feedback in their heavily modified version
of \DUrole{sw}{GADGET 3} (see \DUrole{citet}{Springel2005b} for the previous version
\DUrole{sw}{GADGET 2}) for the FIRE (Feedback In Realistic Environments) project
(extending on \DUrole{citet}{Hopkins2011}, \DUrole{citet}{Hopkins2012} and
\DUrole{citet}{Hopkins2012a}). This modified version - P-SPH - features an improved SPH
“‘pressure-entropy formulation’ which resolves problems in the traditional
treatment of fluid interfaces” \DUrole{citep}{Hopkins2013}.  The stellar feedback model
includes probabilistic star formation, radiation pressure, photoionization,
photoelectric heating, stellar winds and supernovas.  “Stellar winds are
algorithmically nearly identical to SN, except they occur continuously”
\DUrole{citep}{Hopkins2014}.  The SN model injects mass, thermal energy, metallicity
and momentum into the gas, with a larger fraction of energy in momentum, when
the injection site is poorly resolved “accounting for the possibility of an
unresolved Sedov-Taylor phase” \DUrole{citep}{Hopkins2014}. The supernova rate is
directly coupled to the simulated star formation.

\DUrole{citet}{Martizzi2014} simulate the evolution of single supernovas in
inhomogeneous initial conditions with the AMR hydro solver \DUrole{sw}{RAMSES}
\DUrole{citep}{Teyssier2002}. From those simulation a sub-scale model for supernova
feedback is derived and tested for periodic boxes with multiple SNRs. A fixed
amount of energy is injected as kinetic and thermal energy, and mass is added to
the SNR region, a sphere with radius of seven times the cell size and a
metallicity dependent cooling function is used for cooling.

\DUrole{citet}{Torrey2014} implement a AGN (Active Galaxy Nucleus) feedback model in
\DUrole{sw}{Arepo}. Mass loss from stellar processes is modelled via “a stochastic
process” that is used “to launch wind particles out of star-forming regions”.

Our scheme is similar to the method of \DUrole{citet}{Hill2012} with random driving.

\section{Overview%
  \label{overview}%
}

In the rest of the thesis we first describe our methods in chapter
\DUrole{ref}{methods}. The supernova feedback method we implement for \DUrole{sw}{Arepo} is
described in section \DUrole{ref}{our-implementation}. Then follows an overview of the
software setup used to run the simulations, plot results and generate this
document in section \DUrole{ref}{software}. In section \DUrole{ref}{plots} we specify in more
detail how the plots were created.

The actual simulations run for testing the implementation of our scheme are
discussed in section \DUrole{ref}{simulations}. In chapter
\DUrole{ref}{discussion-of-results} we discuss their outcome.

Finally give a summary in chapter \DUrole{ref}{conclusion} and an outlook
for possible following work in section \DUrole{ref}{further-work}.

\chapter{Methods%
  \label{methods}%
}

\section{Our implementation%
  \label{our-implementation}%
}

SN event times are assumed to be Poisson distributed with the SN rate
$\lambda$ or its inverse, the SN period. In the current implementation
the SN rate stays constant, but could be varied.

The arrival time, i. e. the time between events, follows the exponential
distribution, with the mean time between events $\mu$. The distribution is
defined by the cumulative distribution function (CDF) \DUrole{citep}{Knuth1998}
\begin{equation}
F(x) = \begin{cases}
1 - e^{-x/\mu} & x \ge 0 \\
0 & x < 0
\end{cases}
\end{equation}
Sampling from a uniform distribution U on the interval $[0, 1)$ (i.e. the
range of the GSL function \DUrole{function}{gsl\_rng\_uniform}), a exponential
distributed random variable X can be calculated: \DUrole{citep}{Knuth1998}
\begin{equation}
X = -\mu \ln \left( 1 - U \right)
\end{equation}
The SN code is called once every timestep. Only if the current simulation time
is later than the time calculated for the next event a supernova is created.
The time for the next event is updated after each spawned supernova to the sum
of the current simulation time and an exponential distributed random variable
with $\mu = \lambda^{-1}$. The supernova rate is therefore generally lower than
specified, as the timestep sizes are discrete. As in the current implementation
only one supernova can occur per timestep, a large timestep can suppress a
number of events that should happen. To improve accuracy the maximum timestep
size should be smaller than the SN period.

The SN period is specified with the runtime parameter \emph{SNEPeriodInYears} in
units of years.

The location of SN events are sampled uniformly in the simulation domain.  This
could be improved by spawning a subset of the SN only in dense or star forming
regions, e.g. at the position of so-called sink particles \DUrole{citep}{Bate1995}.

The state of the “Pseudo-random number generator” (PRNG)\DUfootnotemark{id1}{id2}{1} is saved in the
restart files and restored after restarting from those. When restarting from
snapshot files, the PRNG state is not available, so it is (re-)initialized from
the seed.  Setting a different one is advisable for each subsequent restart from
a snapshot to avoid setting of SN at the same location twice, as reusing a seed
causes the same sequence of random numbers and possibly the same sequence of
supernova parameters.
\DUfootnotetext{id2}{id1}{1}{%
A PRNG doesn’t provide true random values, but
generates numbers deterministically, but numbers generated have certain
statistical properties.
}

SN are approximated as point explosions, starting in the Taylor-Sedov phase.
Voronoi cells with their generating point positioned in a sphere around the SN
are determined, where the SN remnant radius is iteratively increased by
$\Delta R = \SI{0.01}{pc}$ until the sphere includes a total mass of at
least $M_{target}$ (specified with the runtime parameter \emph{SNETargetMass})
and at least $N_{target}$ cells (specified with the runtime parameter
\emph{SNEMinimalParticleNumber}). For that on each process the mass, the volume
(which is needed later for the calculation of the mean density of the SN remnant
cells) and number of gas particles in the sphere is calculated, summed up and
sent to all processes with \DUrole{function}{MPI\_Allreduce}.

When run with periodic boundary conditions, SN occurrences near the domain
boundary could include particles on the other side of the domain. To simplify
the implementation, such SN occurrences are ignored, and a new SN location
chosen. A correct implementation, allowing such overlaps, could be considered
for further work.

The density of the cells in the SN remnant sphere is set to the mean density of
the cells in the SN sphere and each cell receives a corresponding amount
$E_{cell} = E_{SN} \frac{V_{cell}}{V_{SN}}$ of the SN energy
$E_{SN}$ (where $V_{SN}$ is the sum of the volume of the cells
inside the SN sphere, as determined above), which is added to the internal
energy \emph{UTherm}. After that the entropy is recalculated when needed.  In
principle additional mass, metallicity and a fraction of the energy as momentum
can be added at this point as well, though this isn’t implemented at the moment.

$M_{target}$ should be chosen large enough to avoid overpressurizing the
remnant, but small enough so that the remnant is hot enough to avoid
overcooling.  A minimum cell count $N_{target}$ ensures that in regions of
high density, where only one or a few cells are enough to reach the target mass,
the energy is still spread over multiple cells.

The iterative approach to find a suitable remnant radius could be improved by a
more refined search algorithm, e.g. making use of the nearest neighbour search
tree, but the current runtime is negligible even for more than a few million
cells. The choice of $\Delta R = \SI{0.01}{pc}$ is arbitrary, but should
ensure a small error in the optimal number of cells.

This scheme is similar to the method of \DUrole{citet}{Hill2012} with random driving.

\section{Software%
  \label{software}%
}

In this chapter we discuss the software setup that was used in the creation of
this thesis.

\subsection{Simulation setup%
  \label{simulation-setup}%
}

A simulation run is split into several parts:
\begin{itemize}

\item Pre-Processing
\begin{itemize}

\item Generating configuration files from templates via the \DUrole{file}{mktmpl.py} script

\item Generating the initial condition file and compiling \DUrole{sw}{Arepo}

\end{itemize}

\item Running \DUrole{sw}{Arepo} either locally or via a job submission script on a
computing cluster

\item Post-Processing
\begin{itemize}

\item Generating a cache file with aggregated values

\item Generating plots

\end{itemize}

\end{itemize}

The different steps are mediated via shell and \DUrole{sw}{Python} scripts.

\DUrole{sw}{Arepo} is mainly written in standard C, with some additional parts in
Fortran. \DUrole{sw}{Make} is used as the build system. It can be compiled with quite a few
different compile time options specified in a separate file. C preprocessor
definitions are used to enable or disable complete sections of code, with some
of the options excluding others.

The Makefile of \DUrole{sw}{Arepo} also specifies linker and compiler options, which can
be adjusted for the machine and computing cluster the software is run on via the
SYSTYPE option. As the resulting binary is therefore highly dependent on the
compile time options, it makes sense to consider the compile and Makefile flags
as part of the simulation configuration.

Runtime options are specified in a parameter file. An initial description of the
problem is provided in a so-called initial condition file (or IC file). The
formats supported are the same as in \DUrole{sw}{GADGET 2}.

The \DUrole{sw}{Python} package \DUrole{sw}{jelly}, written mainly by Eduard Bopp for his master
thesis \DUrole{citep}{Bopp2014}, handles the compilation of Arepo, generation of
initial conditions and (optionally) the execution of the simulation. As most of
the simulations are run on a distributed cluster, which defers execution to a
separate scheduler, the later feature was mostly not used for this thesis.

In the post-processing step, snapshot data and projections of the data generated
by Arepo are read in and several plots are generated using \DUrole{sw}{Python} and the
\DUrole{sw}{matplotlib} plotting package. The projection files are read in via a
routine in \DUrole{file}{utils.py}. For reading and parsing the snapshot data
\DUrole{sw}{SOAPP}, a module created by Eduard Bopp, Jon Ramsey and Mei Sasaki, is
used. Data is mainly stored in \DUrole{sw}{numpy} arrays internally, as native array
operations are to slow for larger datasets.

To reduce the chance of errors in unit conversion the \DUrole{sw}{Python} package
\DUrole{sw}{Pint} is used, which makes available a consistent unit system for
\DUrole{sw}{Python}, support conversion between different units and keeps track of
units in calculations. It also interacts nicely with \DUrole{sw}{numpy}. As \DUrole{sw}{SOAPP}
and \DUrole{sw}{Jelly} don’t have support for \DUrole{sw}{Pint} yet, but expect and read in
native units, \DUrole{sw}{Pint} units are applied manually after reading in or
converted to native \DUrole{sw}{Arepo} units before writing data.

Configuration files and scripts for different simulations have large parts in
common. Only a few parameters vary and are sometimes used in several places. To
make use of this fact, config files and scripts are generated from a common
template. The templates reside in the \emph{templates} directory. The \DUrole{sw}{Python}
package \DUrole{sw}{Jinja2} is used as the templating engine, as it is flexible, fast
and widely used.

The \DUrole{file}{mktmpl.py} script (\DUrole{ref}{code:mktmpl-py}) applies the templating
parameters specified in YAML (“YAML Ain’t Markup Language”) input files (where
later specified files overwrite parameters in earlier ones) using Jinja2 from
the \DUrole{file}{templates} directory and writes them to an output directory.

From there the simulations are prepared using the \DUrole{file}{prepare.sh} script
(\DUrole{ref}{code:prepare-sh}), calling mainly \DUrole{file}{sim.py} (\DUrole{ref}{code:sim-py}),
which sets up a randomly placed distribution of mesh generating points, and
optionally adds energy for simulating a supernova explosion for the later
described test case (see \DUrole{ref}{simulations}),  and also compiles Arepo
with the given configuration in \DUrole{file}{Config.sh} (\DUrole{ref}{code:Config-sh}). An
injection radius of 0 forces the injection into a single cell.

\DUrole{sw}{Arepo} is compiled for each simulation independently, as the binary is
highly dependent on the simulation configuration. It is also later archived, as
e.g.  the locally installed libraries and the compiler, also influence the final
binary.

The simulation itself can be run via one of the run scripts, or via the
\DUrole{file}{run\_mpi.sh} script (\DUrole{ref}{code:run-mpi-sh}) can be submitted to
a job scheduler on a computing cluster (the template for this file can be adjusted for different
cluster configurations, e.g. for the Milkyway cluster).

Arepo generates several different files in the \DUrole{file}{output} directory,
snapshot files, which contain the physical data of the grid points, projection
and slice images, restart files for continuing a run, and several other files,
mainly for diagnostic purposes. Arepo also outputs log data to the terminal,
which is written to a logfile using the \DUrole{sw}{tee} utility.

The image and snap shot files are processed first by the
\DUrole{file}{cache\_quantities.py} script (\DUrole{ref}{code:cache-quantities-py}), writing
aggregated data, (the mean and average values, their standard deviation, etc.)
for different particle data fields, to the \emph{quantities.yaml} file, to speed up
the plot generation.  After that plots are generated using \DUrole{file}{plots.sh}
(\DUrole{ref}{code:plots-sh}) and several plotting \DUrole{sw}{Python} scripts.

A detailed list of files in the project, including the source code, can be found
in the Appendix \DUrole{ref}{appendixsource}.

\subsection{Other%
  \label{other}%
}

Final simulation runs are compressed with \DUrole{sw}{bzip2}, \DUrole{sw}{gzip} or \DUrole{sw}{lzma}
in \DUrole{sw}{tar} files.

Some packages, that are not commonly available on most machines, were compiled
manually. Those include \DUrole{sw}{ccache}, a C object file cache to reduce
compilation time, \DUrole{sw}{GSL}, the GNU Scientific Library, \DUrole{sw}{hwloc},
\DUrole{sw}{openssl}, \DUrole{sw}{pv}, and \DUrole{sw}{Python}.

Especially the \DUrole{sw}{Python} packages were installed manually, as system packages often
contain old versions. A virtual environment was set up using \DUrole{sw}{virtualenv}
and packages were installed mainly \DUrole{sw}{pip} (or \DUrole{sw}{easy\_setup} for packages
that don’t support \DUrole{sw}{pip}).

Most of the code written for this thesis was edited with the \DUrole{sw}{vim} text
editor on Debian and Arch Linux. For editing \DUrole{sw}{Python} code also the (partly
proprietary) editor \DUrole{sw}{PyCharm}, and for C code the editor \DUrole{sw}{Eclipse} was
used.

Doctests, unit tests directly attached to the implementation,
for some of the \DUrole{sw}{Python} modules are included. They can be run with \DUrole{sw}{nosetests}.

The code was managed in \DUrole{sw}{Git} repositories, citations with \DUrole{sw}{Zotero}.

This thesis itself is written in ReSTructuredText, a simple markup language of
the \DUrole{sw}{docutils} package. It is transformed with the \DUrole{sw}{Python} package
\DUrole{sw}{docutils} to \DUrole{latex}{\LaTeX}, which than is compiled to a PDF file with
\DUrole{latex}{\XeLaTeX} of the \DUrole{latex}{\XeTeX} fork of \DUrole{latex}{\TeX}, which has
support for Unicode and modern fonts. The arXiv version uses \DUrole{sw}{pdflatex}
instead.

\DUrole{sw}{Scons} is used as the build system.

\section{Plots%
  \label{plots}%
}

The plots in this thesis are generated with \DUrole{sw}{matplotlib}. Post-processing
for the arXiv version of this paper uses \DUrole{sw}{GNU parallel} \DUrole{citep}{Tange2011a}.

\subsection{Slices and projection plots%
  \label{slices-and-projection-plots}%
}

Arepo can calculate an image of the simulation domain, either a slice through
the domain, or integrating the simulated quantifies along an axis via ray
tracing, weighted by the density of the intersected cells. The gradient estimate
between Voronoi cells can be optionally included.  The image files are stored as
a flatted array of the pixel values in the unformatted binary convention of
Fortran, similarly to blocks in the snapshot format 1 and 2 of \DUrole{sw}{GADGET} and
\DUrole{sw}{Arepo} \DUrole{citep}{Springel2005}. Arepo supports the creation of plots from
snapshot files, but it is more efficient to create them during the simulation,
when the data is in memory and the non-trivial Voronoi tessellation is already
done.

We use these images as the basis of slice and projection plots. Both are
generated with the \emph{plot.py} script (see \DUrole{ref}{code:plot-py}). All plots are
parallel to the xy-plane with the gradient disabled. Slices through the domain
have $z = h/2$, where $h$ is the box size, i.e. the slices are in
the midplane.

Likely due to a bug in \DUrole{sw}{Arepo} the first pixel of the image files contain
invalid data. Instead the second pixel value is used for it.

\subsection{Histograms%
  \label{histograms}%
}

Histograms for various quantities are created with the \DUrole{file}{plot.py} script
(\DUrole{ref}{code:plot-py}). The bins have logarithmically constant size. Plotted is
the mass-weighted probability function (PDF) normalized to the total mass in the
box, including sinks particles.

\subsection{Radius of supernovae%
  \label{radius-of-supernovae}%
}

A combined plot of the supernovae radii at each snapshot time is created with
the \DUrole{file}{plot\_max.py} script (\DUrole{ref}{code:plot-max-py}). The cells are sorted by
density and radial distance of the grid generating points to the centre of the
simulation domain is averaged for the ten heaviest cells to estimate the
supernova radius. To correct for the initial size of the SN remnant, i.e. the
injection radius, the plotted time is the simulation time plus the time where
the Sedov-Taylor solution reaches the injection radius, i.e. the curves are
shifted to the right.

The analytic Sedov-Taylor solution for the shock front radius and the analytic
momentum conserving solution for the snowplow phase after the time
$t_{rad}$ (see \DUrole{ref}{the-sedov-taylor-blast-wave-solution}) is also
plotted.

The standard deviation of the averaged values is added as lighter coloured areas
in the backgrounds as an estimation of the error.

\subsection{Velocity dispersion%
  \label{velocity-dispersion}%
}

The mass-weighted velocity dispersion for each coordinate $k$ and
different chemical species $c$ is
\begin{equation}
\sigma_{k,c} = \sqrt\frac{\sum\limits_i (v_{i,k} - \bar{v_k})^2
               \times m_{i,c}}{\sum\limits_i m_{i,c}}
\end{equation}
where $m_{i,c}$ is the mass, $v_{i,k}$ is the k-th coordinate of the
velocity of the gas particle $i$ and the k-th coordinate of the
component-wise mean velocity is $\bar{v_k}$. Sink particles are ignored.
The mass in a chemical species is determined the product of the abundance and
the mass in the cell.

The velocity dispersions for each cardinal direction are averaged to derive an
average 1d mass-weighted velocity dispersion, similar to \DUrole{citet}{Gatto2014}.

\section{Simulation on Milkyway%
  \label{simulation-on-milkyway}%
}

Simulations were mainly run on the supercomputing MilkyWay cluster (which is
supported by the DFG via SFB 881 (sub-project Z2) and SPP 1573) , an extension
of the cluster JUDGE in Jülich. For running simulations a job scheduler is used,
with each user of the cluster registering requests for running a MPI program
with a certain amount of resources (CPUs, machines and RAM) needed. A job
scheduler runs as many jobs in parallel as possible under these constrains.

With the cluster used by several research groups, a waiting time of a few hours
to several days were customary at the time of writing for a medium number of
cores (about 30-200).

Development of the simulation setup and plotting scripts was done in parallel
with running the simulations, results iteratively informing software changes.

The configurations for the different simulations are included in the
configuration files in Appendix \DUrole{ref}{appendixsource}.

\section{Simulations%
  \label{simulations}%
}

As the implementation is mainly intended for use with cooling and chemistry, we
focus on testing the interaction with SGChem
\DUrole{citep}{Glover2007a,Glover2007,Glover2012a,Glover2012,Smith2014}, which provides
cooling and chemical network modelling of \DUrole{latex}{\h, \hp, \htwo} and \DUrole{latex}{\co}.
Photochemistry uses Treecol \DUrole{citep}{Clark2012} for radiative transfer.

Additional heating is provided by an UV Field of
$\num{1.7}\times\SI{1.2e-4}{erg.cm^{-2}.s^{-1}.sr^{-1}}$ or 1.7
\DUrole{citet}{Habing1968} units.  The cosmic ray ionization rate of H I is set to
$\SI{3e-17}{s^{-1}}$.

The \DUrole{sw}{Python} script \emph{sim.py} (see \DUrole{ref}{code:sim-py}) creates the initial
conditions file. It also compiles Arepo at the same time. The \DUrole{sw}{Python}
package \DUrole{sw}{Jelly} provides both functionalities and can also run Arepo on a
single machine. As the simulations were mainly run on a supercomputing cluster
\DUrole{file}{sim.py} doesn’t use this feature.

The positions of the mesh generating points for the initial conditions are
randomly sampled, with the coordinates each drawn independently from the
pseudo-random number generator (PRNG) “Mersenne Twister” \DUrole{citep}{Matsumoto1998},
which is the standard PRNG of the random module of recent \DUrole{sw}{Python} versions
\DUrole{citep}{Pythondocumentation}. Many PRNGs have a so-called “seed” that is used to
initialize its internal state. Two invocations with the same seed and the same
number already of values already drawn, generate the same output.  This property
is useful for the generation of ICs, as they can be later reproduced by simply
running the script again, when the seed is set before drawing the first random
number to a known value. It can be specified via the \emph{–seed} parameter for
\DUrole{sw}{sim.py}, and is fixed when invoked by the preparation script \emph{prepare.sh}
(see \DUrole{ref}{code:prepare-sh}).

The simulations are run in a so-called periodic box, where only a small volume
is represented, which repeats periodically and infinitely in each cardinal
direction. The simulation domain is a cube of given size, where the length of
its side is the \emph{box size}. Particles moving out of the domain are moved to the
opposite side of the simulation domain, and particles on the boundary interact
hydrodynamically with so-called mirror particles, that duplicate the properties
of particles on the other side of the box. As the gravitational forces are more
far reaching than the hydrodynamical interactions between particles, the
contributions of neighbouring virtual images of the simulation domain must be
factored in. The so called “Ewald correction” \DUrole{citep}{Hernquist1991} provides
correction terms for the gravitational forces.

This makes it possible to only simulate a representative volume of a larger or
even infinite system, that ideally behaves isotropically or homogeneously on larger
scales, e.g. a star forming region in a galaxy, or a galaxy cluster in a
cosmological simulation.

The density and internal energy of all particles are set to the same value. The
initial chemical abundances of the particles are set by the parameter file. This
all in all provides an initially homogeneous density medium on a random,
unrelaxed mesh. Optionally particles laying in a sphere around the centre of the
simulation domain can be given a different internal energy amount, i.e. can be
set to a different temperature than the background. In single SN runs
self-gravity of gas cells is disabled, in other runs it is enabled.

The initial background temperature is set to approximate equilibrium
temperatures for the given densities and chemical abundances, which are taken
from simulation runs with feedback turned off.

\subsection{Isolated supernovae%
  \label{isolated-supernovae}%
}

As described earlier, instead of injecting the supernova energy into a single
cell we spread it over a larger spherical region.

Runs with injection radii of $\SI{0.5}{pc}$, $\SI{1}{pc}$,
$\SI{2}{pc}$, $\SI{5}{pc}$, $\SI{7.5}{pc}$ and
$\SI{10}{pc}$ were done, each for background densities of
$\SI{1}{m_{H} / cm^3}$ and $\SI{100}{m_{H} / cm^3}$ and with and
without chemistry and cooling enabled. The runs included 2 million cells,
without refinement. Additionally runs with energy injected into a single cell
are included for comparison.

The energy injection was done during initial condition creation, with the SN
feedback code disabled, to verify our approach is valid regardless of
implementation details. Similarly to the IC creation in the Feedback section
\DUrole{ref}{feedback} the particle positions are chosen randomly from a uniform
distribution.

Arepo expects the specific energy (energy per unit mass) in the IC input for the
internal energy. As we only know the initial density of the cells, the volume
would need to be calculated from the Voronoi tessellation. As the output of the
Voronoi tessellation library \DUrole{sw}{Qhull} in the \DUrole{sw}{Python} library \DUrole{sw}{scipy}
doesn’t include the volumes of individual cells and no other library for this
use case is known to the author, a simpler approach to estimate the volume was
chosen. Instead of calculating the aggregate volume of the cells with the cell
centre inside the injection sphere, the actual volume of the sphere was used to
calculate the specific energy. This provedto be problematic as it leads to a
too high amount of energy injected (see chapter \DUrole{ref}{discussion-of-results} for a discussion).

This approach is slightly different from that implemented for the supernova
feedback mechanism in Arepo as described in section \DUrole{ref}{our-implementation},
as the energy density is not necessarily constant. Also an error in the
amount of energy in the supernova is introduced, as the volume of the injection
sphere is generally not the volume of the cells inside this region. For
injection radii of $> \SI{1}{pc}$ in the cases below, or in general when
the number of particles inside the supernova is large enough, this error seems
to be small, although calculating the supernova energy after importing in Arepo
might be useful to see the difference with this approach more accurately.

For the single cell injection, the volume $V_{SN}$ was calculated as
$V_{SN} = \frac{V_{box}}{n}$, from the box volume $V_{box}$ and the
particle number $n$. This again is different from the approach used in the
actual feedback code, but seems to be a better estimate for a small number of
particles than the approach above (See discussion below in section
\DUrole{ref}{single-supernovae}).

Actually doing the Voronoi tessellation would lead to results more similar to
the actual supernova feedback implementation in Arepo. Lacking time, no reruns
of the affected simulations was possible, limiting the significance of these
simulations.

\subsection{Feedback%
  \label{feedback}%
}

For testing the feedback model a simulation of a periodic box with $L =
\SI{400}{pc}$ per side and initially homogeneous density of $\rho _{1} =
\SI{1}{m_{H}/cm^3}$ - a typical value found in the WNM and WIM - was run.\DUfootnotemark{id3}{id4}{2}
\DUfootnotetext{id4}{id3}{2}{%
Simulations with higher initial densities crashed, likely
caused by problems in the sinks implementation at the time of writing.
}

Similar to the approach of \DUrole{citet}{Gatto2014}, we estimate the supernova rate from
the star formation rate using the Kennicutt-Schmidt relation
\DUrole{citep}{Kennicutt1998}. It relates the typical star formation rate surface density
to the total surface gas density $\Sigma _{Gas} \approx L \times \mean{\rho }$,
where $\mean{\rho }$ is the mean density in the box.
\begin{align}
\Sigma _{SFR} & = \num{2.5 +- 0.7} \times 10^{-10} \times \left( \frac{\Sigma _{gas}}{\si{M_{\odot}.pc^{-2}}} \right)^{\num{1.4 +- 0.15}} \times \si{M_{\odot}.yr^{-1}.pc^{-2}} \\
& \approx \num{2.5} \times 10^{-10} \times \left( \frac{\Sigma _{gas}}{\si{M_{\odot}.pc^{-2}}} \right)^{\num{1.4}} \times \si{M_{\odot}.yr^{-1}.pc^{-2}}
\end{align}
From about $\SI{100}{M_\odot}$ of gas one heavy star forms, that will
explode as a type II supernova \DUrole{citep}{Chabrier2003a,Gatto2014}. If we ignore the
contribution from Type Ia supernovas (and other less energetic SN types), we get
a supernova rate of
\begin{equation}
SNR \approx \frac{\Sigma _{SFR} \times L^2}{\SI{100}{M_{\odot}}}
\end{equation}
For the given box size L and the density $\rho _{1}$ we
get $SNR(\rho _{1}) \approx \SI{9.78}{Myr^{-1}}$.

The mean time between supernova explosions, $SNP = 1 / SFR$, must be
specified in the parameter file. For our density the period between SN events is
\begin{equation}
SNP(\rho _{1}) \approx \SI{0.10}{Myr}
\end{equation}
As \DUrole{sw}{SGChem} can’t handle nonphysically large densities, the creation of sinks was
enabled, which effectively creates an upper bound for the density during the
runs, as cells of large enough mass are converted to sink particles. A
refinement and derefinement criterion for a target mass of approximately
$\SI{1.0}{M_\odot}$ was applied. The size of the SNR was mainly determined
by the minimal particle number of six, with the mass target of only
$\approx \SI{1.6}{M_\odot}$, to stay near the resolution of the
simulation, a conservative choice, as the results of section
\DUrole{ref}{isolated-supernovae} are of limited usefulness.

Sink particles were created when a number density of $\SI{2e8}{cm^{-3}}$ was
reached. The accretion radius of the sink particles was set to 0.3 internal
length units or approximately $\SI{9.7e-3}{pc}$. This corresponds to a
region containing about $\SI{18.7}{M_\odot}$, or about 10-20 particles at the
resolution where sink creation was happening, i.e. the sink particles were of an
effective size comparable to the local mesh resolution.

\clearpage

\setlength{\DUtablewidth}{\linewidth}
\begin{longtable}[c]{|p{0.406\DUtablewidth}|p{0.113\DUtablewidth}|p{0.113\DUtablewidth}|p{0.162\DUtablewidth}|p{0.162\DUtablewidth}|}
\caption{Simulation parameters for single SN}\\
\hline

Name
 & 
SN Radius
 & 
SN Particles
 & 
Chemistry?
 & 
Density
 \\
\hline

single\_radius\_0
 &  & 
1
 & 
yes
 & 
\DUrole{latex}{\SI{1}{u/cm^3}}
 \\
\hline

single\_radius\_0.5
 & 
0.5 pc
 & 
2
 & 
yes
 & 
\DUrole{latex}{\SI{1}{u/cm^3}}
 \\
\hline

single\_radius\_1
 & 
1.0 pc
 & 
4
 & 
yes
 & 
\DUrole{latex}{\SI{1}{u/cm^3}}
 \\
\hline

single\_radius\_2
 & 
2.0 pc
 & 
27
 & 
yes
 & 
\DUrole{latex}{\SI{1}{u/cm^3}}
 \\
\hline

single\_radius\_5
 & 
5.0 pc
 & 
522
 & 
yes
 & 
\DUrole{latex}{\SI{1}{u/cm^3}}
 \\
\hline

single\_radius\_7.5
 & 
7.5 pc
 & 
1761
 & 
yes
 & 
\DUrole{latex}{\SI{1}{u/cm^3}}
 \\
\hline

single\_radius\_10
 & 
10 pc
 & 
4233
 & 
yes
 & 
\DUrole{latex}{\SI{1}{u/cm^3}}
 \\
\hline

single\_radius\_0\_nochem
 &  & 
1
 & 
no
 & 
\DUrole{latex}{\SI{1}{u/cm^3}}
 \\
\hline

single\_radius\_0.5\_nochem
 & 
0.5 pc
 & 
2
 & 
no
 & 
\DUrole{latex}{\SI{1}{u/cm^3}}
 \\
\hline

single\_radius\_1\_nochem
 & 
1.0 pc
 & 
4
 & 
no
 & 
\DUrole{latex}{\SI{1}{u/cm^3}}
 \\
\hline

single\_radius\_2\_nochem
 & 
2.0 pc
 & 
27
 & 
no
 & 
\DUrole{latex}{\SI{1}{u/cm^3}}
 \\
\hline

single\_radius\_5\_nochem
 & 
5.0 pc
 & 
522
 & 
no
 & 
\DUrole{latex}{\SI{1}{u/cm^3}}
 \\
\hline

single\_radius\_7.5\_nochem
 & 
7.5 pc
 & 
1761
 & 
no
 & 
\DUrole{latex}{\SI{1}{u/cm^3}}
 \\
\hline

single\_radius\_10\_nochem
 & 
10 pc
 & 
4233
 & 
no
 & 
\DUrole{latex}{\SI{1}{u/cm^3}}
 \\
\hline

single\_radius\_0\_100u
 &  & 
1
 & 
yes
 & 
\DUrole{latex}{\SI{100}{u/cm^3}}
 \\
\hline

single\_radius\_0.5\_100u
 & 
0.5 pc
 & 
2
 & 
yes
 & 
\DUrole{latex}{\SI{100}{u/cm^3}}
 \\
\hline

single\_radius\_1\_100u
 & 
1.0 pc
 & 
4
 & 
yes
 & 
\DUrole{latex}{\SI{100}{u/cm^3}}
 \\
\hline

single\_radius\_2\_100u
 & 
2.0 pc
 & 
27
 & 
yes
 & 
\DUrole{latex}{\SI{100}{u/cm^3}}
 \\
\hline

single\_radius\_5\_100u
 & 
5.0 pc
 & 
522
 & 
yes
 & 
\DUrole{latex}{\SI{100}{u/cm^3}}
 \\
\hline

single\_radius\_7.5\_100u
 & 
7.5 pc
 & 
1761
 & 
yes
 & 
\DUrole{latex}{\SI{100}{u/cm^3}}
 \\
\hline

single\_radius\_10\_100u
 & 
10 pc
 & 
4233
 & 
yes
 & 
\DUrole{latex}{\SI{100}{u/cm^3}}
 \\
\hline

single\_radius\_0\_nochem\_100u
 &  & 
1
 & 
no
 & 
\DUrole{latex}{\SI{100}{u/cm^3}}
 \\
\hline

single\_radius\_0.5\_nochem\_100u
 & 
0.5 pc
 & 
2
 & 
no
 & 
\DUrole{latex}{\SI{100}{u/cm^3}}
 \\
\hline

single\_radius\_1\_nochem\_100u
 & 
1.0 pc
 & 
4
 & 
no
 & 
\DUrole{latex}{\SI{100}{u/cm^3}}
 \\
\hline

single\_radius\_2\_nochem\_100u
 & 
2.0 pc
 & 
27
 & 
no
 & 
\DUrole{latex}{\SI{100}{u/cm^3}}
 \\
\hline

single\_radius\_5\_nochem\_100u
 & 
5.0 pc
 & 
522
 & 
no
 & 
\DUrole{latex}{\SI{100}{u/cm^3}}
 \\
\hline

single\_radius\_7.5\_nochem\_100u
 & 
7.5 pc
 & 
1761
 & 
no
 & 
\DUrole{latex}{\SI{100}{u/cm^3}}
 \\
\hline

single\_radius\_10\_nochem\_100u
 & 
10 pc
 & 
4233
 & 
no
 & 
\DUrole{latex}{\SI{100}{u/cm^3}}
 \\
\hline
\end{longtable}

\clearpage

\setlength{\DUtablewidth}{\linewidth}
\begin{longtable}[c]{|p{0.406\DUtablewidth}|p{0.113\DUtablewidth}|p{0.113\DUtablewidth}|p{0.162\DUtablewidth}|p{0.162\DUtablewidth}|}
\caption{Simulation parameters for feedback}\\
\hline

Name
 & 
Target mass
 & 
Min. \# of part.
 & 
Chemistry?
 & 
Density
 \\
\hline

single\_radius\_0
 & 
\DUrole{latex}{\SI{1.6}{M_{\odot}}}
 & 
6
 & 
yes
 & 
\DUrole{latex}{\SI{1}{u/cm^3}}
 \\
\hline
\end{longtable}

\chapter{Discussion of results%
  \label{discussion-of-results}%
}

\section{Single supernovae%
  \label{single-supernovae}%
}

Fig. \DUrole{ref}{fig:radial-1u-nochem}, Fig. \DUrole{ref}{fig:radial-1u}, Fig.
\DUrole{ref}{fig:radial-100u-nochem} and Fig. \DUrole{ref}{fig:radial-100u} show the radial
distance of the shock wave from the SN centre vs. time. The remnants are smaller
for an initial uniform density of $\SI{100}{u/cm^3}$ compared to an
density of $\SI{1}{u/cm^3}$ at equal times, as expected from formula
\DUrole{ref}{eq:sn-radius}.

As explained in Section \DUrole{ref}{isolated-supernovae}, these tests were set up by
injecting the energy in the initial conditions, rather than by using the
implementation described in Section \DUrole{ref}{our-implementation} this lead to
difficulties with accurately controlling the amount of energy injected, as
previously described. The effect of this is clearly visible in Figs. 4.1-4.4.
The shell radius shell radius for a injection radius of $\SI{0.5}{pc}$ is
much higher than expected and especially in comparison to slightly larger
injection radii, resulting from a too large energy density. The single cell
injection uses a different estimate, better matching the other curves. Those
problem were only found late in the writing process of this thesis. Lacking
time, no reruns of the affected simulations was possible, so the results from
these runs can only be interpreted qualitatively.

In runs with cooling turned off, the curves stay close to the Sedov-Taylor
solution, as expected. Best fitting are the single cell injection and the
$\SI{2}{pc}$ injection radius SN, with the $\SI{1}{pc}$ close to the
larger injection radii. This ordering is puzzling, but might also stem from
errors in the energy density estimate.

When cooling is enabled, the curves are now closer to the pressure modified
snowplow. For $\rho = \SI{100}{u/cm^3}$ the remnants are quite aspherical
(as seen below). This leads to a large error in the radii estimation. Only here
overcooling effects are clearly visible for the largest injection radii
($\SI{7.5}{pc}$ and $\SI{10}{pc}$), with almost no expansion
happening after injection.

\begin{figure}
\noindent\makebox[\textwidth][c]{\includegraphics[width=1.000\linewidth]{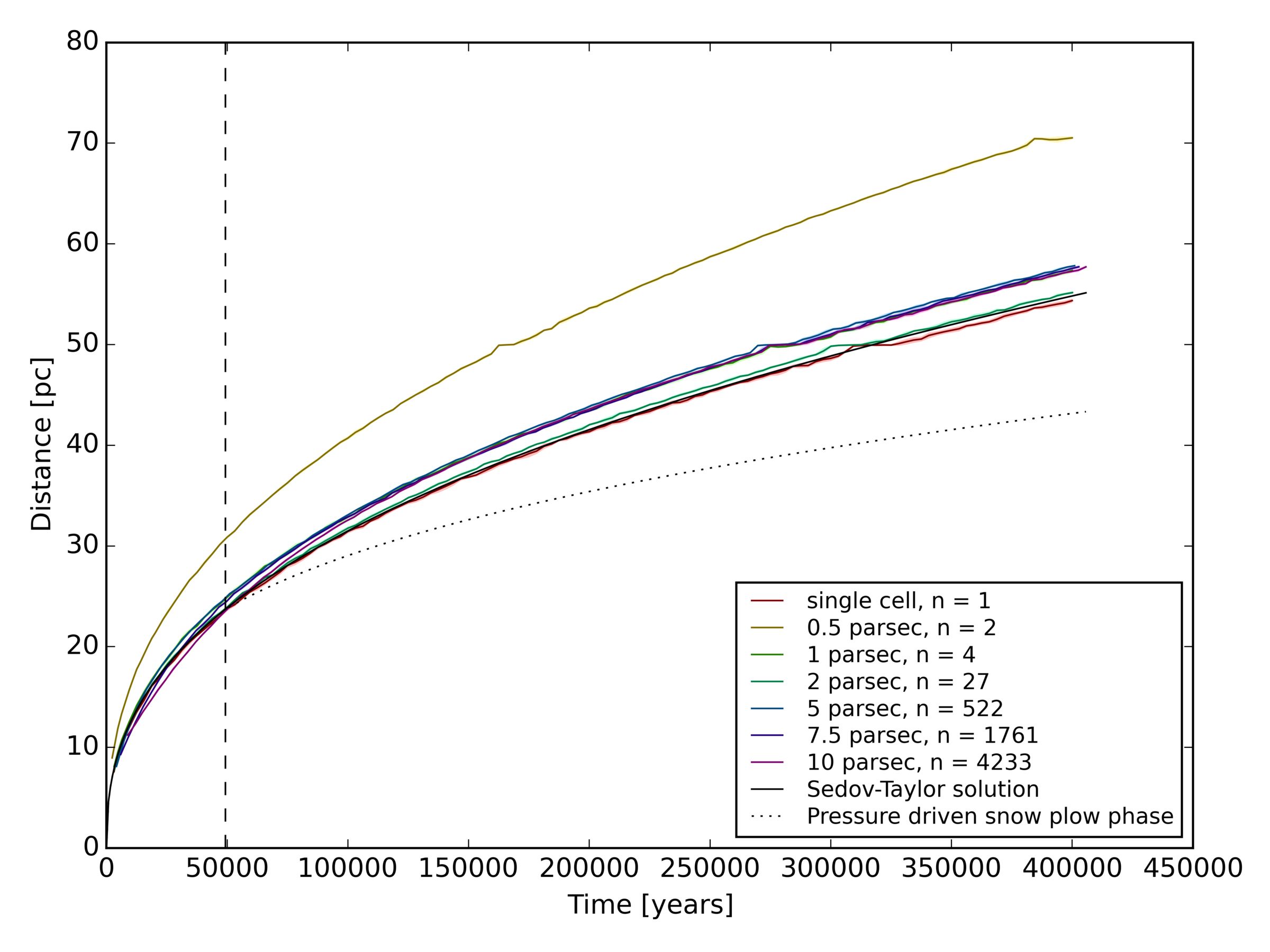}}
\caption{\DUrole{label}{fig:radial-1u-nochem}
Radius of the supernova shell during the Sedov-Taylor and pressure driven
snowplow phase. The background density is $\SI{1}{u/cm^3}$ and
injection energy is $\SI{e51}{erg}$. To compensate for the initial
injection radius, the beginning of the curves are shifted by an amount
estimated from the Sedov-Taylor solution. The vertical dashed line marks the
beginning of the radiative phase, where the Sedov-Taylor phase ends. The
analytic Sedov-Taylor solution and pressure-modified snowplow is overlaid.
Chemistry and cooling is disabled.}
\end{figure}

\begin{figure}
\noindent\makebox[\textwidth][c]{\includegraphics[width=1.000\linewidth]{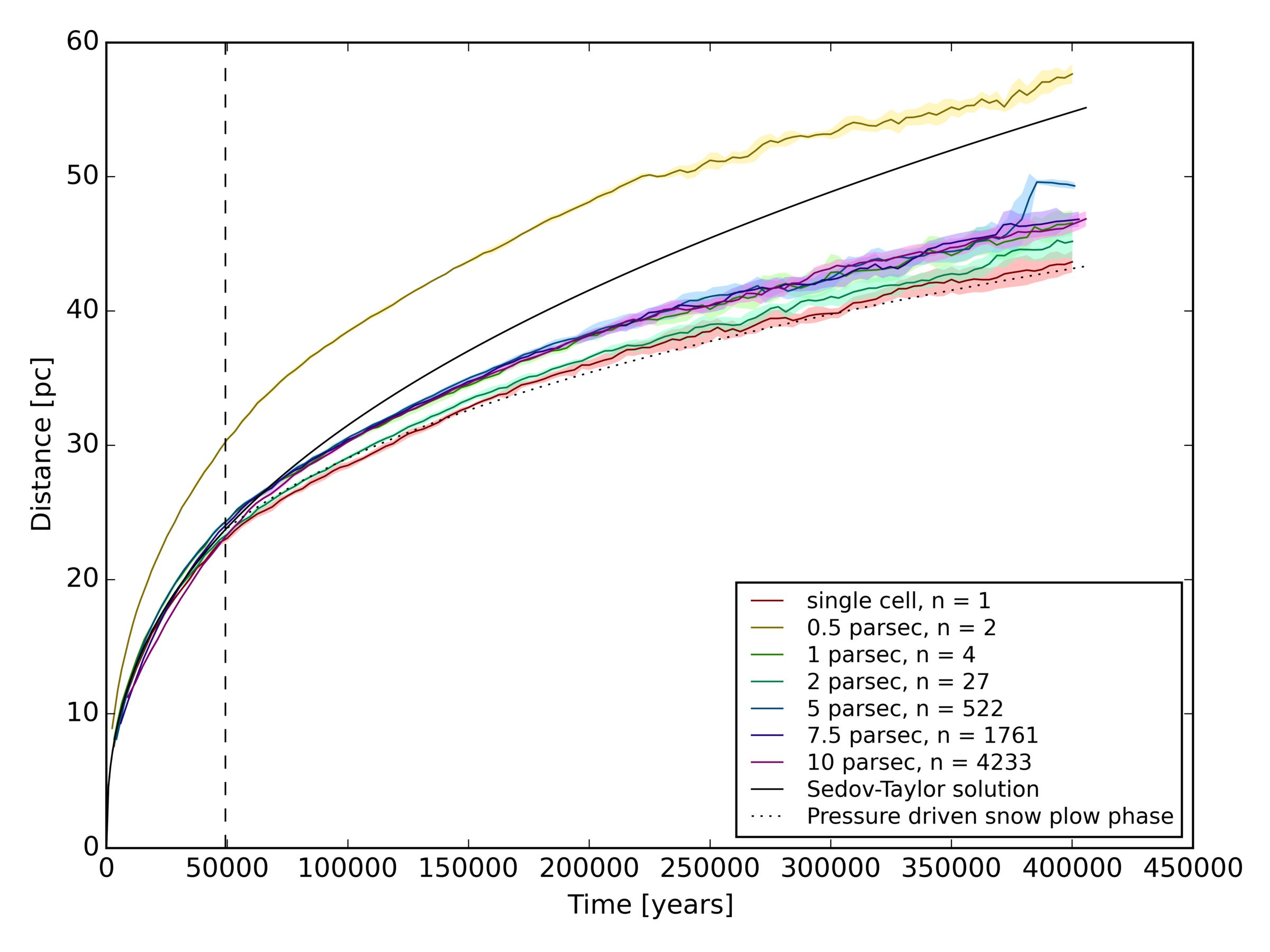}}
\caption{\DUrole{label}{fig:radial-1u} Radius of the supernova shell during the Sedov-Taylor
and pressure driven snowplow phase. The background density is
$\SI{1}{u/cm^3}$ and injection energy is $\SI{e51}{erg}$. To
compensate for the initial injection radius, the beginning of the curves are
shifted by an amount estimated from the Sedov-Taylor solution. The vertical
dashed line marks the beginning of the radiative phase, where the
Sedov-Taylor phase ends. The analytic Sedov-Taylor solution and
pressure-modified snowplow is overlaid. Chemistry and cooling is enabled.}
\end{figure}

\begin{figure}
\noindent\makebox[\textwidth][c]{\includegraphics[width=1.000\linewidth]{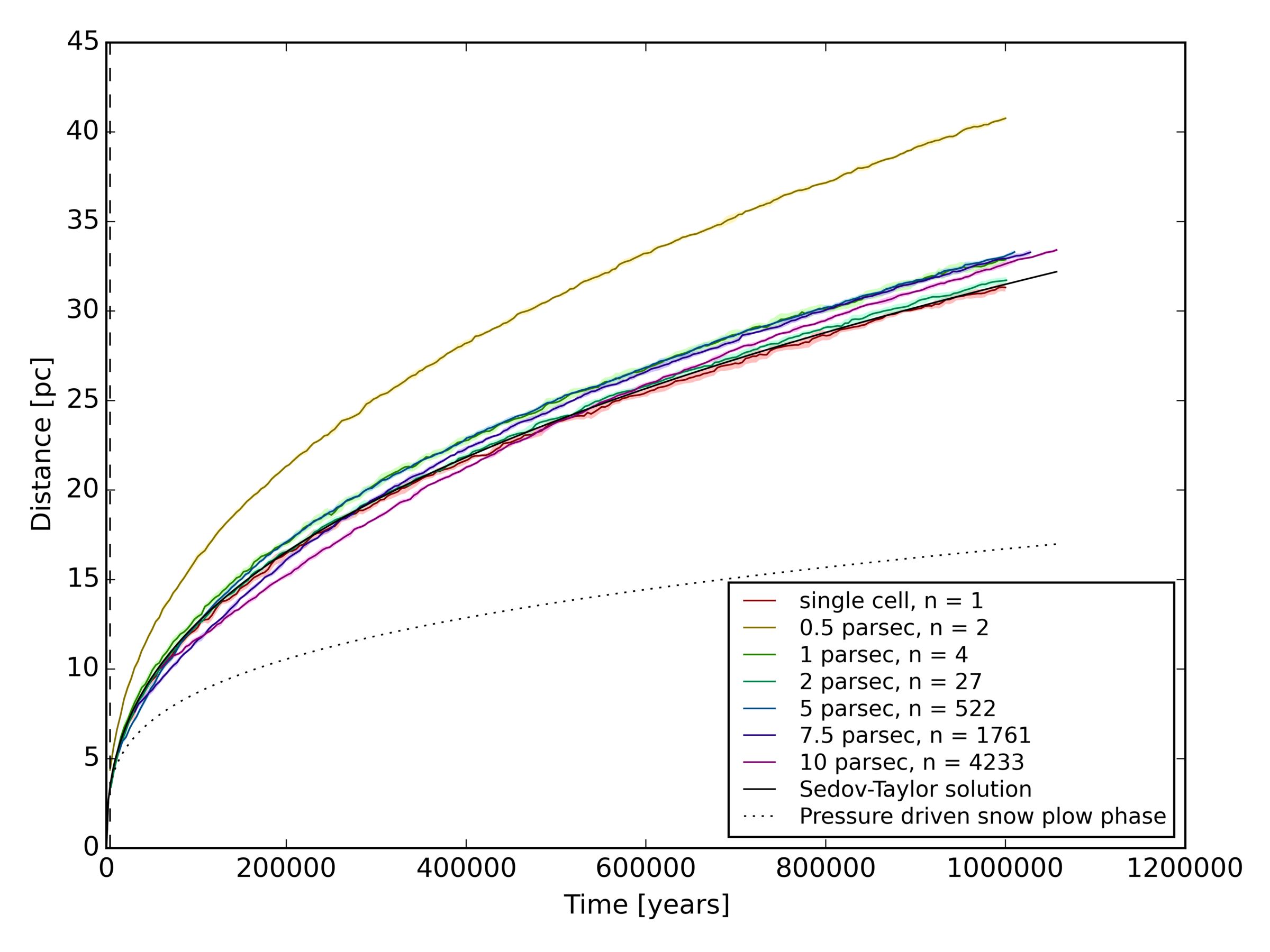}}
\caption{\DUrole{label}{fig:radial-100u-nochem} Radius of the supernova shell during the
Sedov-Taylor and pressure driven snowplow phase. The background density is
$\SI{100}{u/cm^3}$ and injection energy is $\SI{e51}{erg}$. The
vertical dashed line marks the beginning of the radiative phase, where the
Sedov-Taylor phase ends.  The analytic Sedov-Taylor solution and
pressure-modified snowplow is overlaid.  Chemistry and cooling is disabled.}
\end{figure}

\begin{figure}
\noindent\makebox[\textwidth][c]{\includegraphics[width=1.000\linewidth]{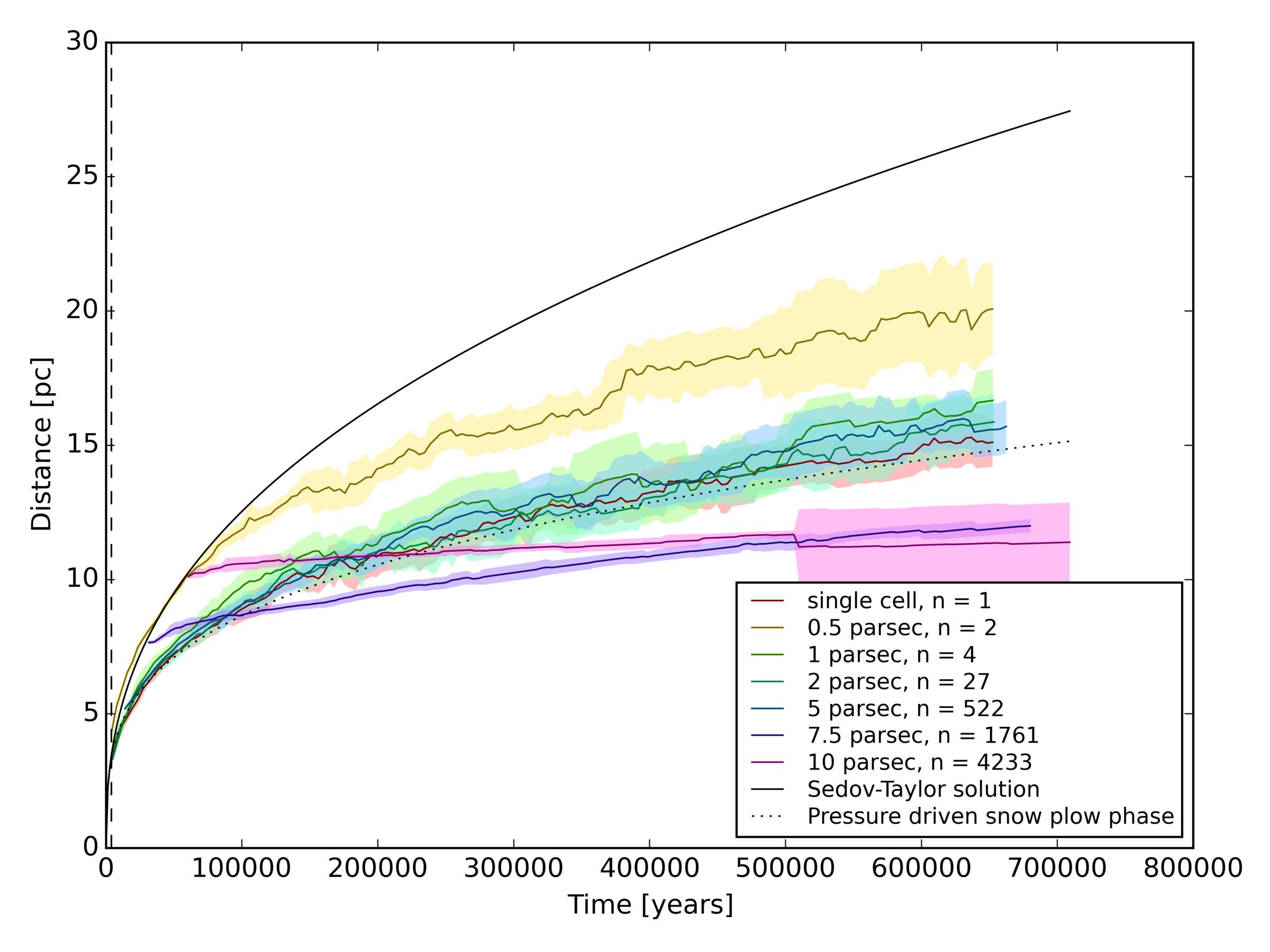}}
\caption{\DUrole{label}{fig:radial-100u}
Radius of the supernova shell during the Sedov-Taylor and pressure driven
snowplow phase. The background density is $\SI{100}{u/cm^3}$ and
injection energy is $\SI{e51}{erg}$. The vertical dashed line marks the
beginning of the radiative phase, where the Sedov-Taylor phase ends.
The
analytic Sedov-Taylor solution and pressure-modified snowplow is overlaid for
comparison.
Chemistry and cooling is enabled.
For injection radii $\SI{7.5}{pc}$ and $\SI{10}{pc}$ almost no
expansion happens, likely caused by overcooling effects.}
\end{figure}

Although the remnant should be perfectly spherical symmetric, when cooling is
disabled small protuberances are visible in the inner regions behind the shock
front. When cooling is enabled, the shell is more aspherical and small dense
regions expand further out than others. This also leads to a high error in the
radii estimates.

The cause of this is likely the low resolution of the initial SN cells. With a
small number of cells the shape of the tessellation, which is a union of
tetrahedra, dominates the initial shape of the supernova and even after rapid
expansion it has an effect on the final shape of the shock front. Also the
actual resolution in the shell is limited and the tessellation stays visible
(e.g. compare Fig. \DUrole{ref}{fig:single-1-density-100u-166}).

With cooling the cause of the filaments could be the
Vishniac instability \DUrole{citep}{Vishniac1983}, although \DUrole{citet}{Michaut2012} find
in two-dimensional simulations “that the supernova remnant returns to a stable
evolution and the Vishniac instability does not lead to the fragmentation of the
shock as predicted by the theory”.

\begin{figure}
  \centering
  \includegraphics[width=0.95\linewidth]{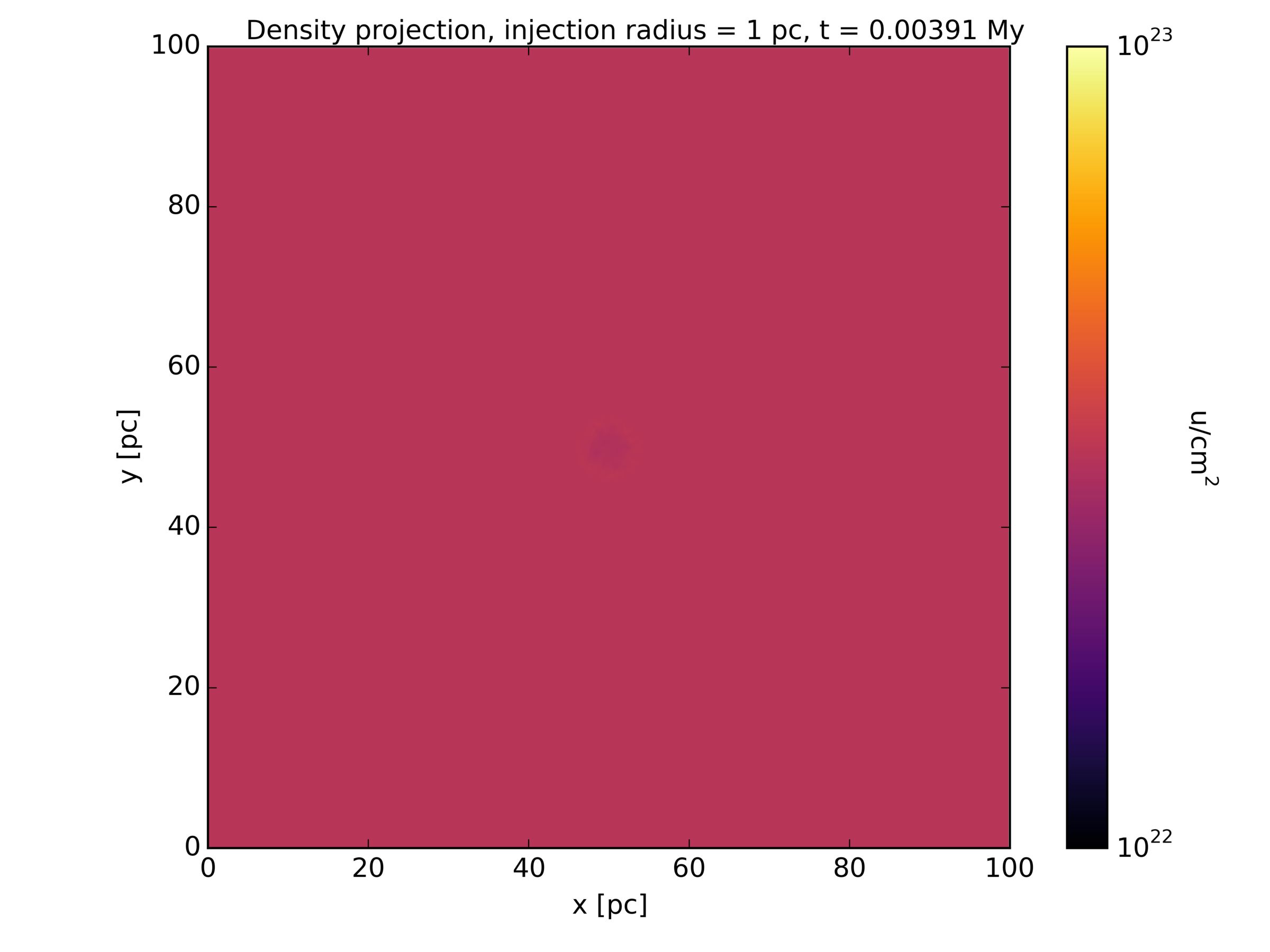}
  \centering
  \includegraphics[width=0.95\linewidth]{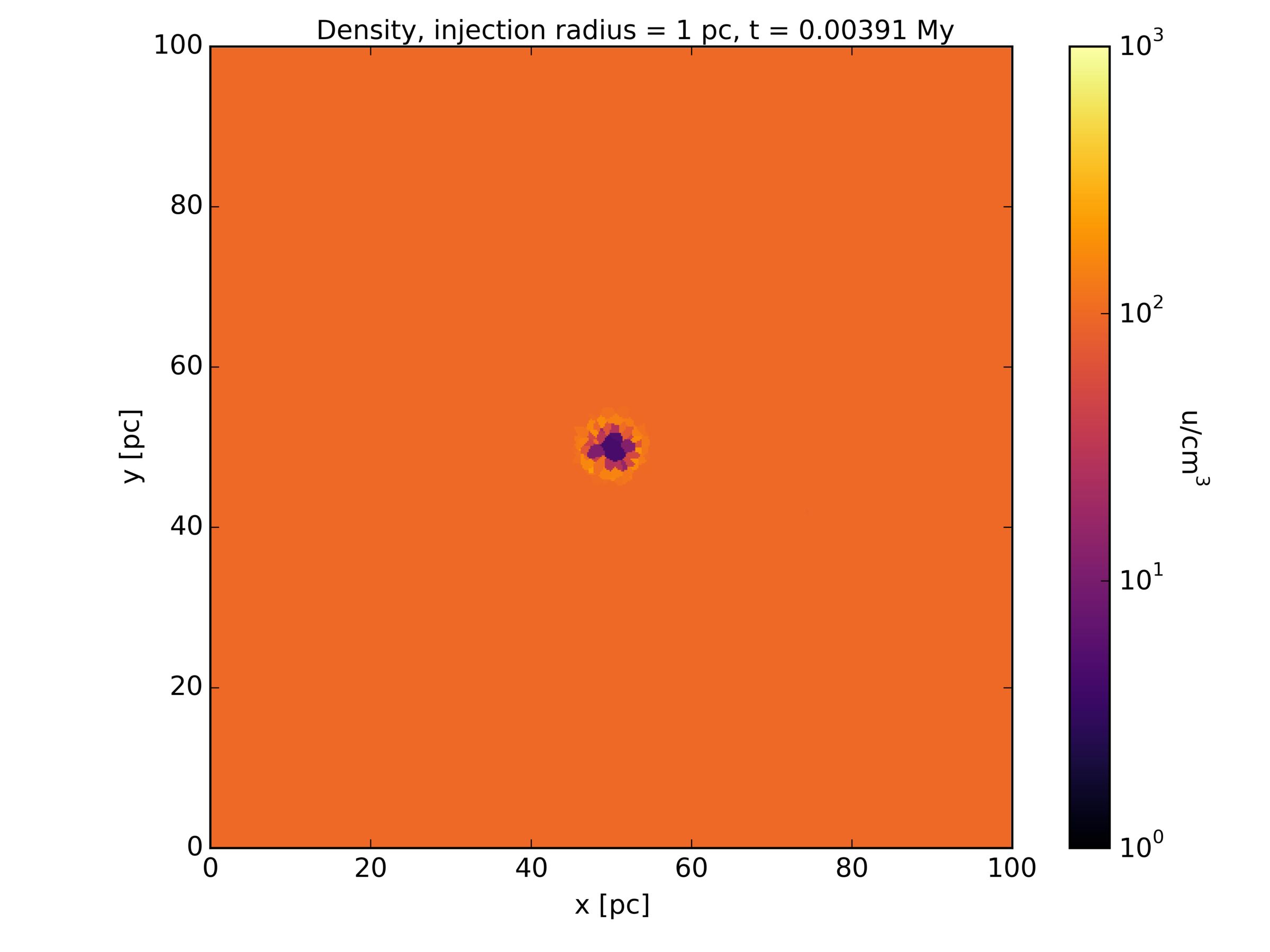}
  \caption{
    Density projection and slice of a single
    supernova remnant just after injection with radius of $\SI{1}{pc}$ and initial
    uniform density of $\SI{100}{u/cm^3}$. Chemistry and cooling is
    enabled.
  }
  \label{fig:single-1-density-100u}
\end{figure}

\begin{figure}
  \centering
  \includegraphics[width=0.95\linewidth]{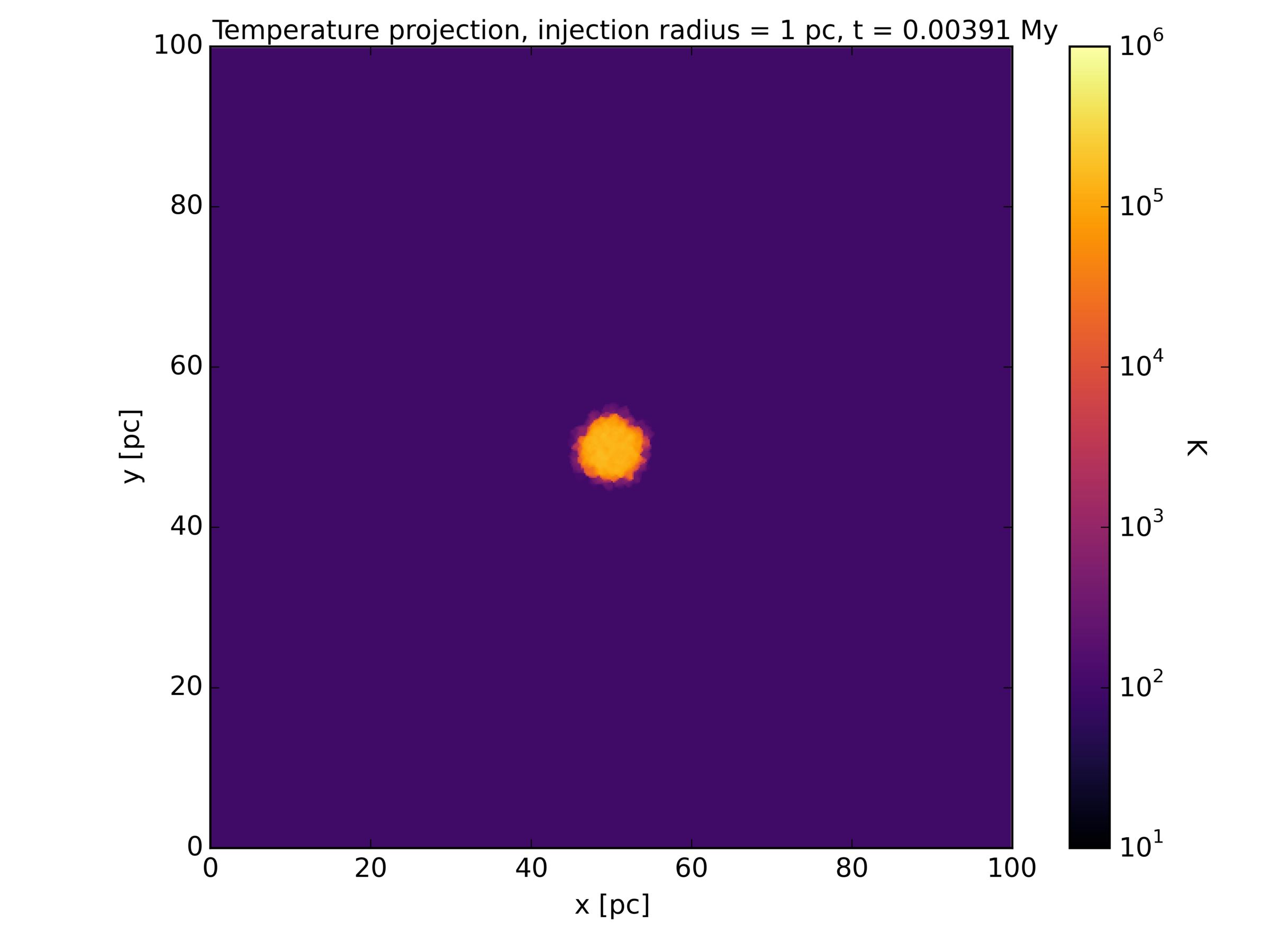}
  \centering
  \includegraphics[width=0.95\linewidth]{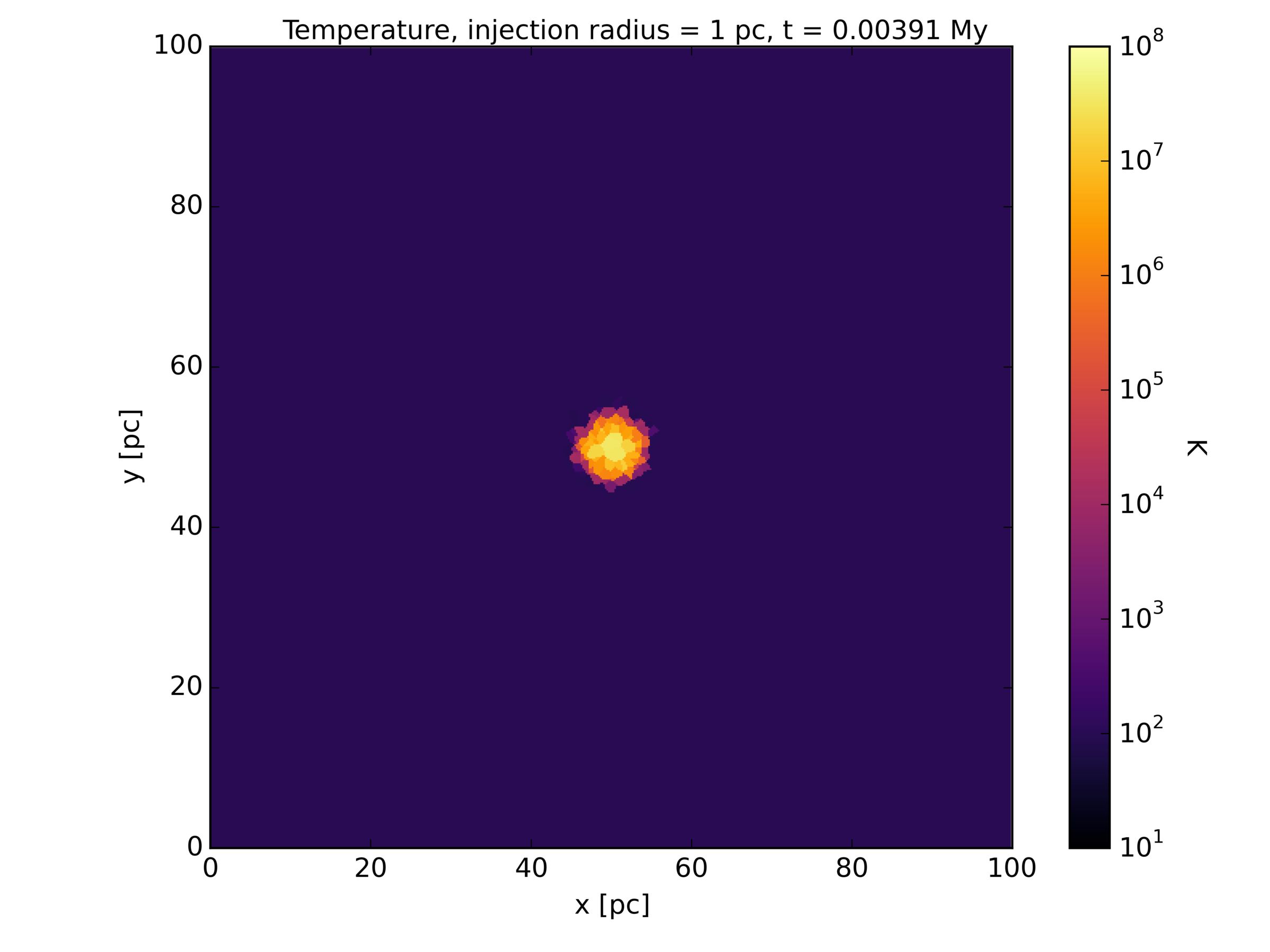}
  \caption{
     Temperature projection and slice of a single
     supernova remnant just after injection with radius of \SI{1}{pc} and initial
     uniform density of $\SI{100}{u/cm^3}$. Chemistry and cooling is
     enabled.
  }
  \label{fig:single-1-temperature-100u}
\end{figure}

 \begin{figure}
   \centering
   \includegraphics[width=0.95\linewidth]{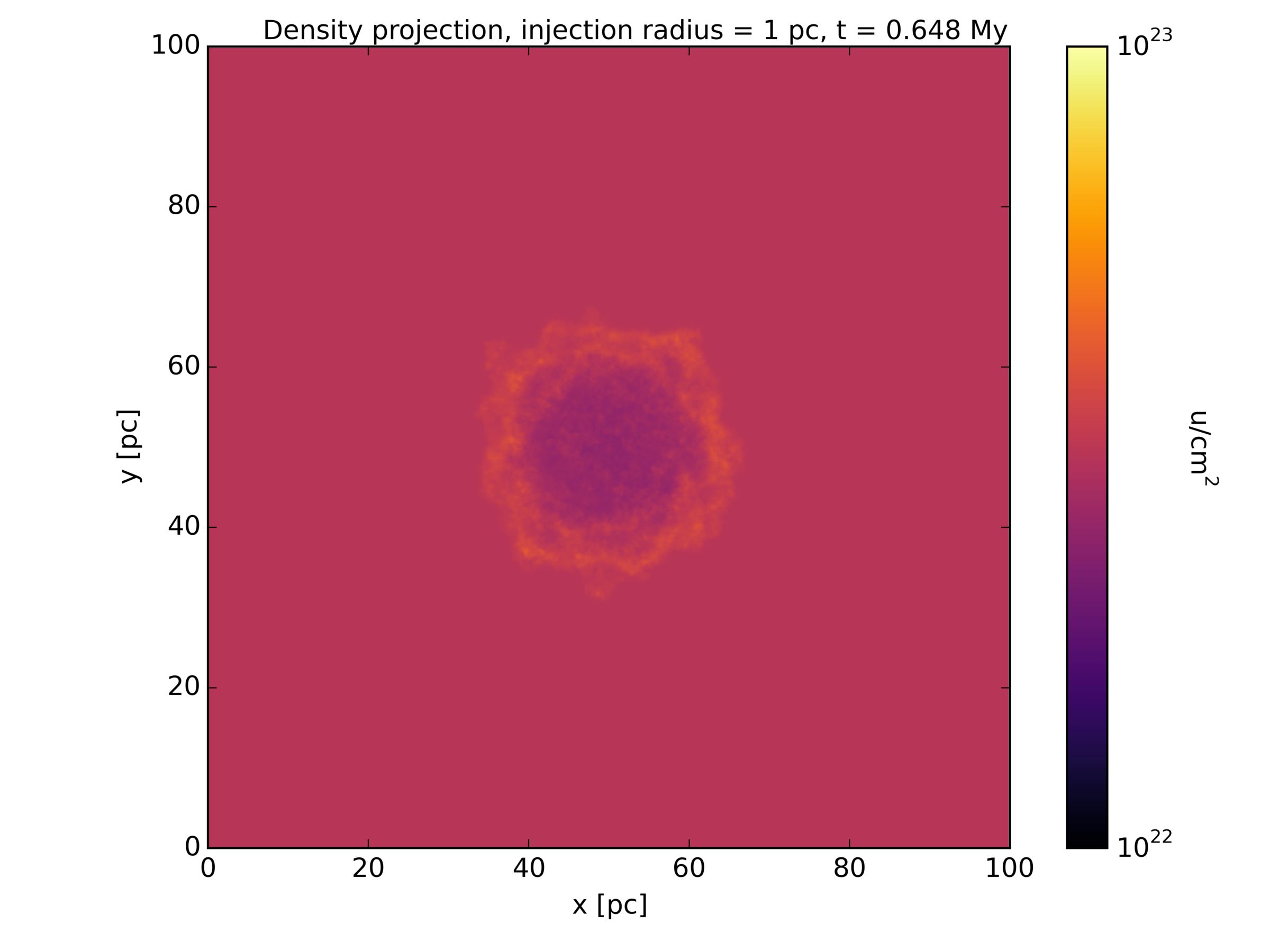}
   \centering
   \includegraphics[width=0.95\linewidth]{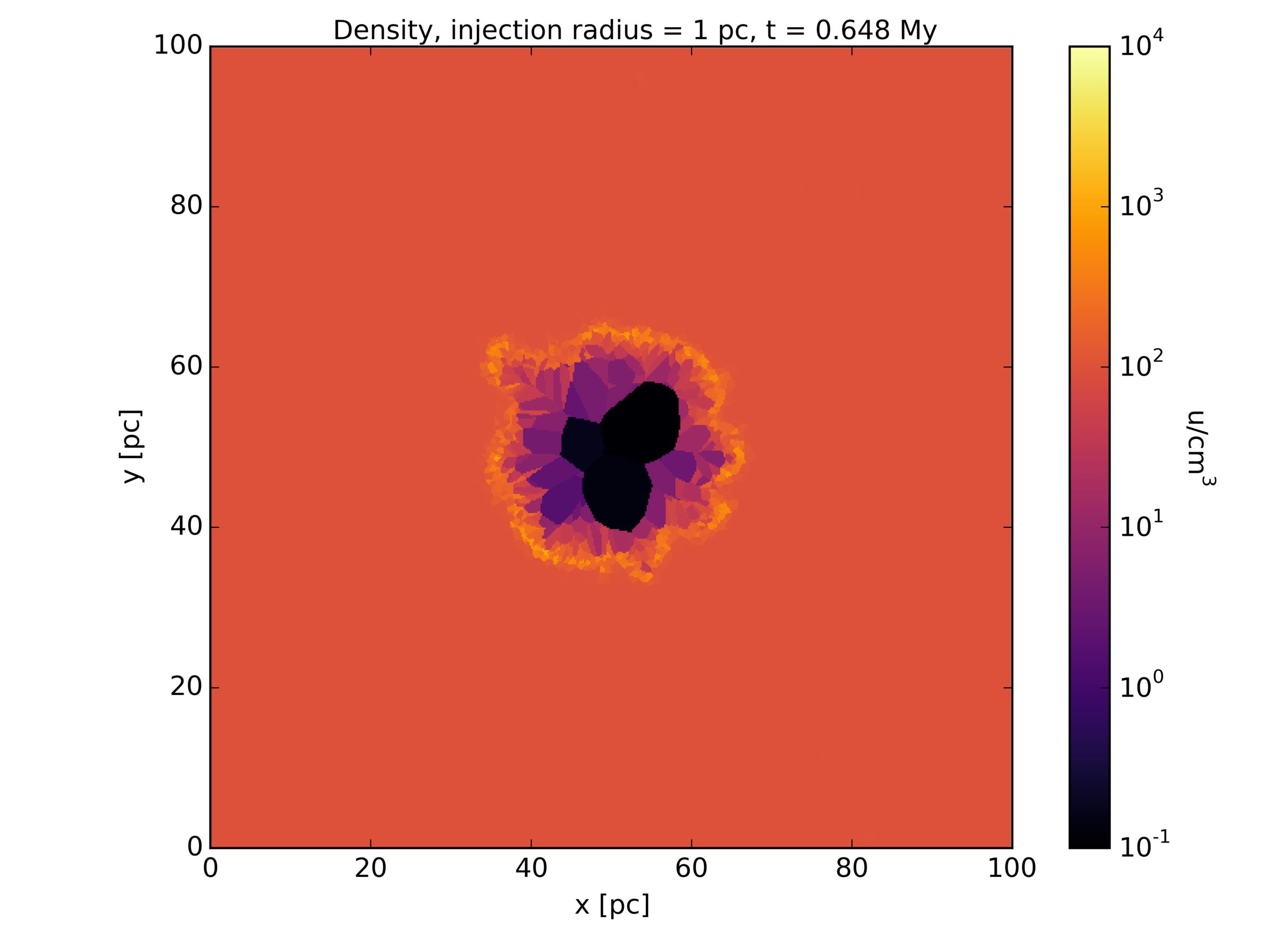}
   \caption{
Density projection and slice of a single supernova remnant with injection
radius of $\SI{1}{pc}$ in the snowplow phase ($t \approx \SI{16.6}{Mys}$) and initial uniform density of $\SI{100}{u/cm^3}$. Chemistry and
cooling is enabled.
The shock is now quite aspherical, likely caused by inhomogeneous cooling
effects.
   }
   \label{fig:single-1-density-100u-166}
 \end{figure}

 \begin{figure}
   \centering
   \includegraphics[width=0.95\linewidth]{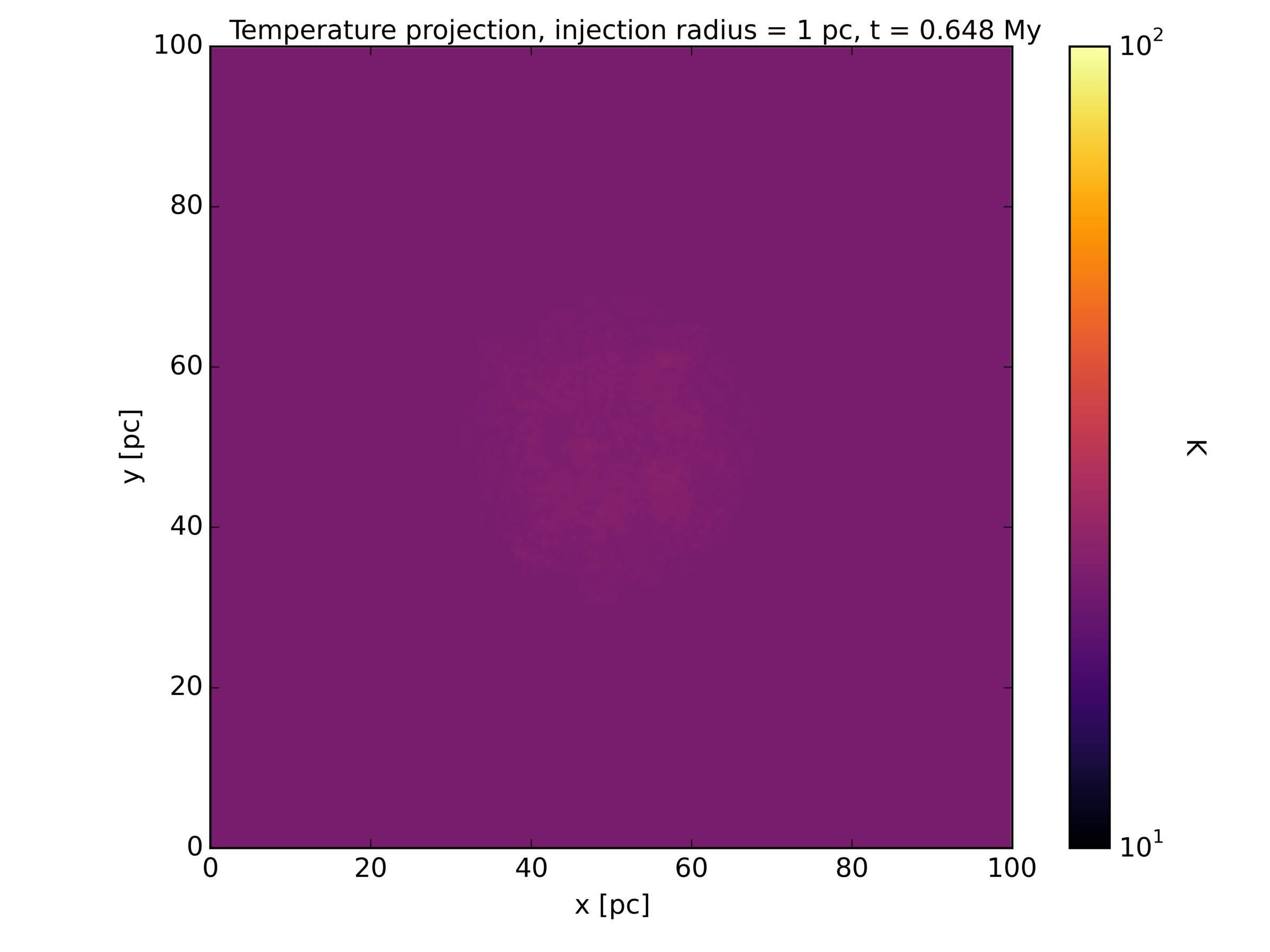}
   \centering
   \includegraphics[width=0.95\linewidth]{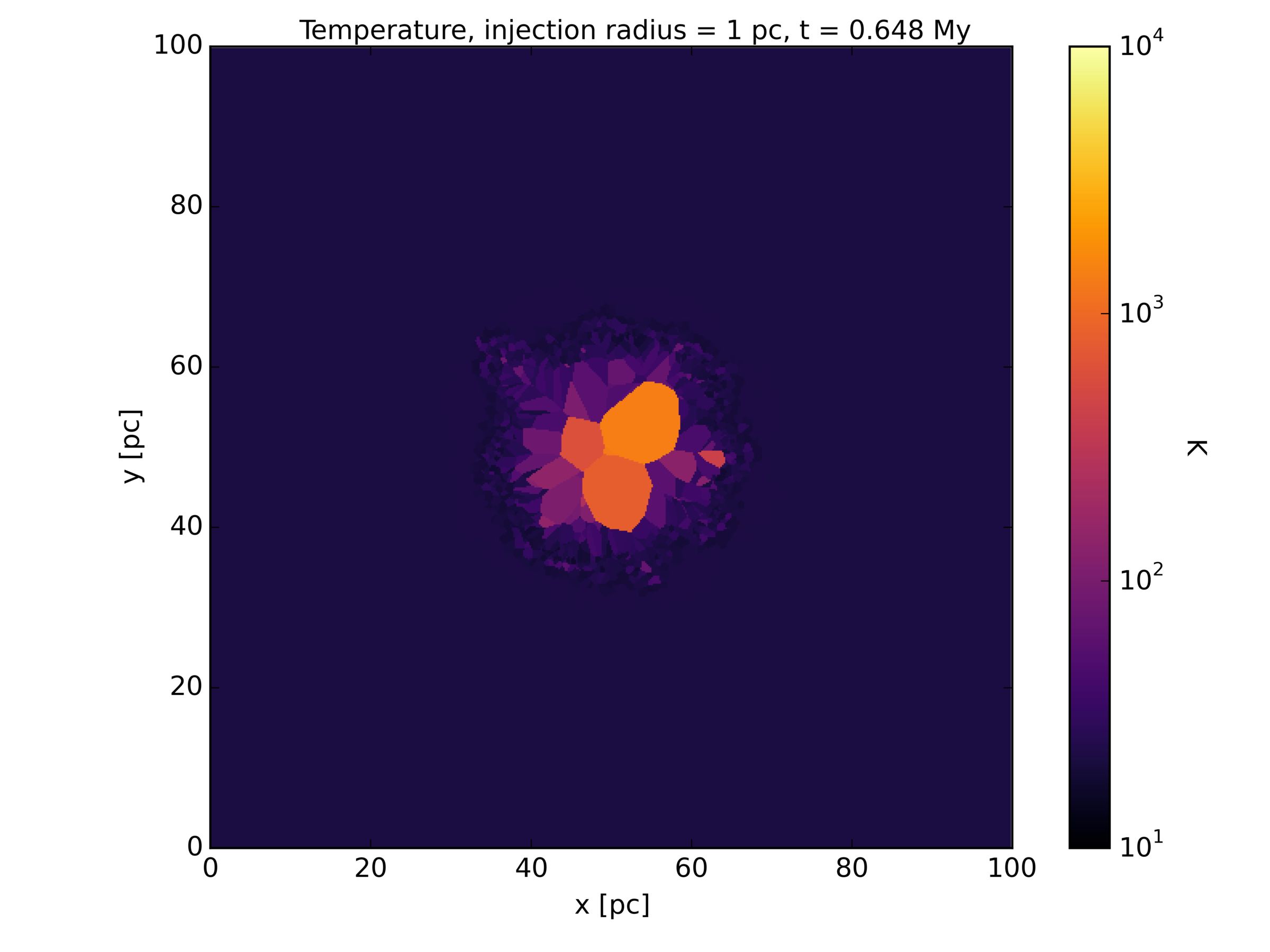}
   \caption{
Density-weighted mean temperature along the line of sight and slice of a single supernova remnant with injection
radius of $\SI{1}{pc}$ in the snowplow phase ($t \approx \SI{16.6}{Mys}$) and initial uniform density of $\SI{100}{u/cm^3}$. Chemistry and
cooling is enabled. The shock now is quite aspherical, likely caused by inhomogeneous cooling
effects.
   }
   \label{fig:single-1-temperature-100u-166}
 \end{figure}

  \begin{figure}
   \centering
   \includegraphics[width=0.95\linewidth]{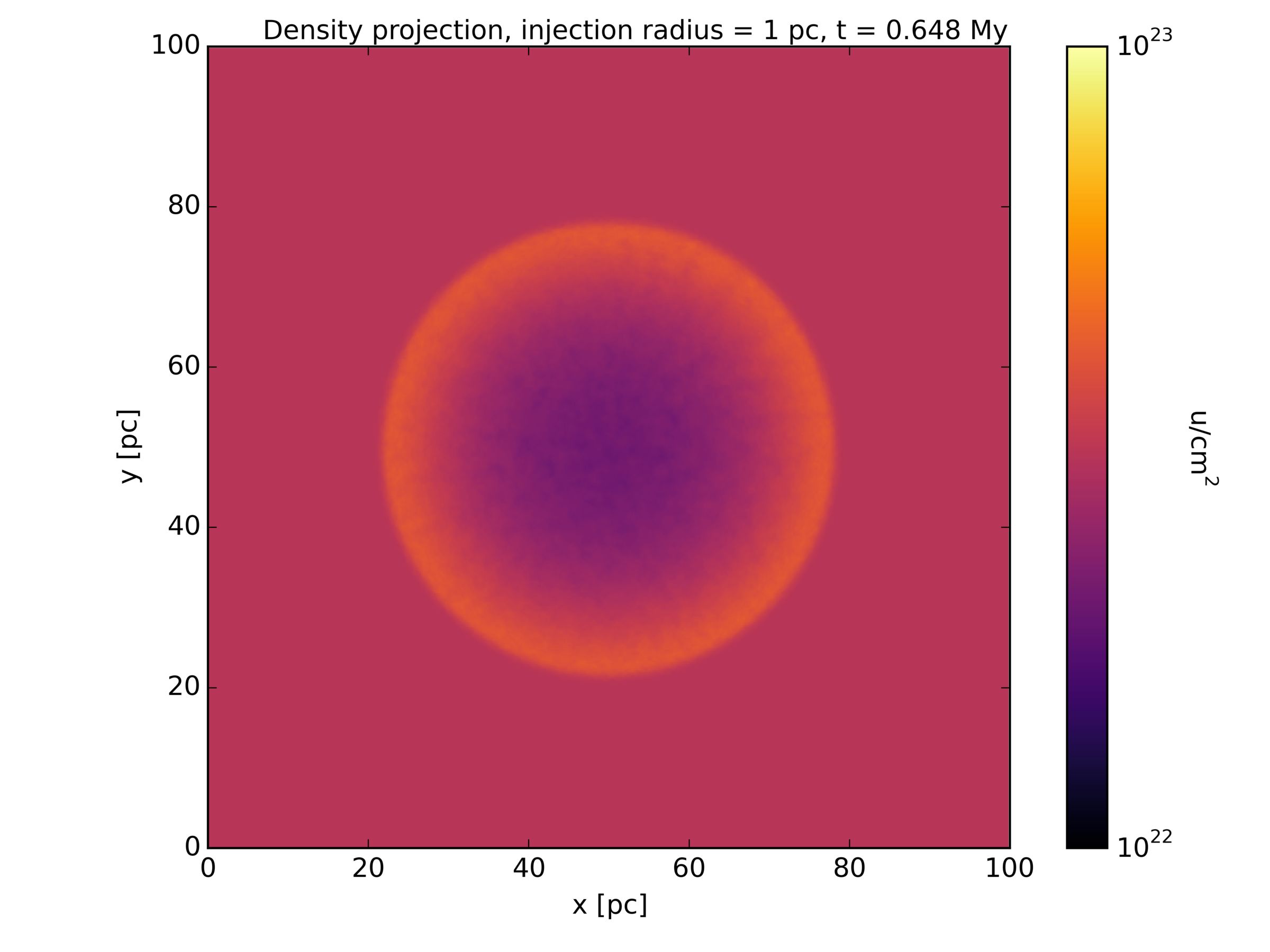}
   \centering
   \includegraphics[width=0.95\linewidth]{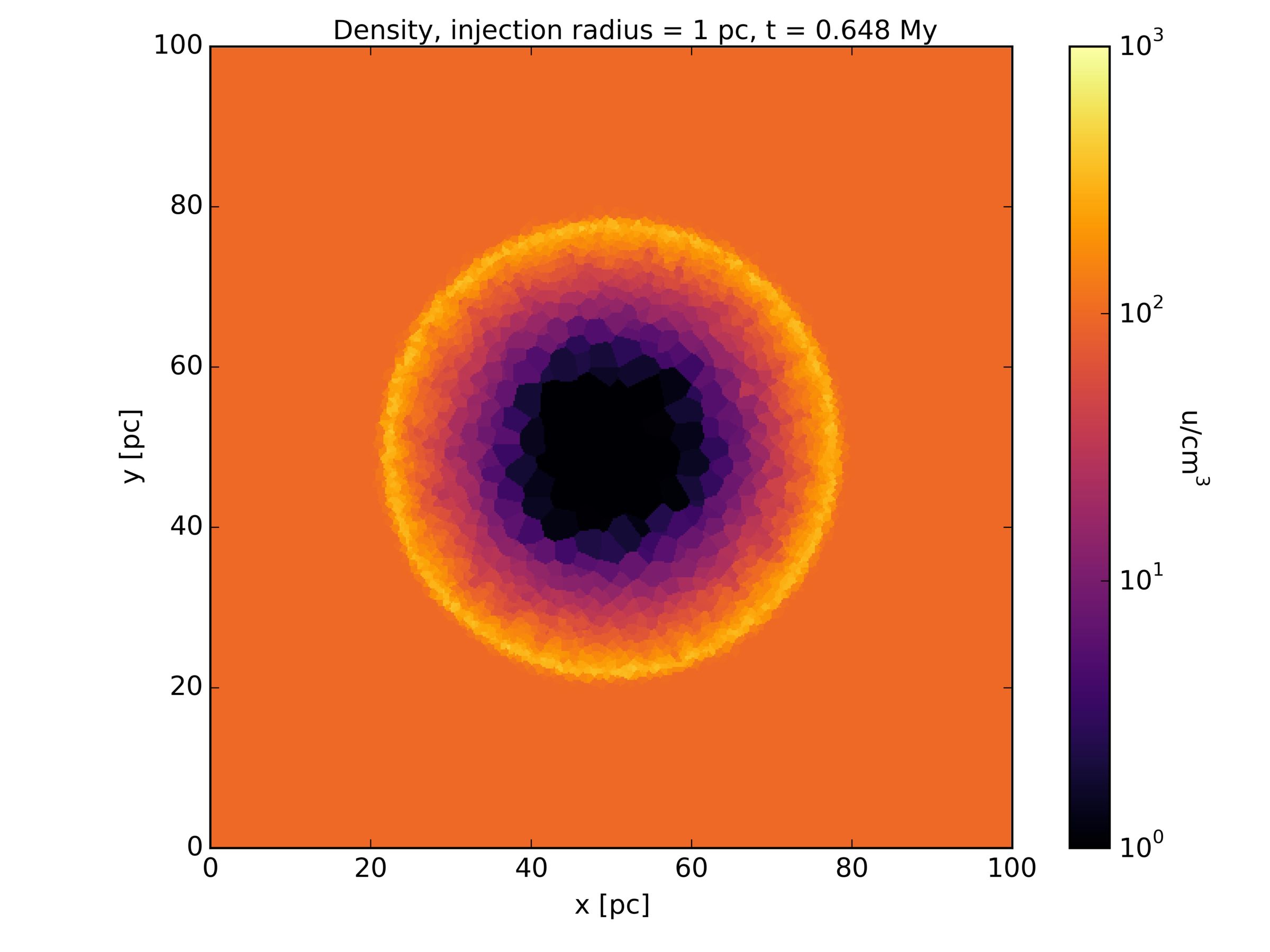}
   \caption{
Density projection and slice of a single supernova remnant with injection
radius of $\SI{1}{pc}$ in the snowplow phase ($t \approx \SI{16.6}{Mys}$) and initial uniform density of $\SI{100}{u/cm^3}$. Chemistry and
cooling is disabled.
   }
   \label{fig:single-1-density-100u-166-nochem}
 \end{figure}

 \begin{figure}
   \centering
   \includegraphics[width=0.95\linewidth]{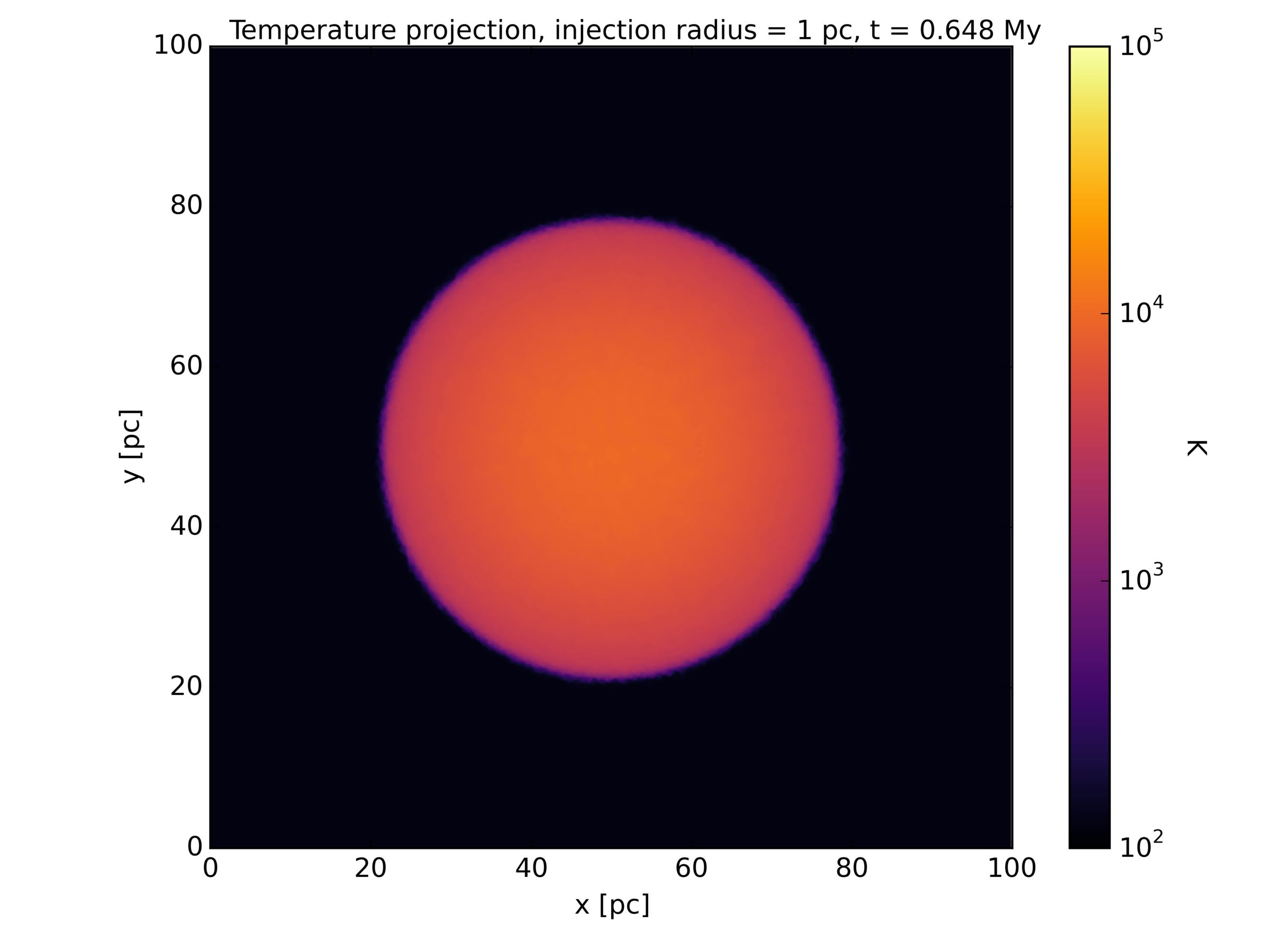}
   \centering
   \includegraphics[width=0.95\linewidth]{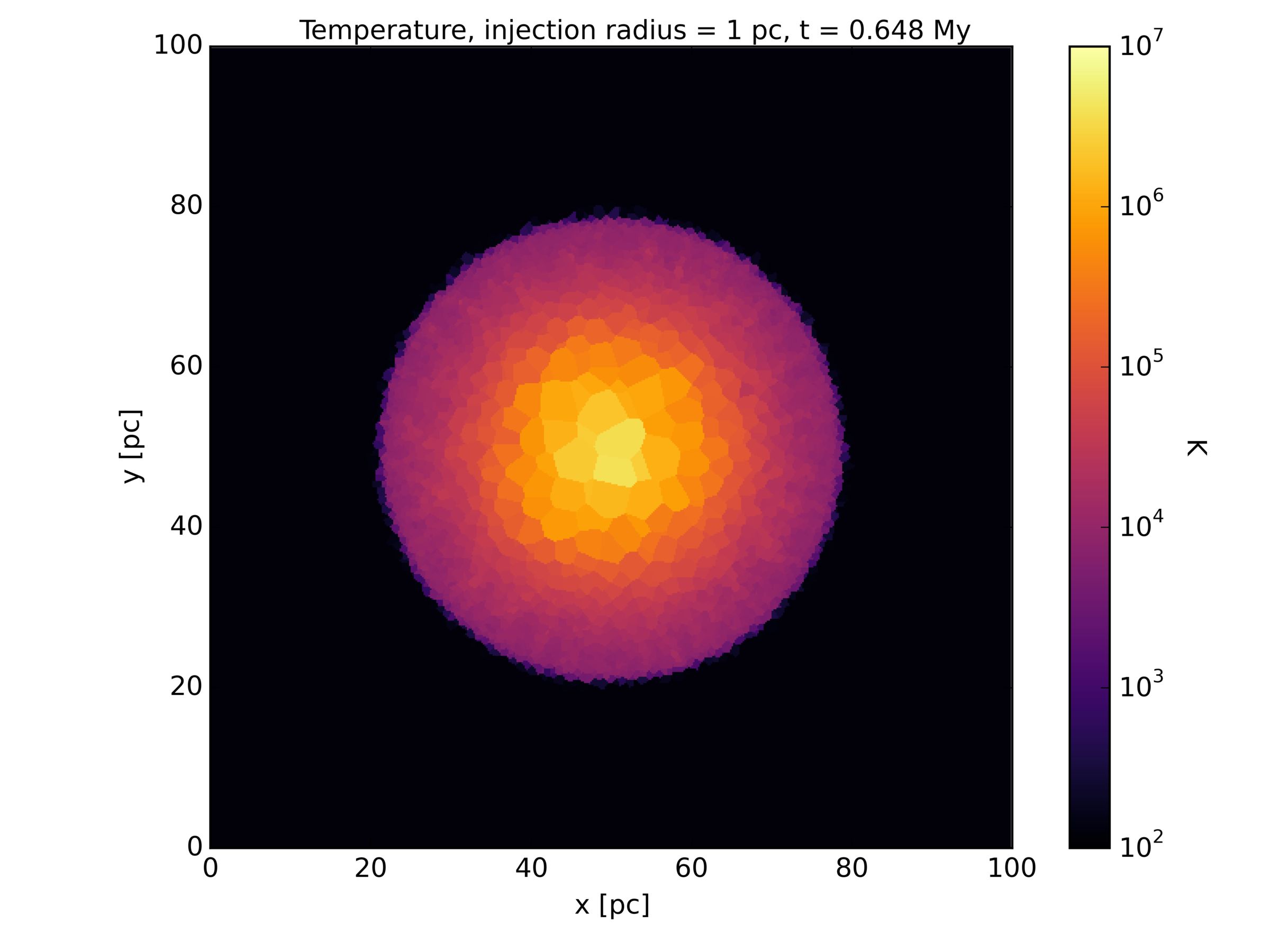}
   \caption{
Density-weighted mean temperature along the line of sight and slice of a single supernova remnant with injection
radius of $\SI{1}{pc}$ in the snowplow phase ($t \approx \SI{16.6}{Mys}$) and initial uniform density of $\SI{100}{u/cm^3}$. Chemistry and
cooling is disabled.
   }
   \label{fig:single-1-temperature-100u-166-nochem}
 \end{figure}

Real supernova remnants though are typical not perfectly spherical symmetric
either (e.g. Cassiopeia A, see Fig. \DUrole{ref}{cassiopeia}). Interaction with the
interstellar medium or asymmetric ejecta, due to MHD jets or Neutrino heating
might explain these features \DUrole{citep}{Fesen2006}.

\begin{figure}
\noindent\makebox[\textwidth][c]{\includegraphics[width=0.950\linewidth]{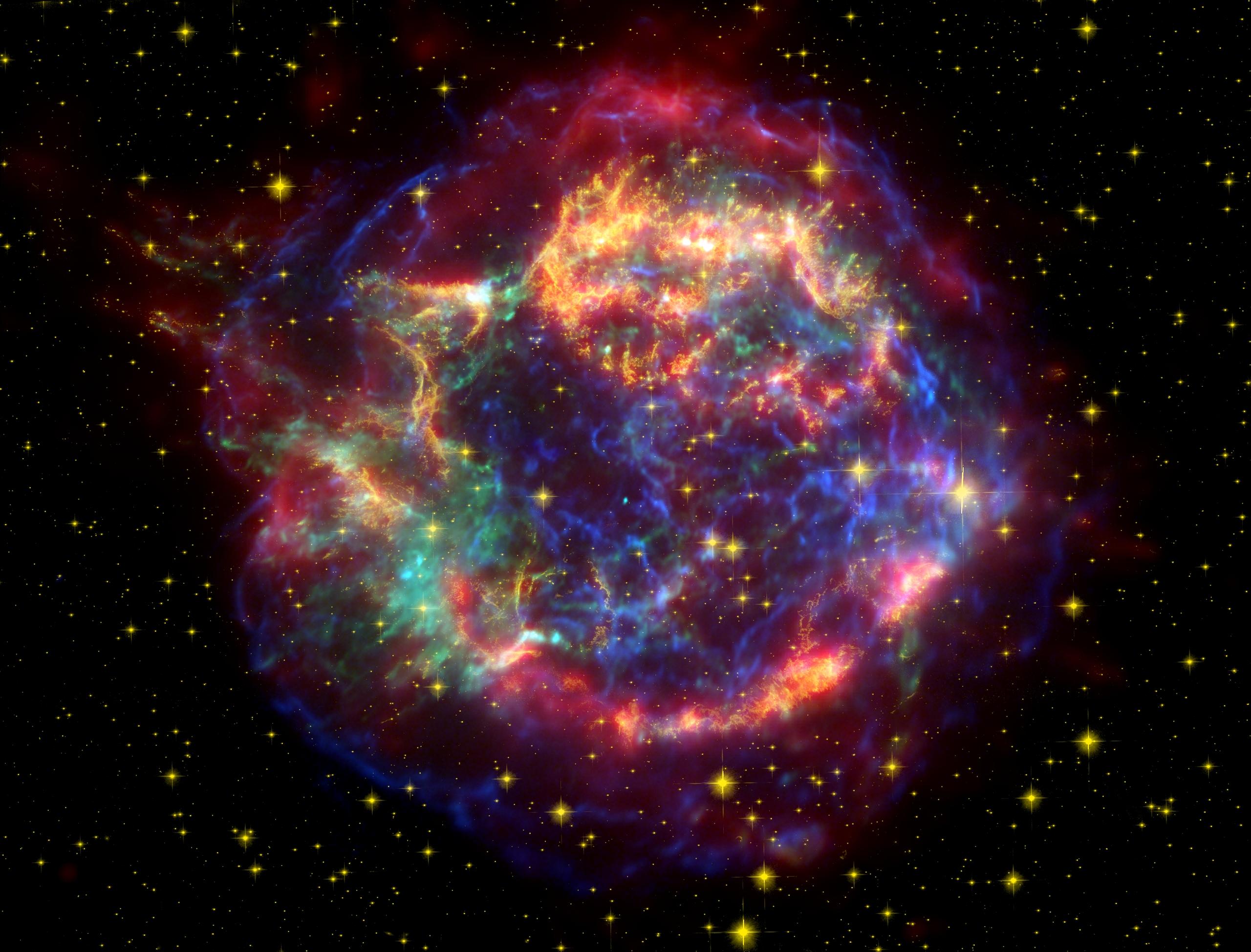}}
\caption{\DUrole{label}{cassiopeia}
A false colour image of Cassiopeia A using images from the Hubble, Spitzer
telescopes and the Chandra X-ray Observatory. Cassiopeia A is a supernova
remnant of a SN that happened around 330 years ago. It is asymmetric, with two
Jets protruding \DUrole{citep}{Fesen2006}. The image is in the public domain.}
\end{figure}

\clearpage

\section{Supernova feedback%
  \label{id5}%
}

A simulation with our feedback implementation was run for about $t = \SI{20}{Mys}$ with cubic periodic box of $\SI{400}{pc}$ per side and an initial density of $\SI{1}{u/cm^3}$.

\begin{figure}
  \centering
  \includegraphics[width=0.95\linewidth]{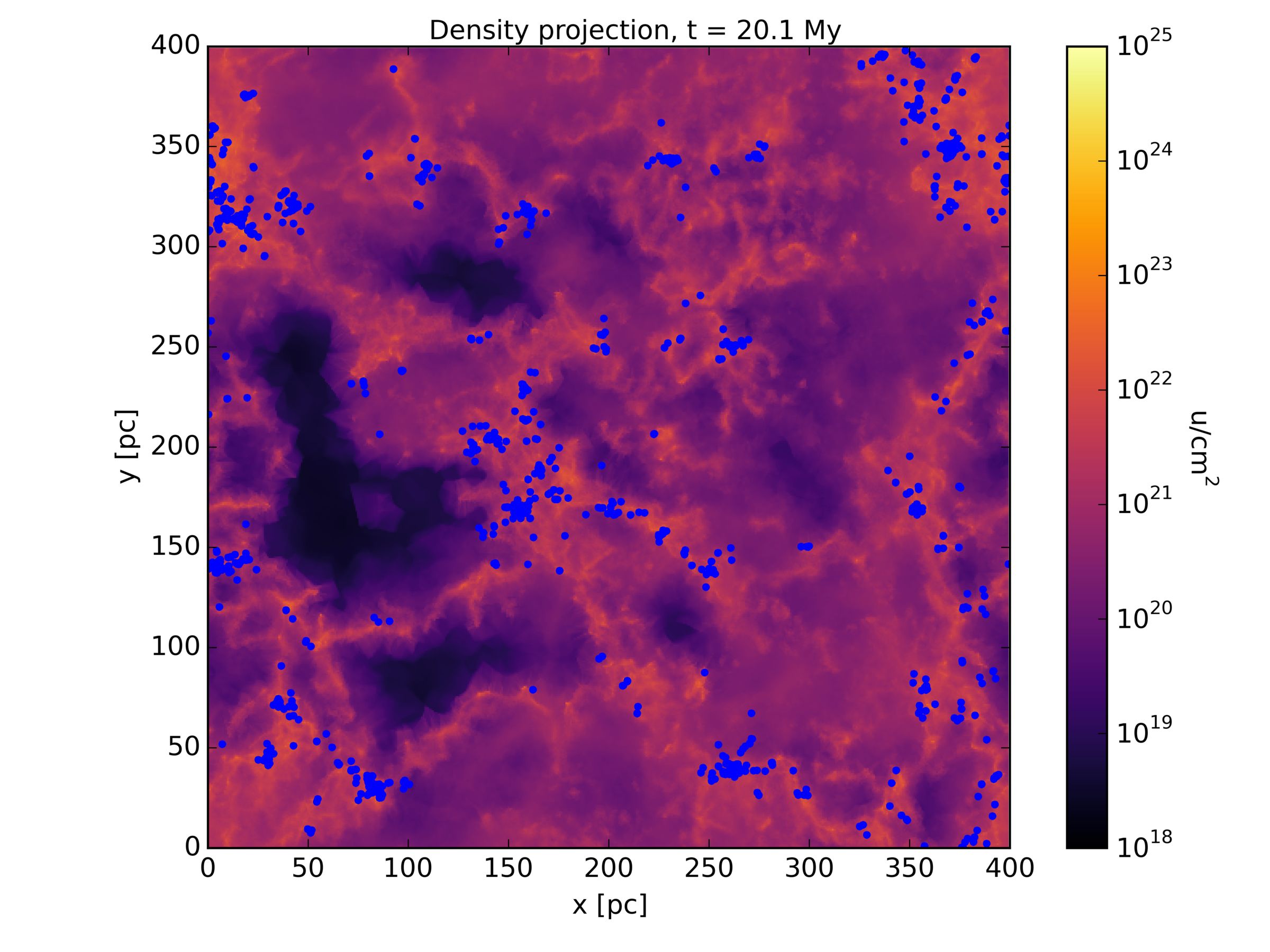}
  \centering
  \includegraphics[width=0.95\linewidth]{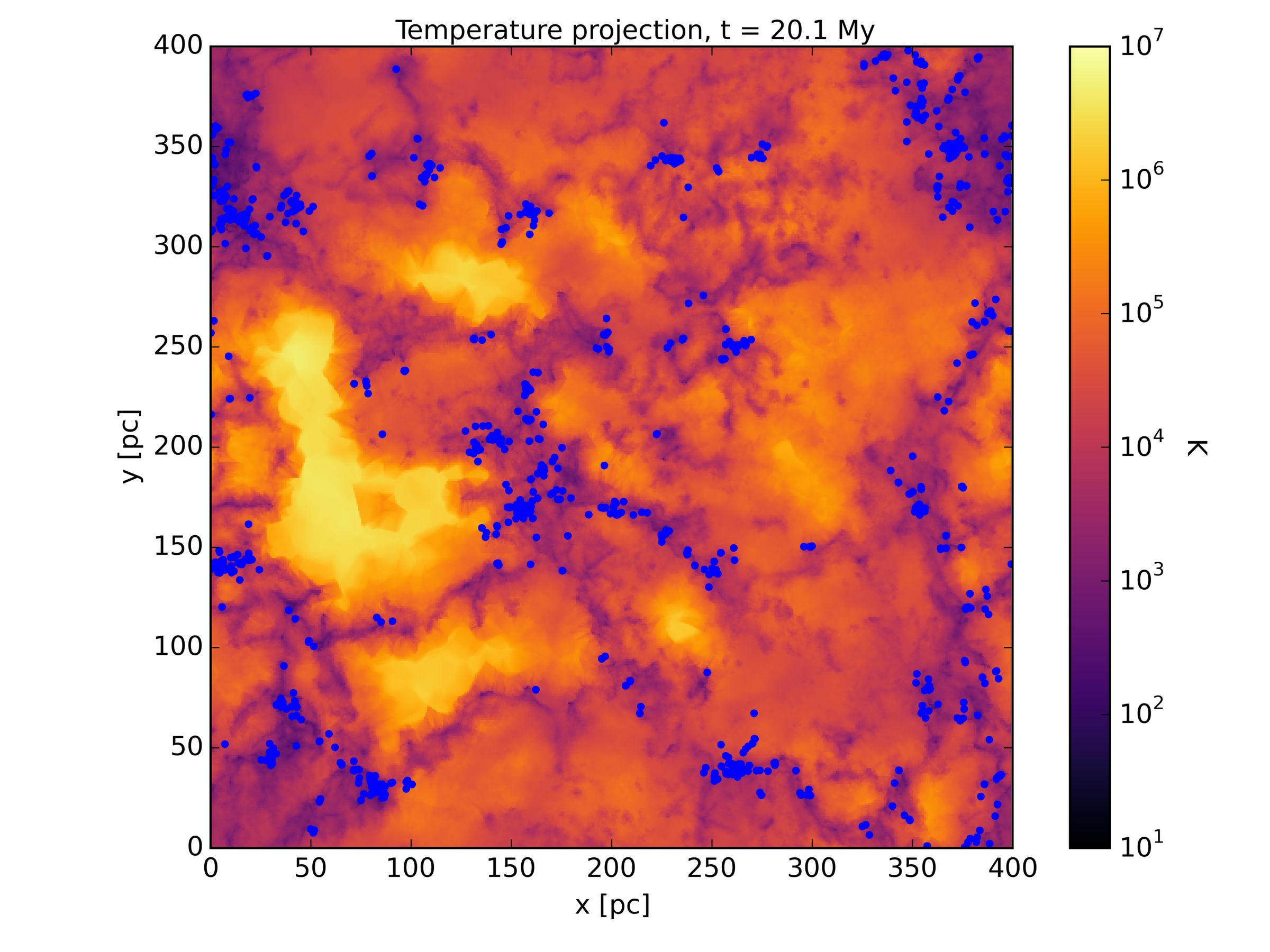}
  \caption{
    Projection of density and temperature with the location of sinks particles
    overlaid; the initial density \SI{1}{u/cm^3}.
  }
  \label{fig:1u-feedback-400pc-density-temp-sinks}
\end{figure}

Triggered by the supernova explosions, the initially uniform gas collapses.
After snapshot 107 ($t = \SI{10.7}{Mys}$) sink particles are formed in the
densest regions, when reaching a minimal number density of about
$\num{2e8}{cm^{-3}}$.  The creation of sink particles limit the upper
density reached in the box and also the smallest cell radius, as the mass of
mesh cells is held constant by the refinement and derefinement criterion. At
snapshot 201 ($t = \SI{20.1}{Mys}$) more than 7.8 \% of the mass is
captured in sink particles, with 1463 sink particles created (Fig.
\DUrole{ref}{fig:1u-feedback-400pc-density-temp-sinks}). At this point the simulation
crashes, likely due to issues in the sink particle implementation. As a
significant amount of gas is already removed from the influence of SN feedback
at this time, we would have stopped the simulation only a few megayears later
anyway.

\begin{figure}
\noindent\makebox[\textwidth][c]{\includegraphics[width=1.000\linewidth]{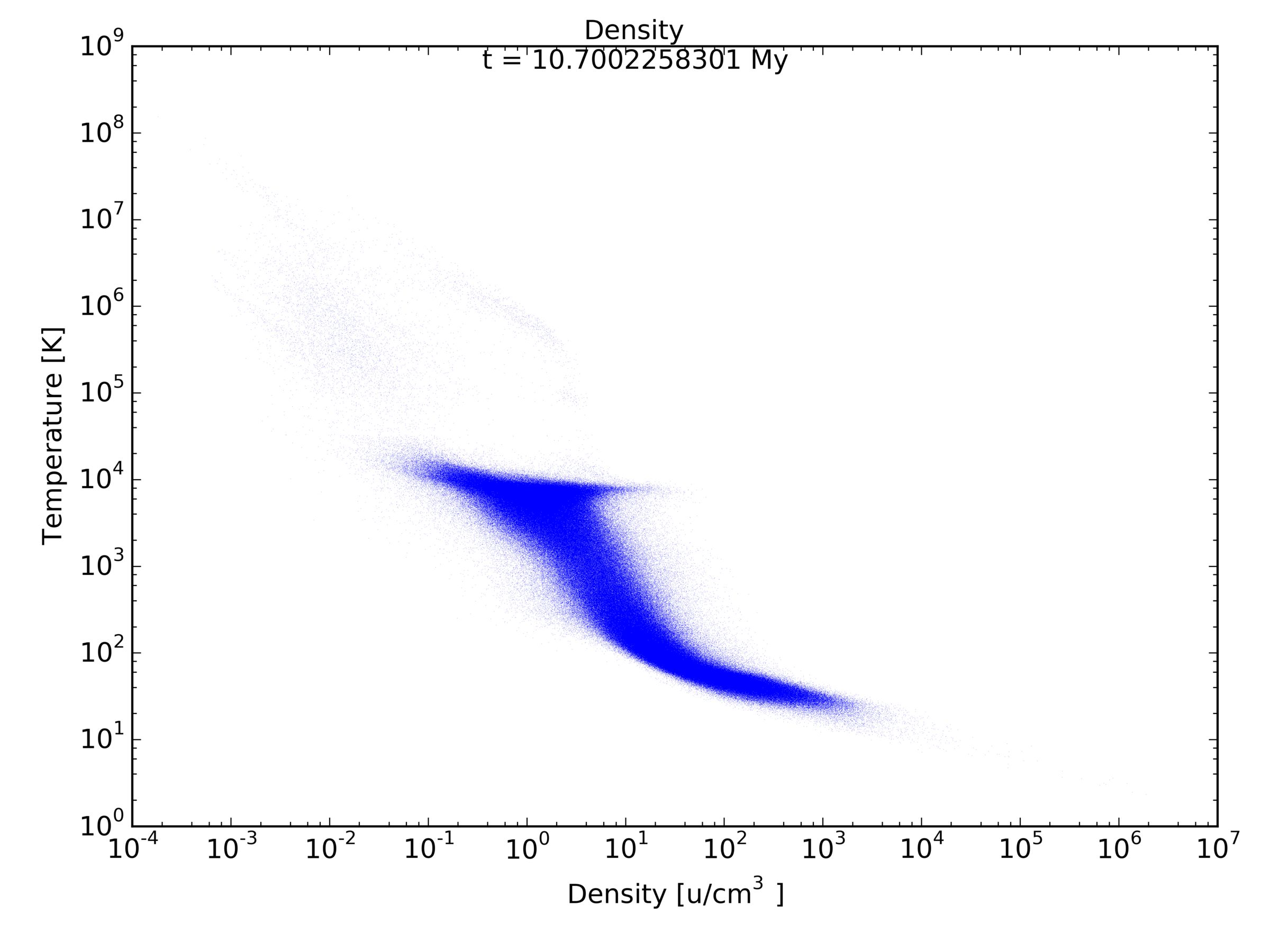}}
\caption{\DUrole{label}{fig:1u-feedback-400pc-scatter-107}
Density vs. temperature just before the creation of sinks.}
\end{figure}

Before the creation of sink particles already some hot, low density gas is present,
and cold, dense gas has begun to form (Fig.
\DUrole{ref}{fig:1u-feedback-400pc-scatter-107}).

\begin{figure}
\noindent\makebox[\textwidth][c]{\includegraphics[width=1.000\linewidth]{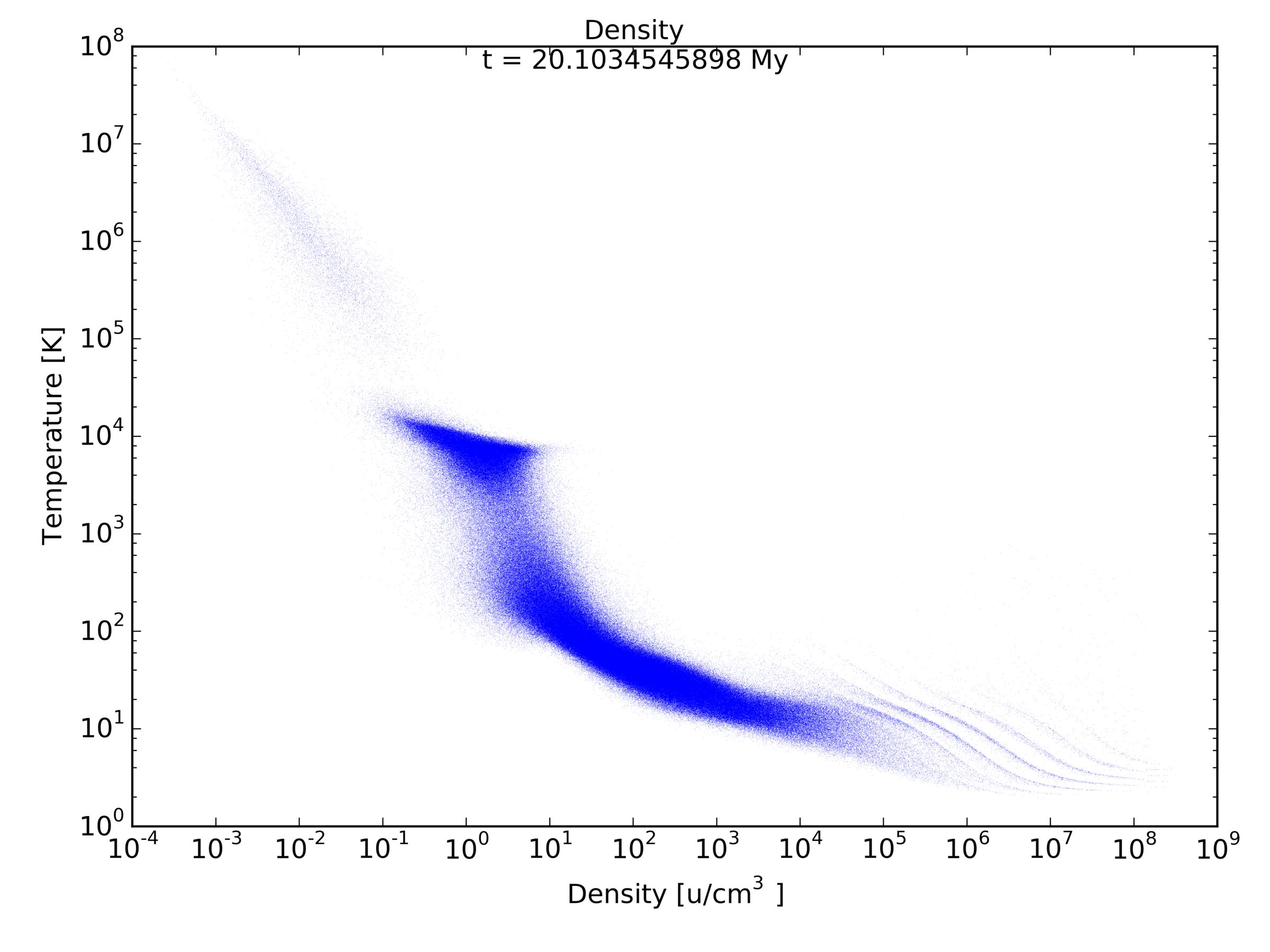}}
\caption{\DUrole{label}{fig:1u-feedback-400pc-scatter-201}
Density vs. temperature at $t = {20.1}{Mys}$.}
\end{figure}

At the end of the simulation, three more or less distinct phases are discernible
in the density-temperature scatter plot (Fig.
\DUrole{ref}{fig:1u-feedback-400pc-scatter-201}): thin gas hotter than
$\SI{e5}{K}$, warm gas at about $\SI{e4}{K}$ and cold, dense gas.
The temperature structure is similar to that of the three-phase medium of
\DUrole{citet}{McKee1977}.

In the coldest parts bands are visible in the density-temperature phase
space. These might be artefacts caused by interaction of the gas with sink
particles, that are created in these densest parts or more likely
differences in shielding and pressure in individual collapsing cores.

\begin{figure}
  \centering
  \includegraphics[width=0.95\linewidth]{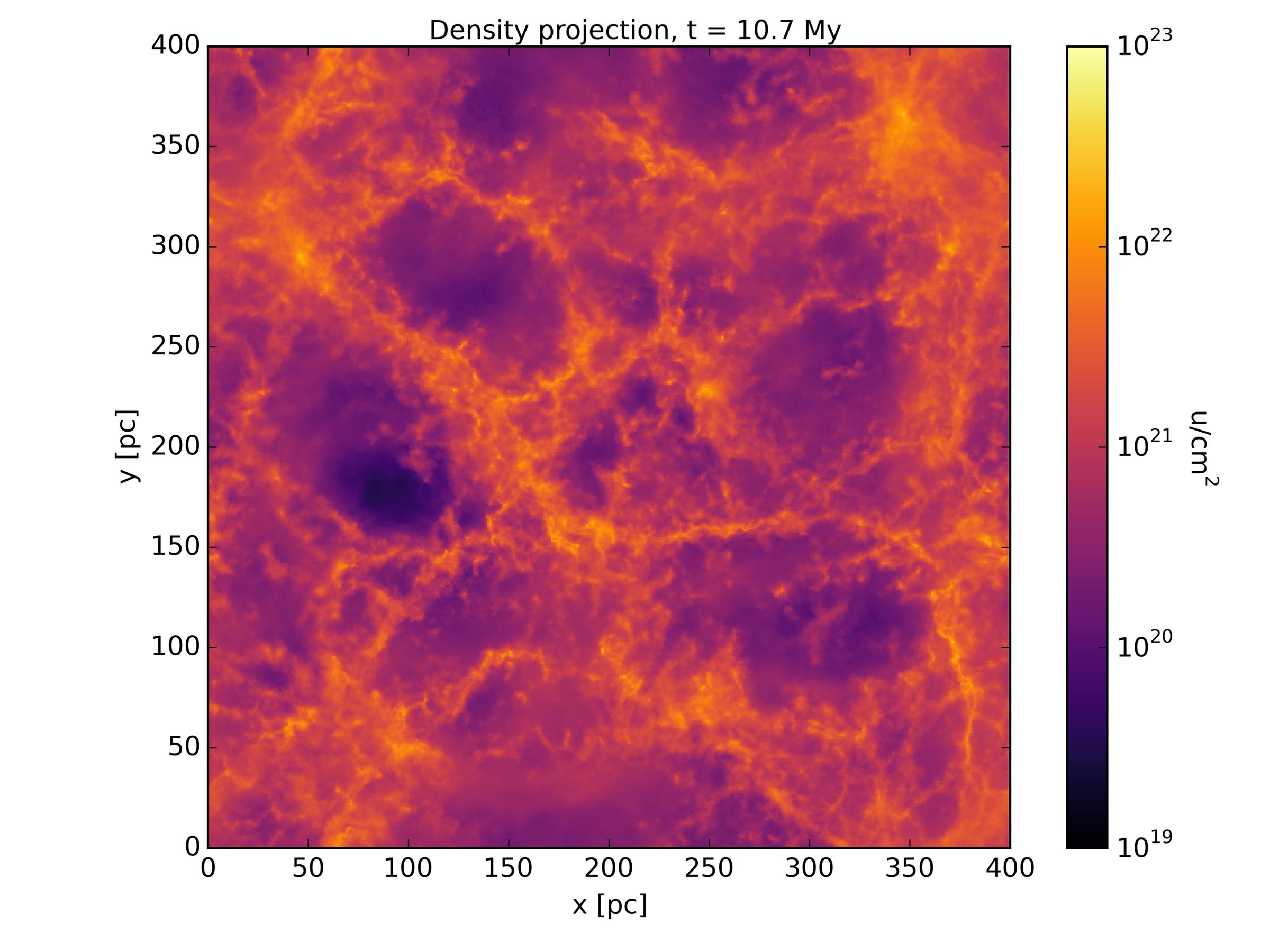}
  \centering
  \includegraphics[width=0.95\linewidth]{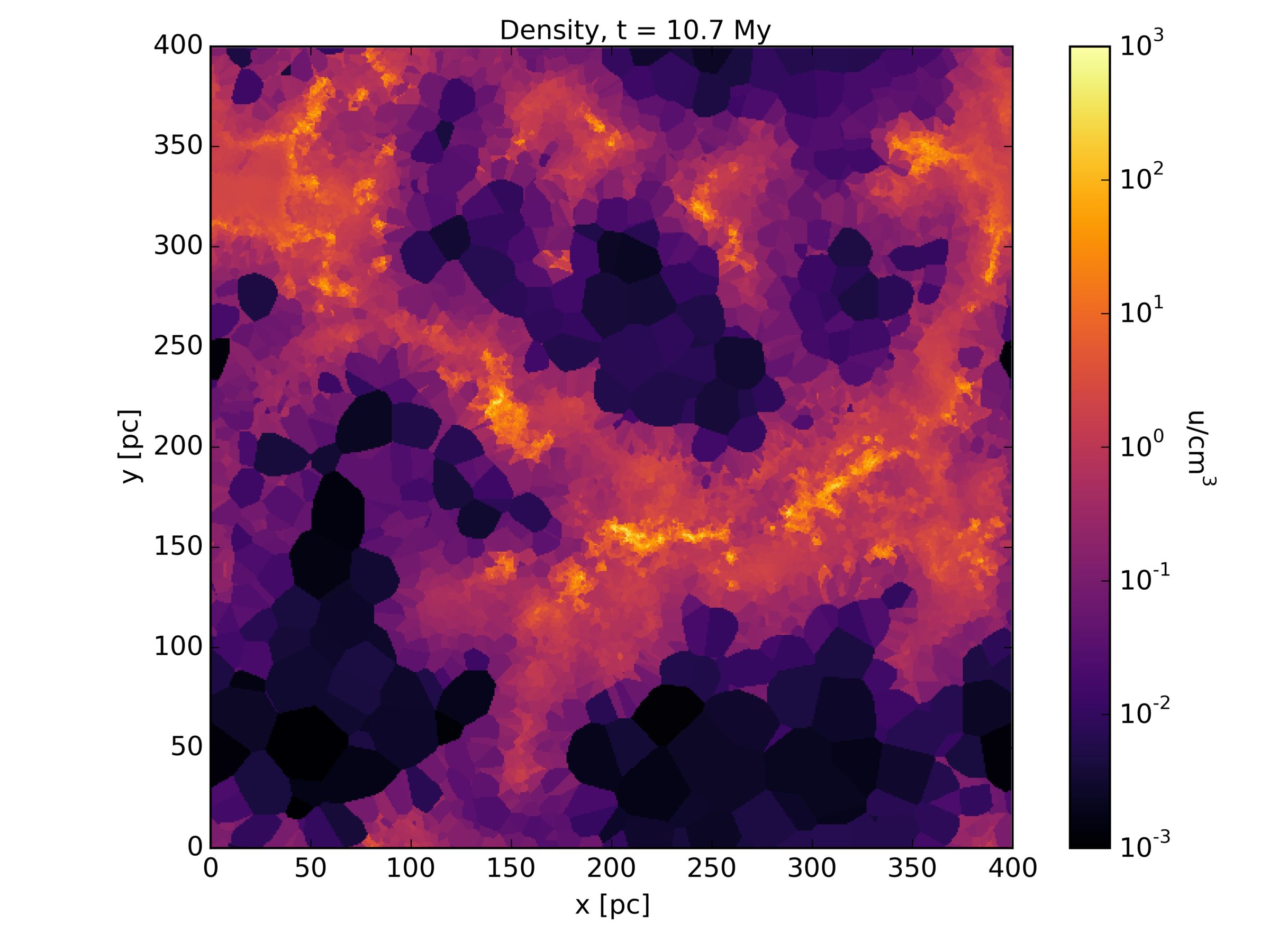}
  \caption{
    Projection and slice of density, with initial density of \SI{1}{u/cm^3}
    before the creation of sink particles ($t = \SI{10.7}{Mys}$).
  }
  \label{fig:1u-feedback-400pc-density-no-sinks}
\end{figure}

\begin{figure}
  \centering
  \includegraphics[width=0.95\linewidth]{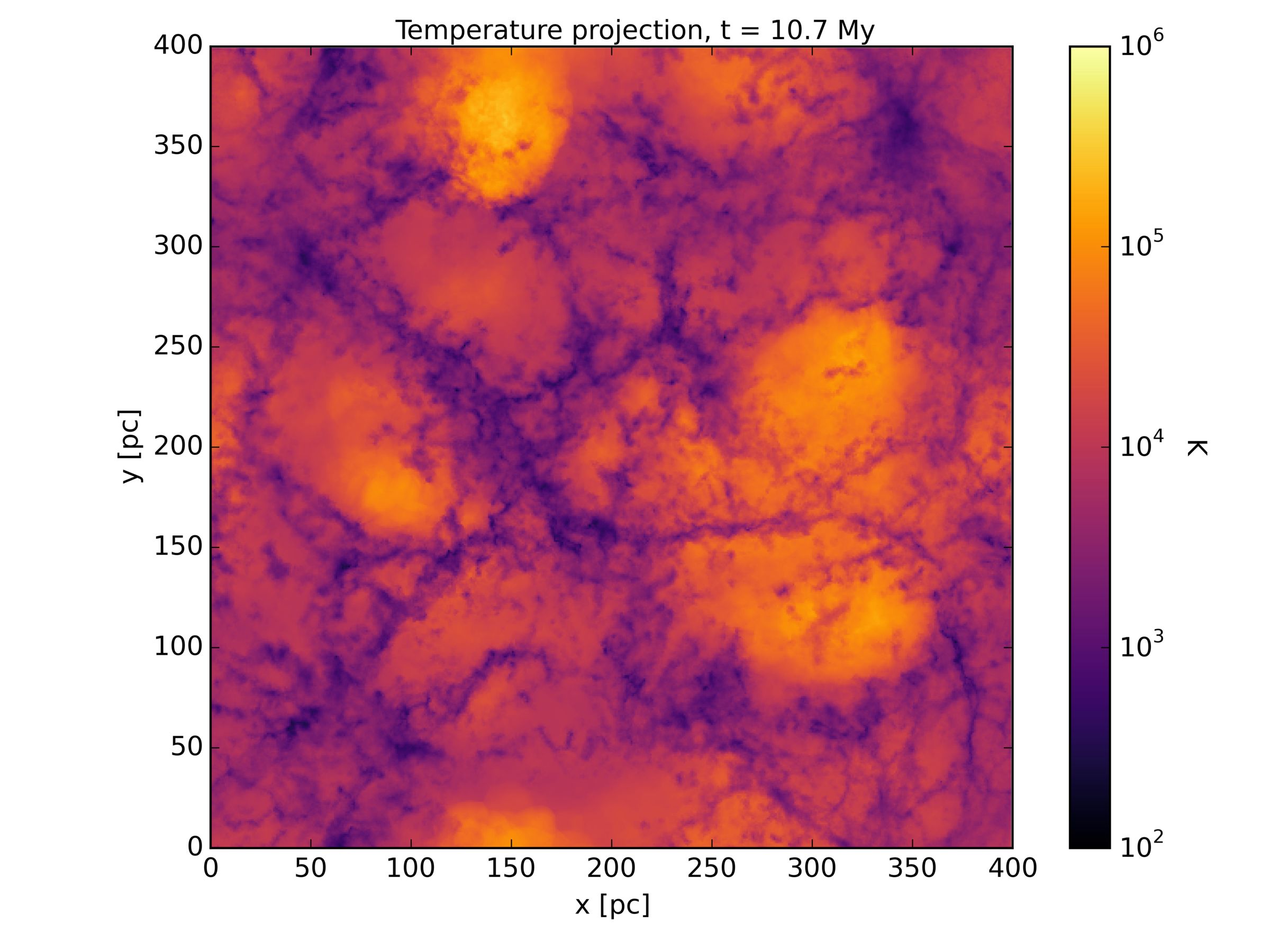}
  \centering
  \includegraphics[width=0.95\linewidth]{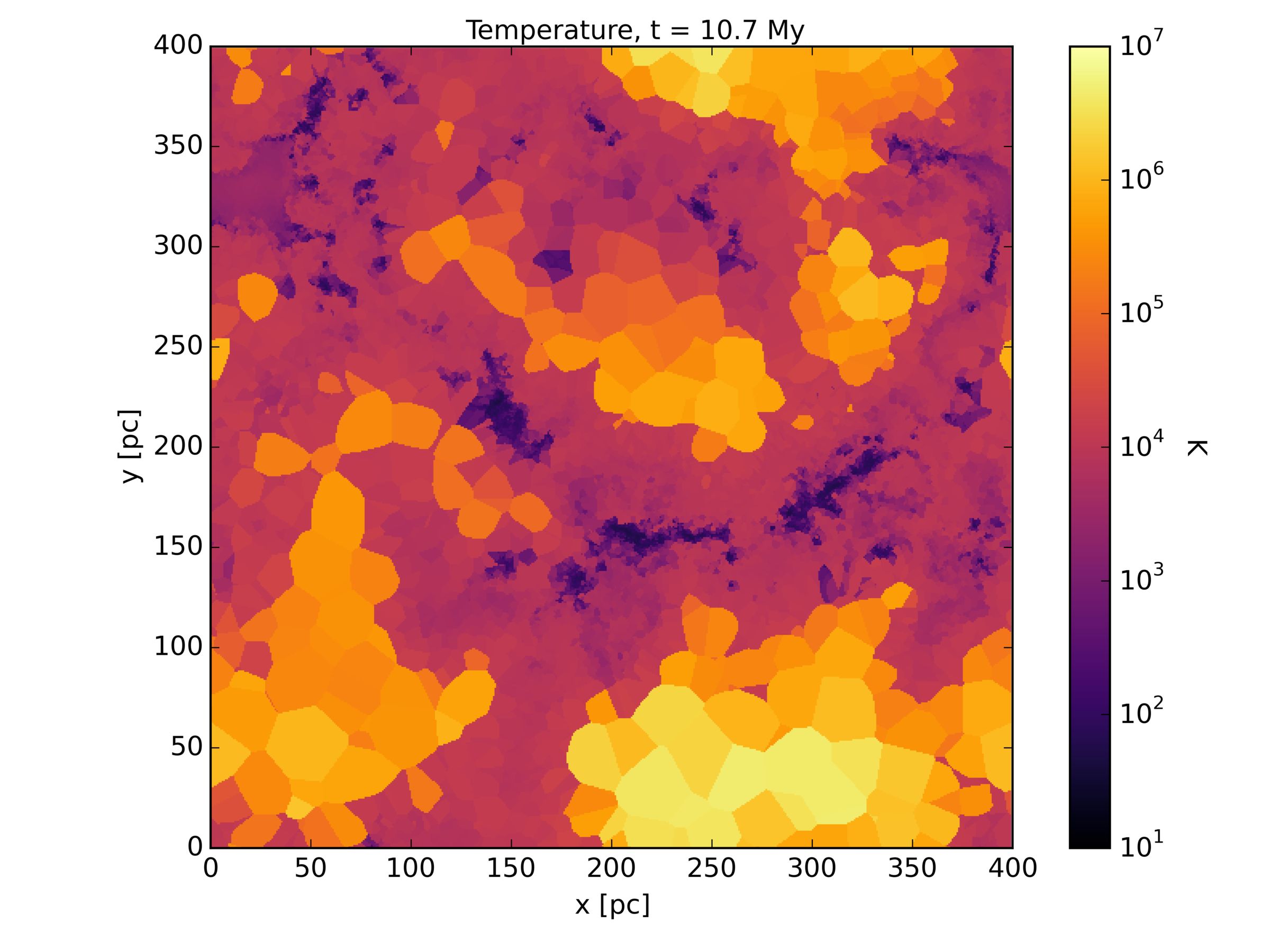}
  \caption{
    Density-weighted mean temperature along the line of sight and slice of temperature, with initial density of
    \SI{1}{u/cm^3} before the creation of sink particles ($t = \SI{10.7}{Mys}$).
  }
  \label{fig:1u-feedback-400pc-temp-no-sinks}
\end{figure}

\begin{figure}
  \centering
  \includegraphics[width=0.95\linewidth]{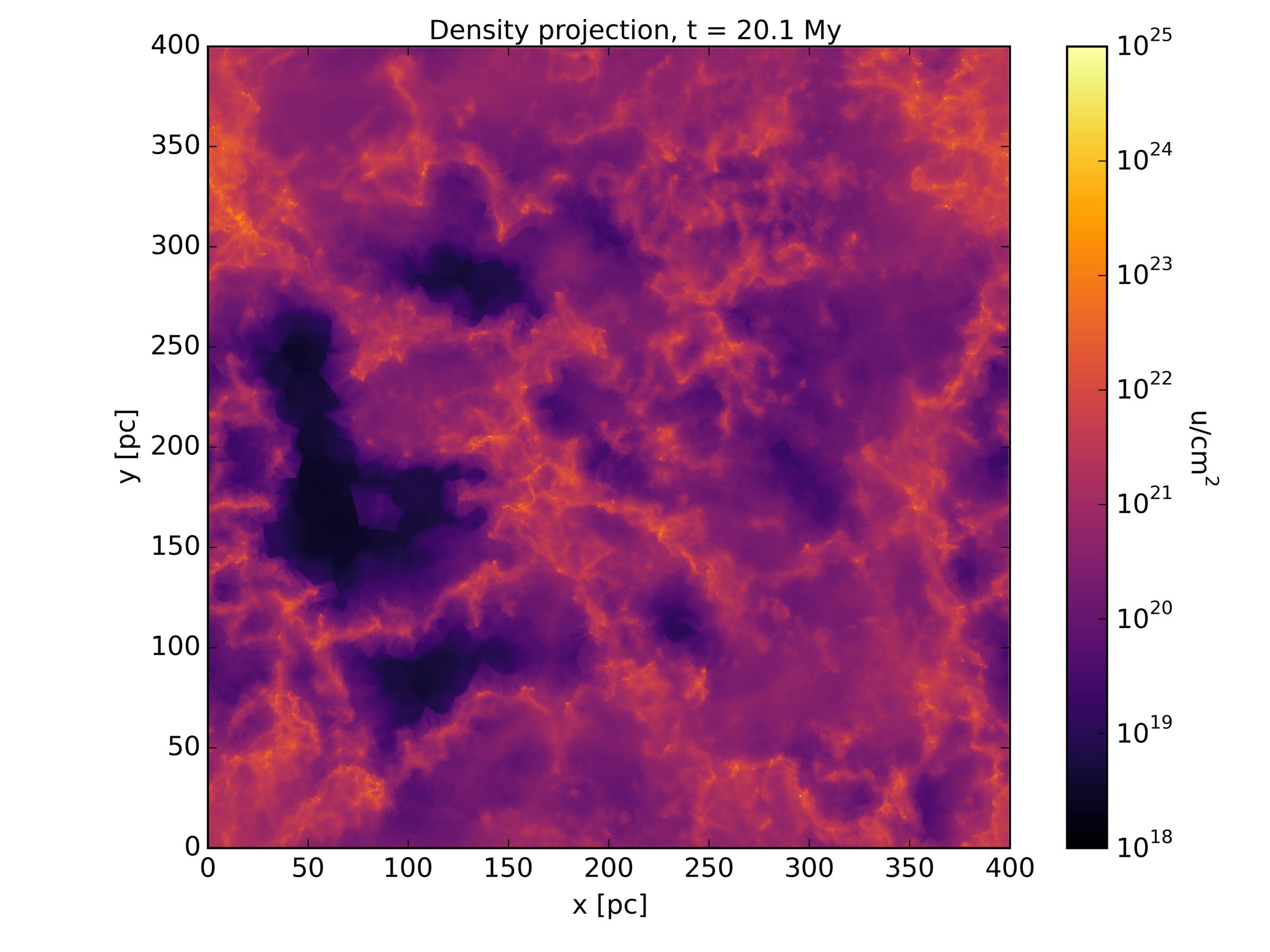}
  \centering
  \includegraphics[width=0.95\linewidth]{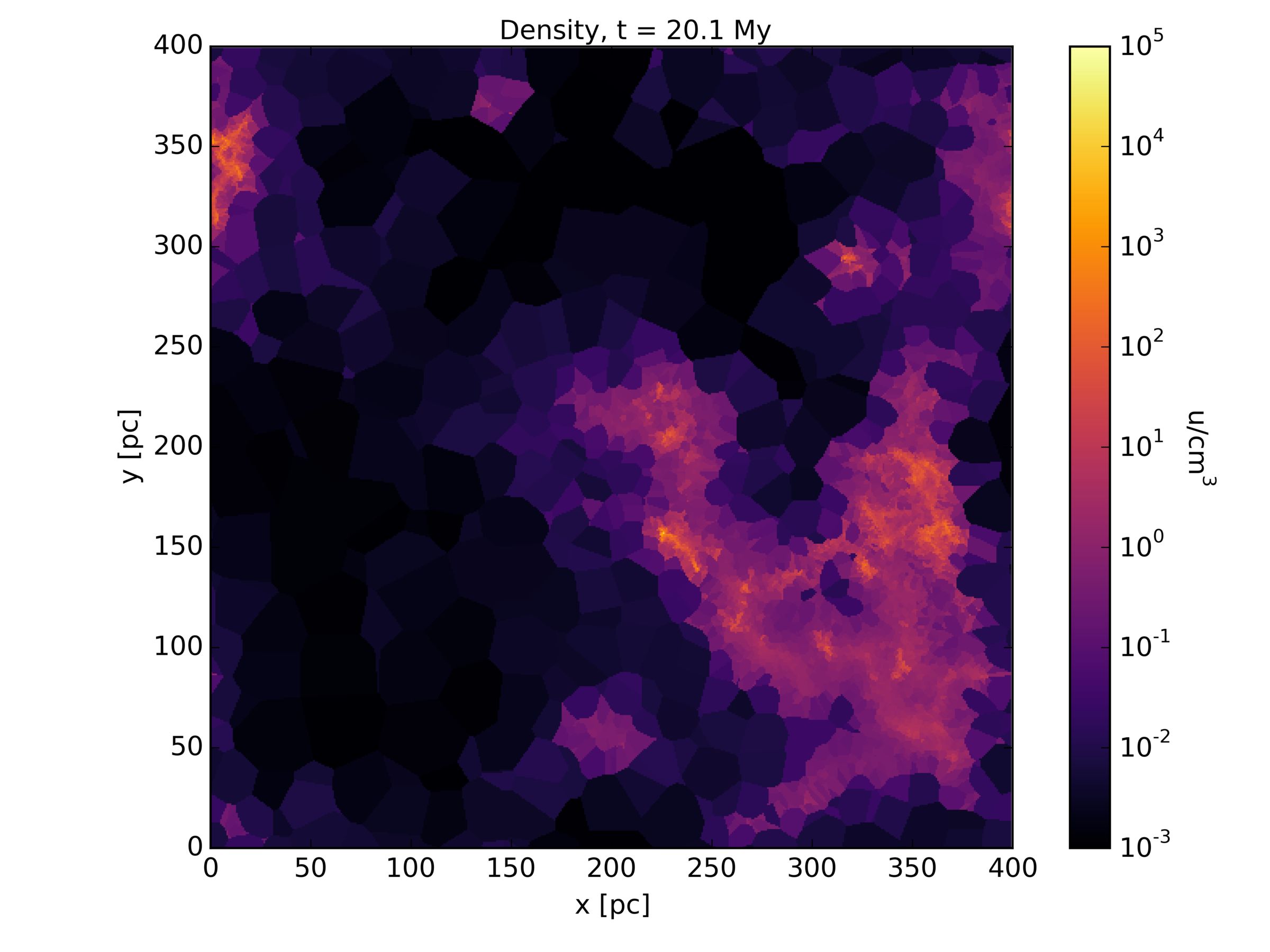}
  \caption{
     Projection and slice of density at the end of the simulation ($t = \SI{20.1}{Mys}$), with initial density of \SI{1}{u/cm^3}.
  }
  \label{fig:1u-feedback-400pc-density}
\end{figure}

\begin{figure}
  \centering
  \includegraphics[width=0.95\linewidth]{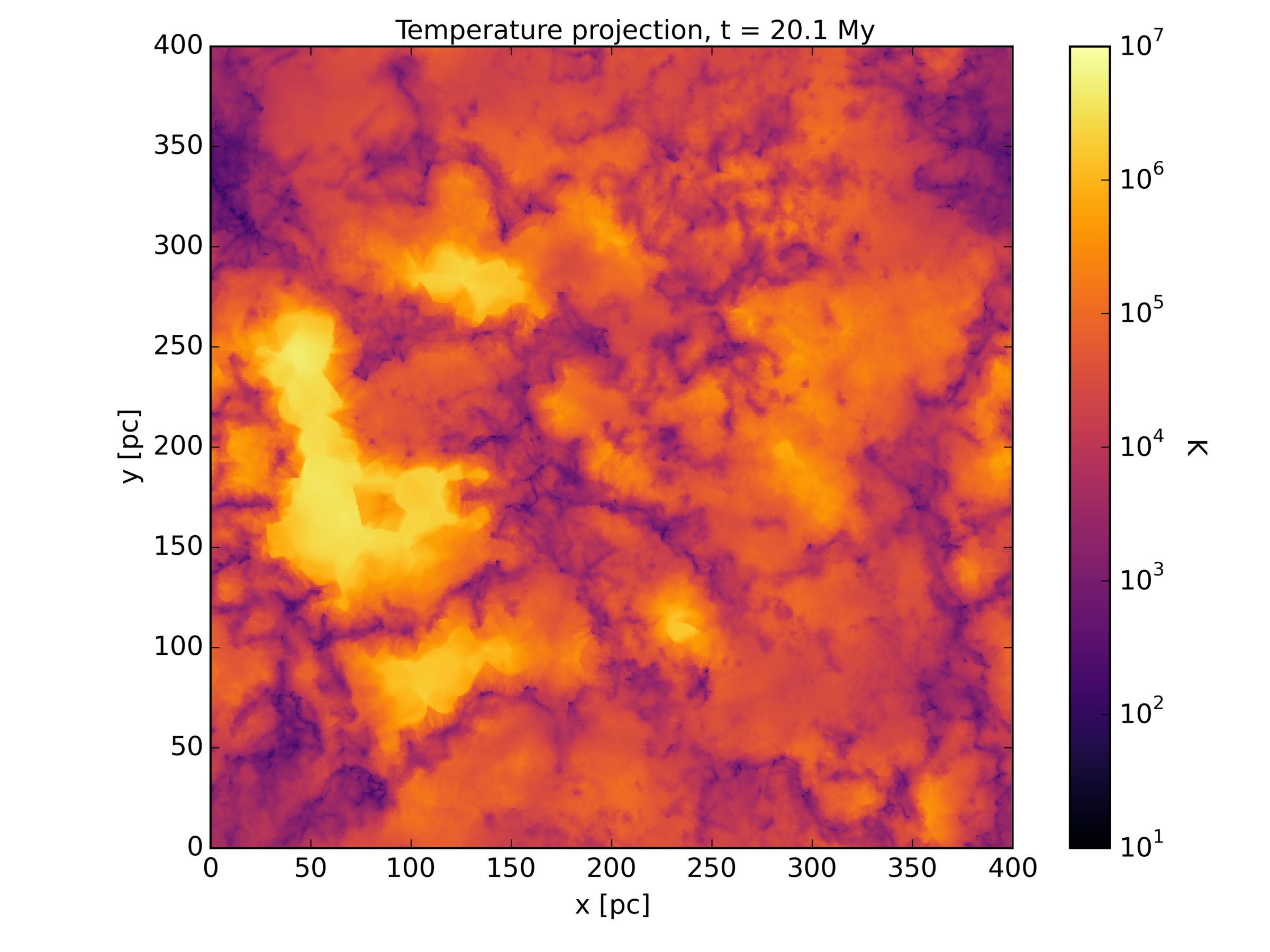}
  \centering
  \includegraphics[width=0.95\linewidth]{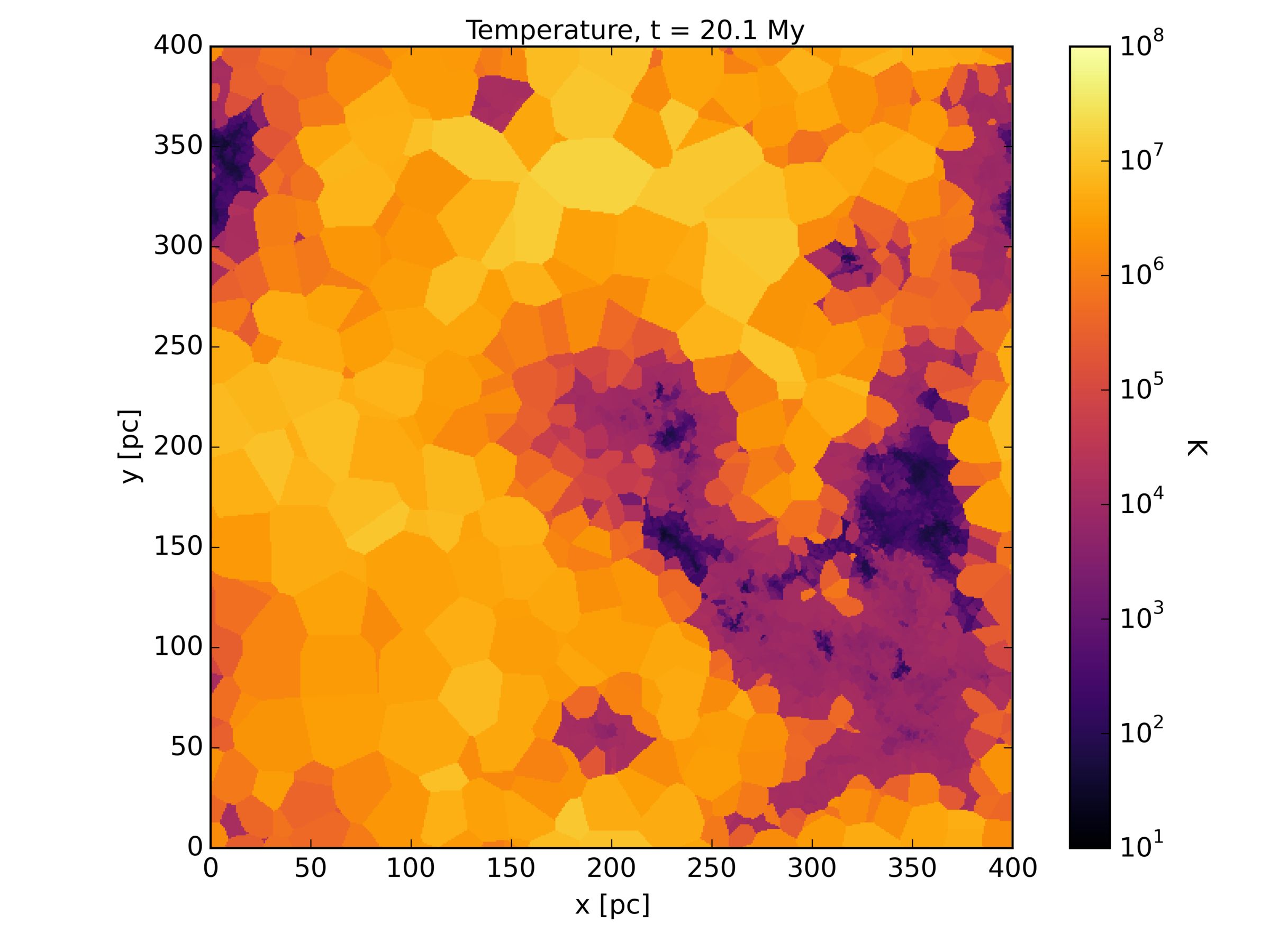}
  \caption{
  Density-weighted mean temperature along the line of sight and slice of temperature at the end of the simulation ($t = \SI{20.1}{Mys}$), with an uniform density of \SI{1}{u/cm^3} at the beginning of the simulation.
  }
  \label{fig:1u-feedback-400pc-temp}
\end{figure}

Large bubbles of hot gas, and structures of cold, dense, collapsed gas are also
visible in the temperature and density plots and slices (Fig.
\DUrole{ref}{fig:1u-feedback-400pc-density-no-sinks},
\DUrole{ref}{fig:1u-feedback-400pc-temp-no-sinks}, \DUrole{ref}{fig:1u-feedback-400pc-density}
and \DUrole{ref}{fig:1u-feedback-400pc-temp}). As the cell mass is held about constant,
the low density gas cells are quite large, reaching a diameter of over
$\SI{50}{pc}$ at the end of the simulation.

\begin{figure}
\noindent\makebox[\textwidth][c]{\includegraphics[width=1.000\linewidth]{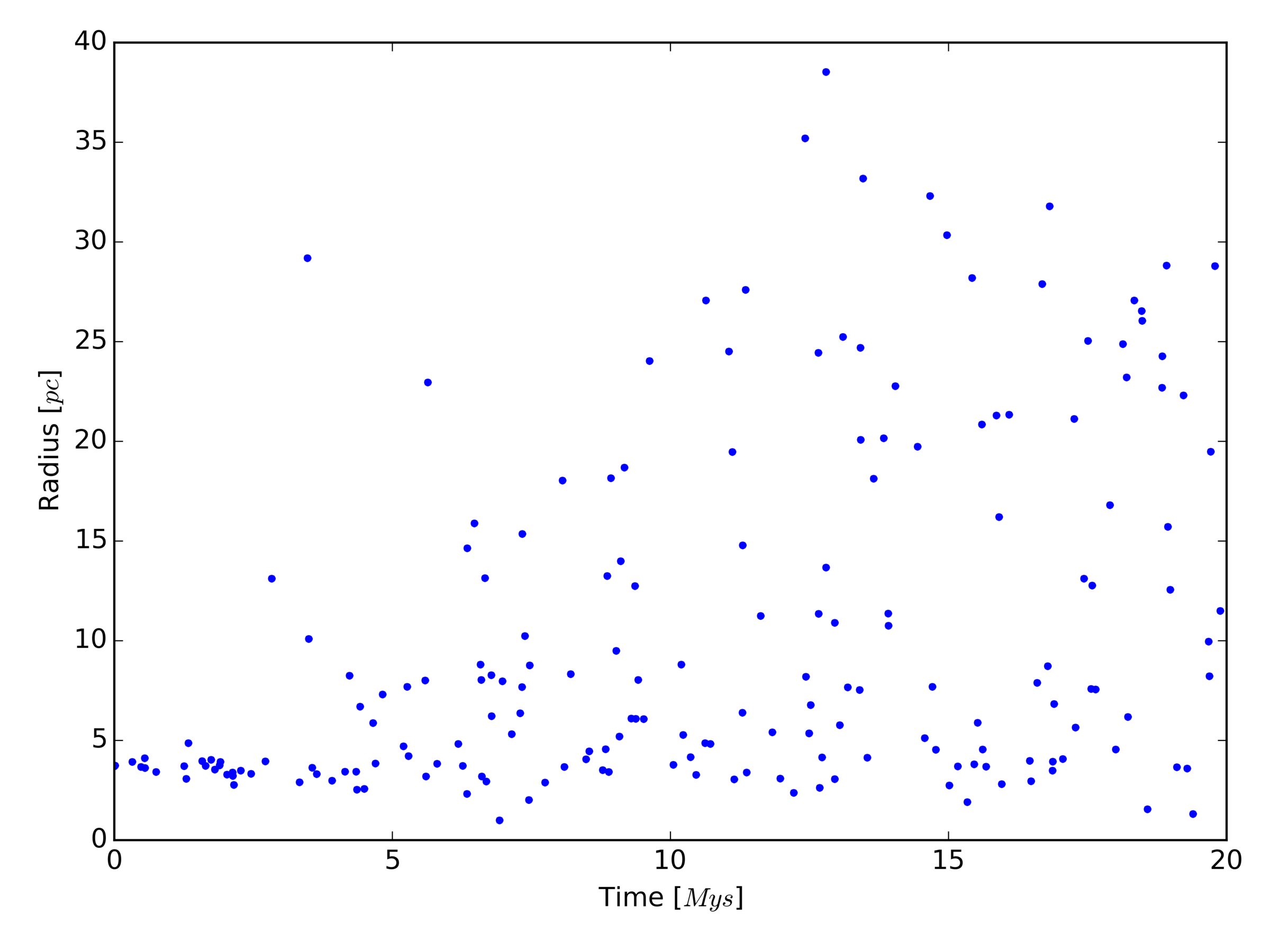}}
\caption{\DUrole{label}{fig:1u-feedback-400pc-sne-radius}
Radii of the supernovas.}
\end{figure}

\begin{figure}
\noindent\makebox[\textwidth][c]{\includegraphics[width=1.000\linewidth]{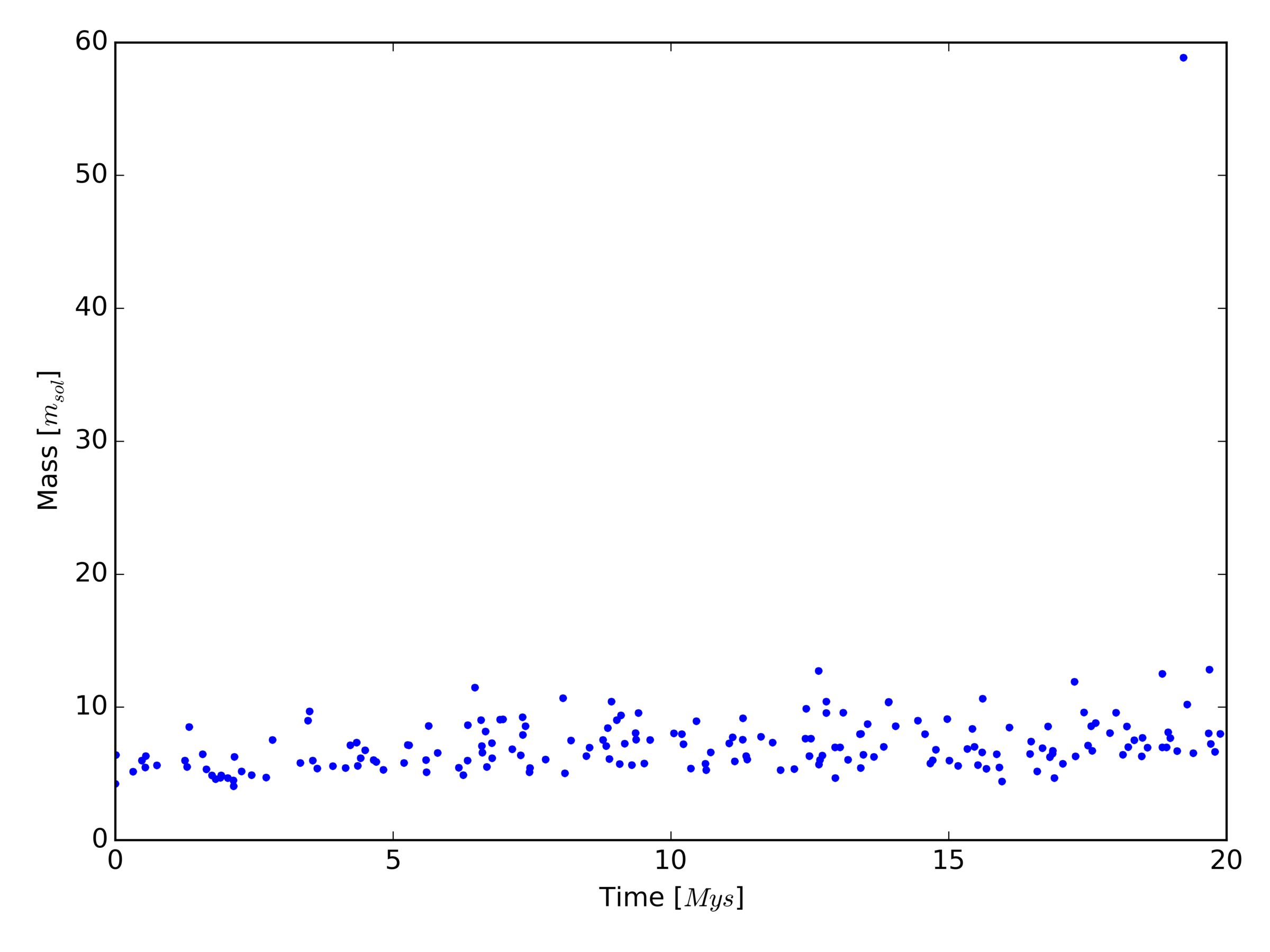}}
\caption{\DUrole{label}{fig:1u-feedback-400pc-sne-mass}
Mass of the supernovas.}
\end{figure}

\begin{figure}
\noindent\makebox[\textwidth][c]{\includegraphics[width=1.000\linewidth]{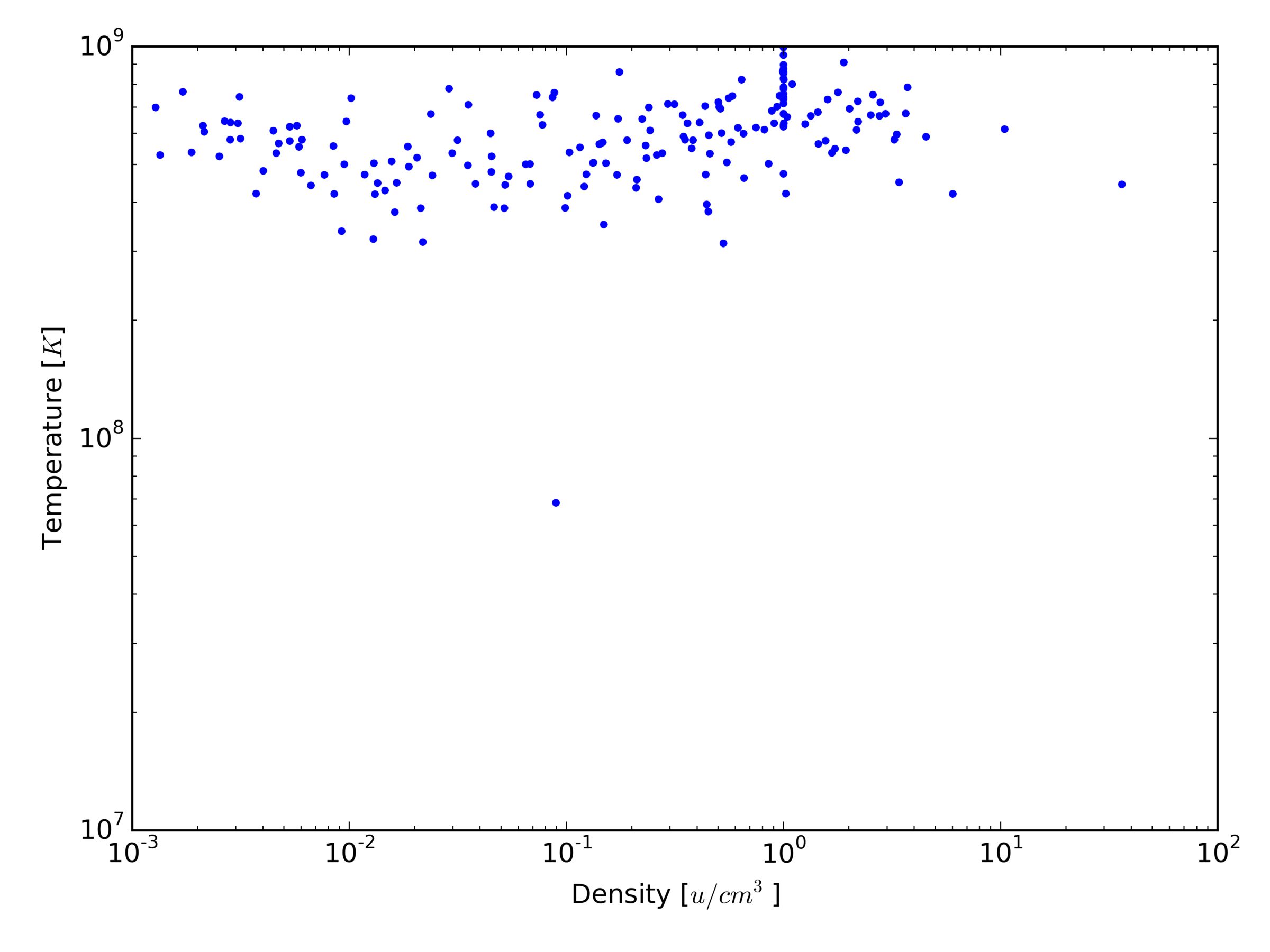}}
\caption{\DUrole{label}{fig:1u-feedback-400pc-sne-temp-vs-density}}
\begin{DUlegend}
Initial temperatures vs. densities of the supernova remnants.
\end{DUlegend}
\end{figure}

Most supernovas were created in thin gas, with all but one mean density of the
SNR under $\SI{15}{u/cm^3}$. As the dense gas only fills a small fraction
of the box, it is unlikely that a supernova is set off near a dense gas cell.
While the radius of the SNR varies widely (about $1$ to $45 \si{pc}$,
see Fig. \DUrole{ref}{fig:1u-feedback-400pc-sne-radius}), the mass is more constrained
($\approx 4 - 15 \si{M_\odot}$, see Fig.
\DUrole{ref}{fig:1u-feedback-400pc-sne-mass}), with one heavy SN, the only one that
might have occurred in a dense region. As the target mass for the SNR was set
quite low, for most the minimal particle number criterion of six cells applied.
Only one encompassed seven particles.

The initial temperature of the SN, estimated from the specific energy density,
shows that all supernovae but one were hot ($T > \SI{4e8}{K}$, see Fig.
\DUrole{ref}{fig:1u-feedback-400pc-sne-temp-vs-density}), so were
likely not affected by overcooling. As the development of individual SN was not
tracked, it is not quite clear though, whether SNR with large initial radii actually
captured the Sedov-Taylor phase correctly.

\begin{figure}
\noindent\makebox[\textwidth][c]{\includegraphics[width=1.000\linewidth]{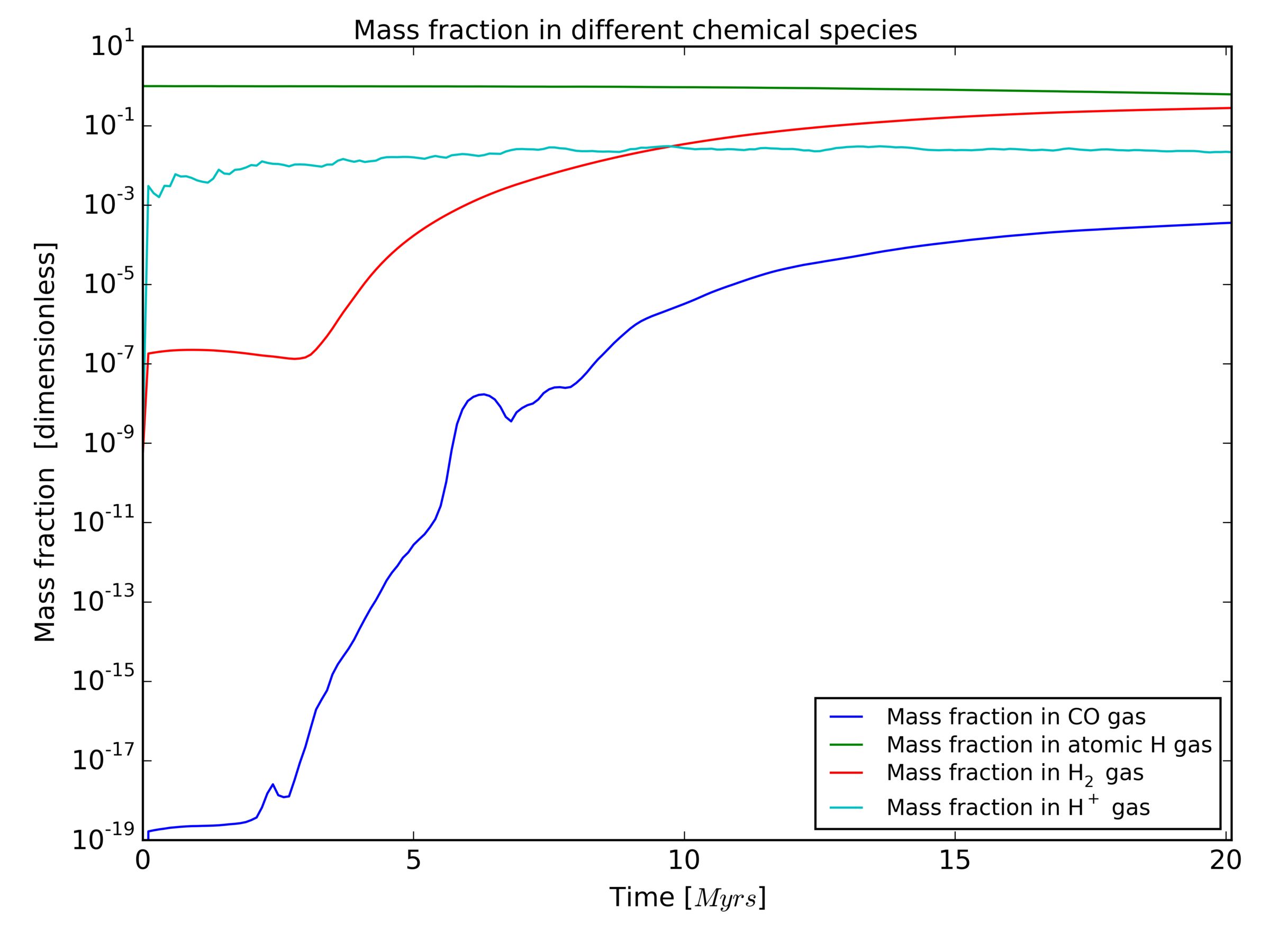}}
\caption{\DUrole{label}{fig:1u-feedback-400pc-fractions-abundances}
Mass fraction of the chemical species, simulation with initial density $\SI{1}{u/cm^3}$. Note that SGChem uses fractional abundances internally.}
\end{figure}

\begin{figure}
\noindent\makebox[\textwidth][c]{\includegraphics[width=1.000\linewidth]{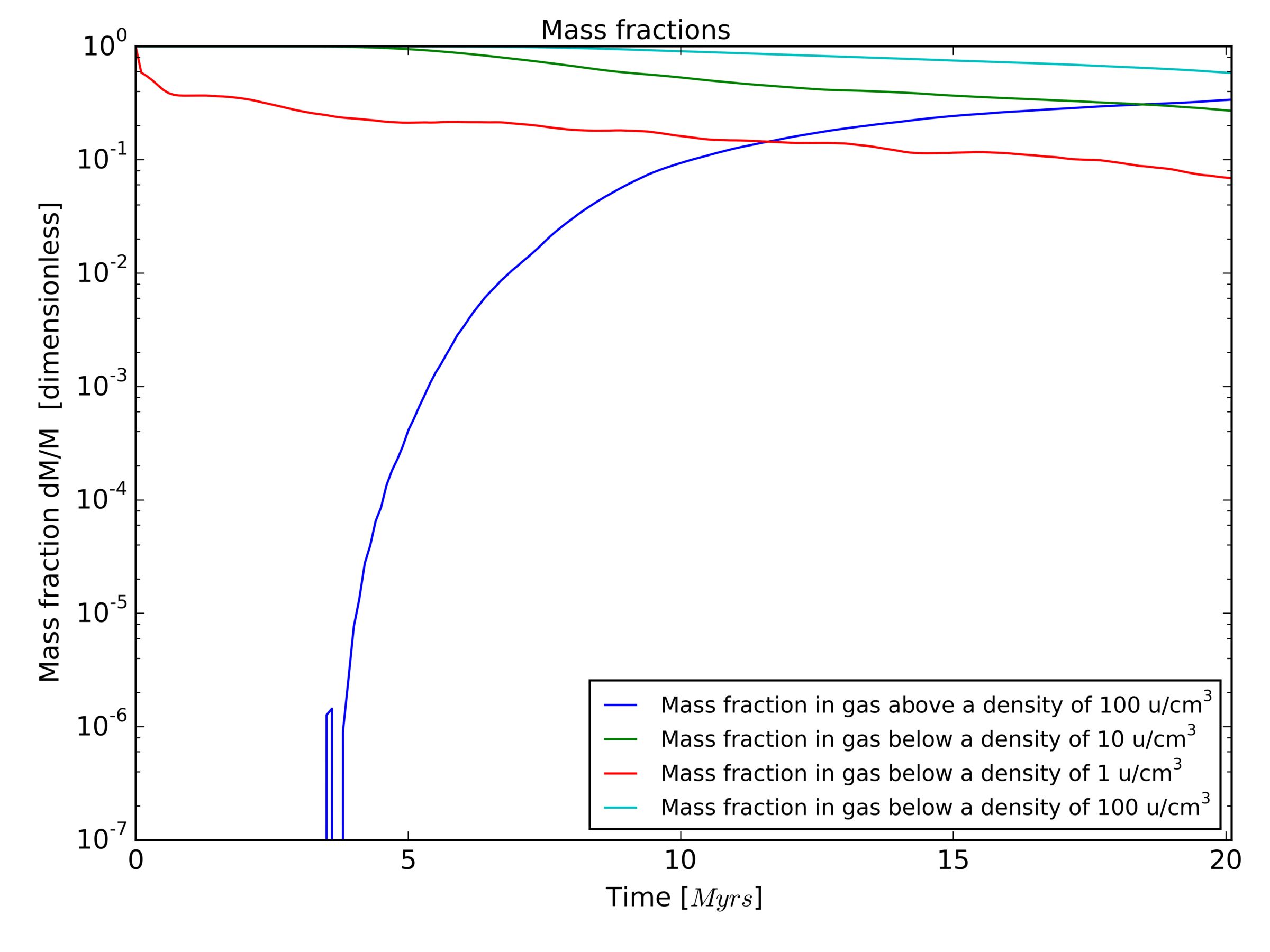}}
\caption{\DUrole{label}{fig:1u-feedback-400pc-fractions}
Mass fraction of the gas below and above certain densities, simulation with initial density $\SI{1}{u/cm^3}$.}
\end{figure}

As the supernovas happen from the beginning of the simulation, a significant
amount of gas is ionized almost from the start. At first almost no \DUrole{latex}{\co}
is present. As dense and cold gas is formed (e.g. see Fig.
\DUrole{ref}{fig:1u-feedback-400pc-fractions}) also \DUrole{latex}{\co} can form, and its
abundance rises quickly, in turn with an increase of molecular hydrogen at about
$\SI{3}{Mys}$ (see also Fig.
\DUrole{ref}{fig:1u-feedback-400pc-fractions-abundances}). While the ionized fraction
soon reaches an approximately constant value of $\approx
\SI{1}{\percent}$, molecular hydrogen and \DUrole{latex}{\co} still have not settled
into equilibrium at the end of the simulation.  As most supernova events are
happening outside of dense clouds, they are not able to disrupt the dense
molecular cores.  E.g. in comparison only half of the gas in the
Milkyway is in dense clouds \DUrole{citep}{Ferriere2001}. \DUrole{citet}{Gatto2014} also see a
larger than 50\% fraction of gas in dense clouds, while peak driving leads to a
filamentary ISM.

Simulations with mixed driving, i.e. setting of a fraction of the SN events inside
dense cores (e.g. at sink particle locations) might be interesting to see if SN
are able to destroy molecular regions, as it is seen e.g. by \DUrole{citet}{Gatto2014}.

\begin{figure}
\noindent\makebox[\textwidth][c]{\includegraphics[width=1.000\linewidth]{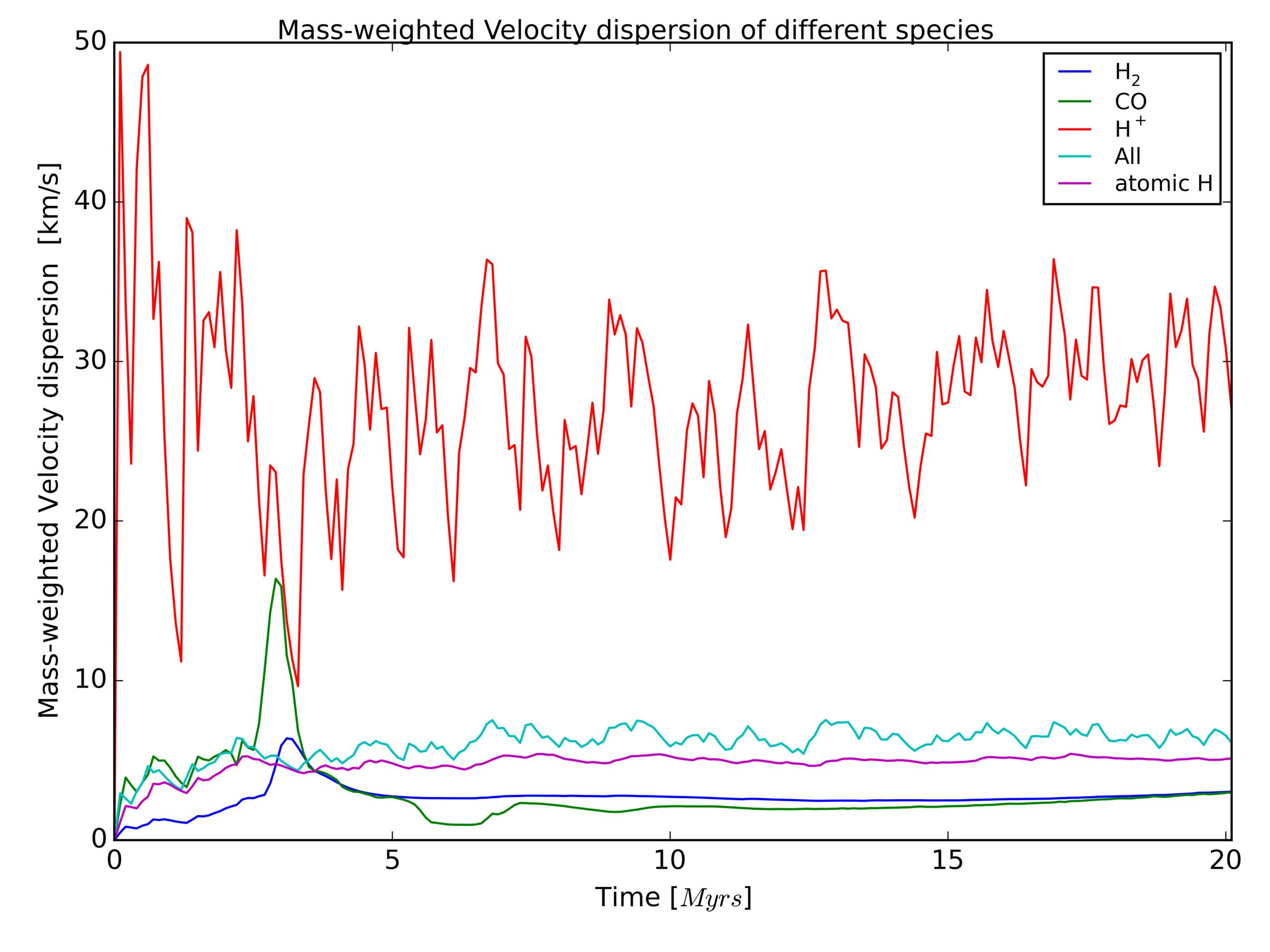}}
\caption{\DUrole{label}{fig:1u-feedback-400pc-velocity-dispersion-species}
Average 1d mass-weighted
velocity dispersion of different chemical species, simulation with initial
density $\SI{1}{u/cm^3}$.}
\end{figure}

The average 1d mass-weighted velocity dispersions (see Fig.
\DUrole{ref}{fig:1u-feedback-400pc-velocity-dispersion-species}) after only a few
megayears reaches $20-50 \si{km/s}$ for ionized hydrogen and $5-8
\si{km/s}$ for other gas. \DUrole{citet}{Gatto2014} report similar values in runs with
$\SI{1}{u/cm^3}$ and random driving with similar event frequency.

\begin{figure}
\noindent\makebox[\textwidth][c]{\includegraphics[width=1.000\linewidth]{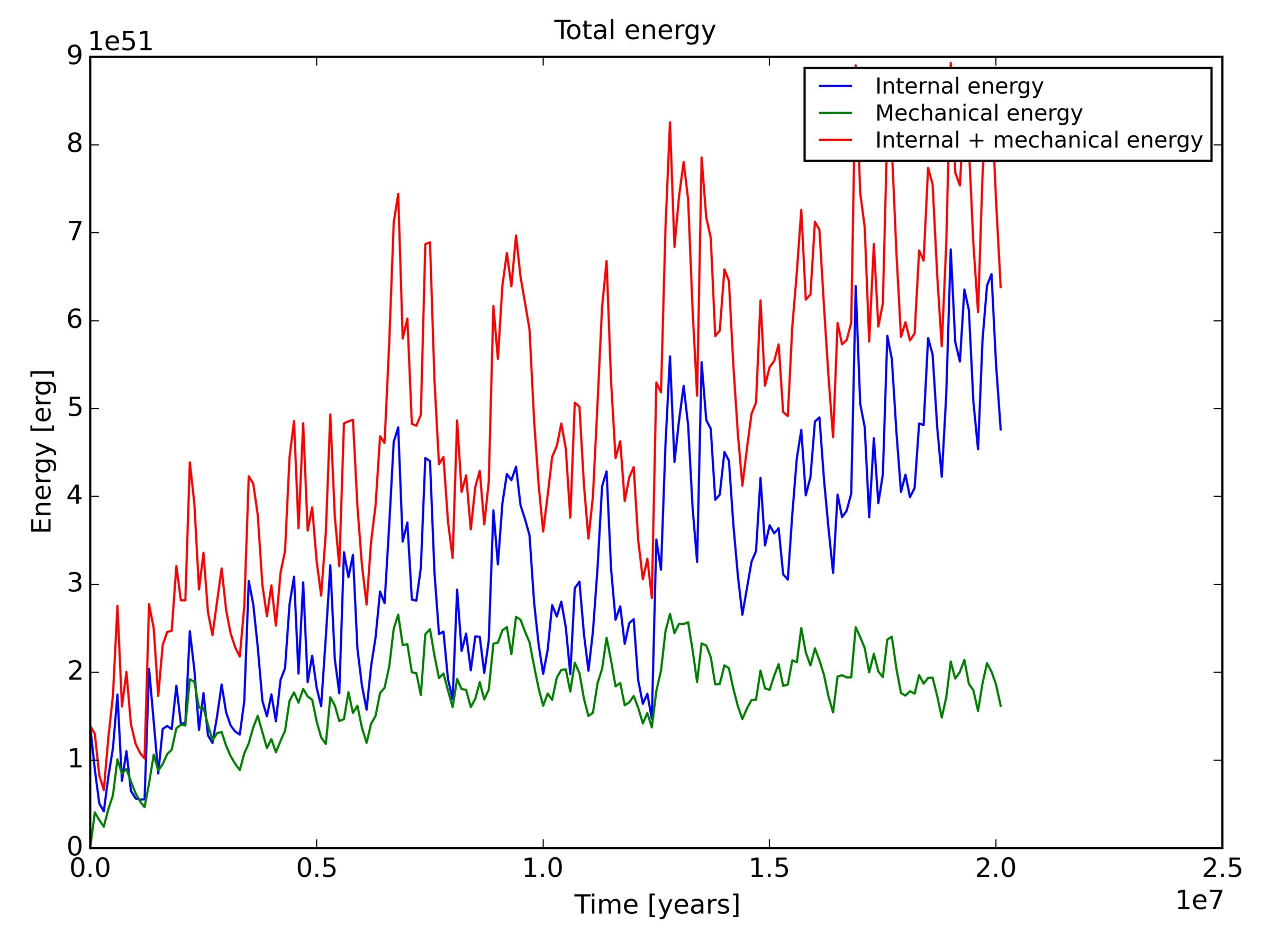}}
\caption{\DUrole{label}{fig:1u-feedback-400pc-total-energy}
Total kinetic and thermal energy
in the box over time, simulation with initial density $\SI{1}{u/cm^3}$.
Energy lost in radiative cooling is not shown.}
\end{figure}

Most of the injected energy gets lost, likely radiated away (see Fig.
\DUrole{ref}{fig:1u-feedback-400pc-total-energy}). While an upward trend is visible for
thermal energy, e.g. the box heats up, the kinetic energy stays roughly in the
range of 1.5 to \DUrole{latex}{\SI{2.5e51}{erg}}. This is consistent with the roughly
constant velocity dispersions.

\begin{figure}
\noindent\makebox[\textwidth][c]{\includegraphics[width=1.000\linewidth]{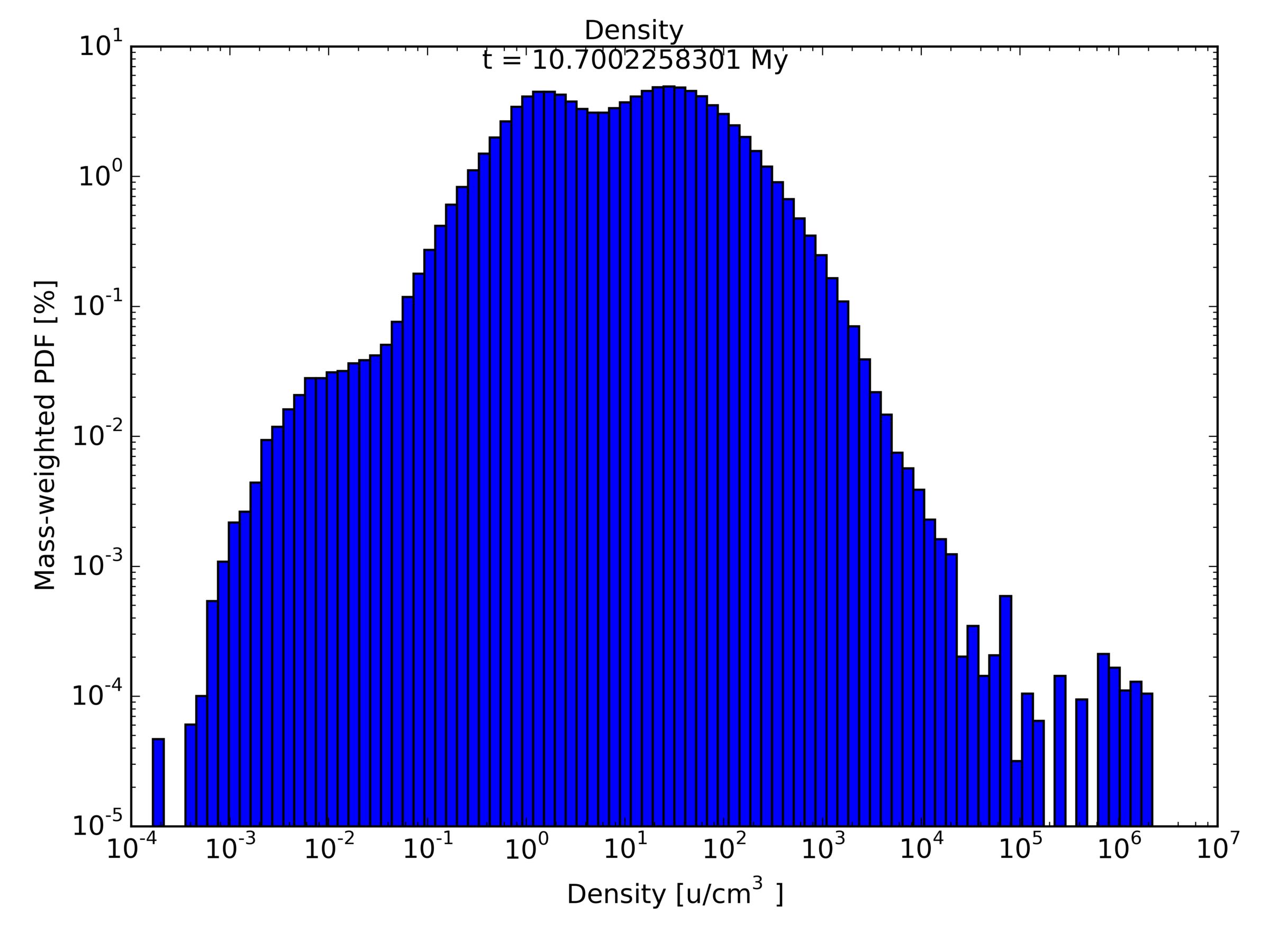}}
\caption{\DUrole{label}{fig:1u-feedback-400pc-hist-density-107}
Mass-weighted density PDF at $t = \SI{10.7}{Mys}$, simulation with initial density $\SI{1}{u/cm^3}$.}
\end{figure}

\begin{figure}
\noindent\makebox[\textwidth][c]{\includegraphics[width=1.000\linewidth]{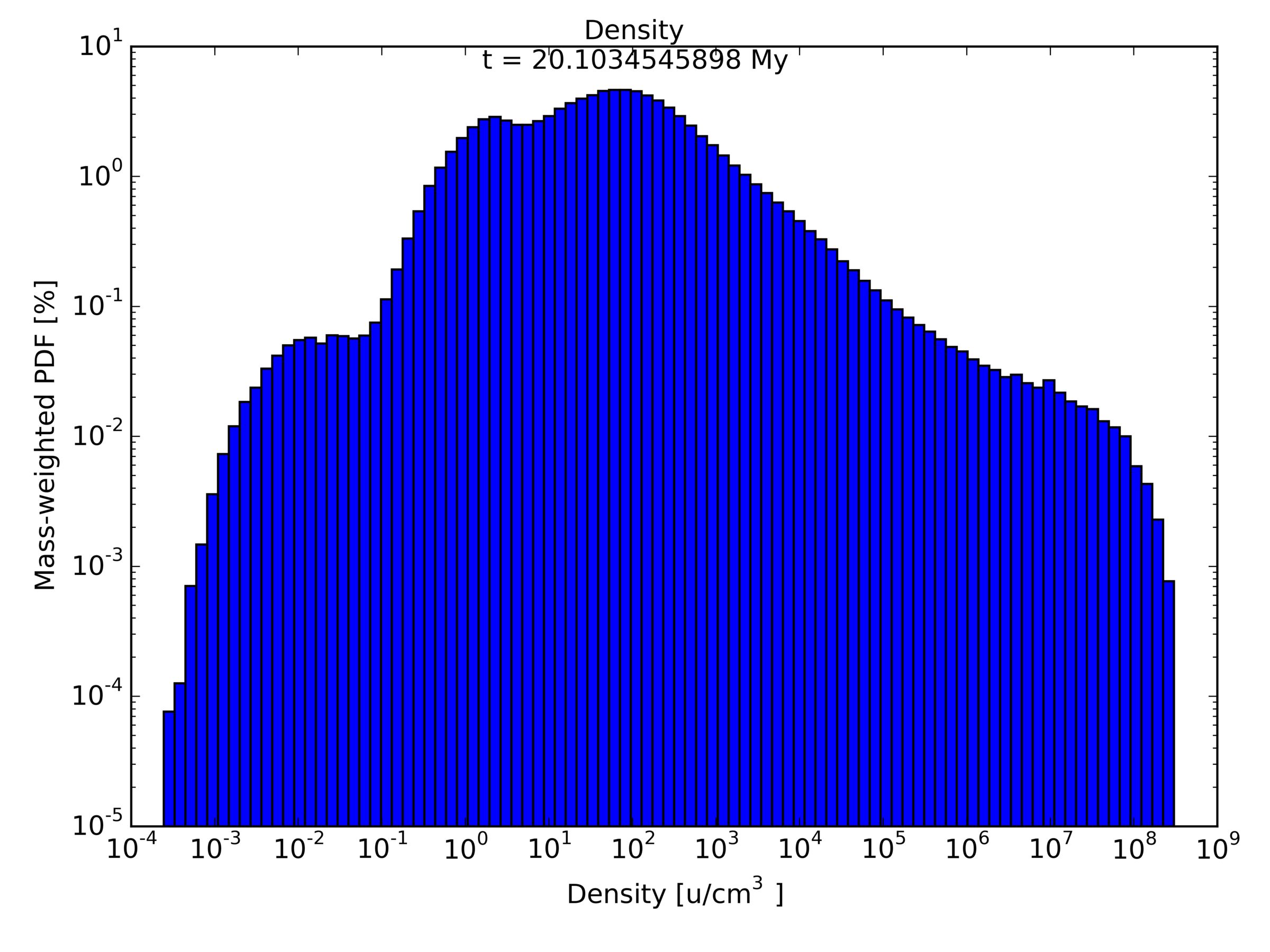}}
\caption{\DUrole{label}{fig:1u-feedback-400pc-hist-density-201}
Mass-weighted density PDF at $t = \SI{20.1}{Mys}$, simulation with initial density $\SI{1}{u/cm^3}$.}
\end{figure}

The mass-weighted density PDF in Fig.
\DUrole{ref}{fig:1u-feedback-400pc-hist-density-107} and Fig.
\DUrole{ref}{fig:1u-feedback-400pc-hist-density-201} show how at the end of the
simulation more than half of the mass has been captured in the high density
fraction. On contrast to the thermal runaway regime that \DUrole{citet}{Gatto2014} see
for random driving, where almost no gas at about $\sim \SI{1}{u/cm^3}$ is
left, in our simulation a significant amount of gas still is in this regime
though, more comparable to the PDF of their mixed driving scheme, but with more
hot gas.

The feedback simulation runs finished at the time of writing has still some quirks:
\newcounter{listcnt0}
\begin{list}{\arabic{listcnt0}.}
{
\usecounter{listcnt0}
\setlength{\rightmargin}{\leftmargin}
}

\item A single supernova was fired out of order and its coordinates are duplicate
of the following one at $t = \SI{12.4914}{Mys}$ and $t =
\SI{12.4913}{Mys}$ respectively. This is likely caused by a bug in the SN
routine. The Anderson-Darling statistic below was calculated from a time
sorted series. The impact of this bug should be small, as only one of 190 SN
is affected.

\item In the summation of the volume and mass of gas in the SN sphere an
out-of-bounds error was introduced while iterating over the particle
properties array, i.e. the each time one element after the end of the static
array was accessed. The impact of this bug on the simulation result is
likely small, as the data of the invalid data section when reinterpreted for
the “particle” structure must produce just the right coordinates to lay
inside a SN region. The bug was fixed in the mean time, but is still present
for the previous results.
\end{list}

As the simulation run took almost a week to calculate on 386 CPUs, because of
limited time at the end of writing the thesis and the cause of problem 1. and 2.
is not yet found, the analysis below is of a simulation run with those problems
uncorrected. The impact on the results are likely small, but a direct
comparison in a replication run will likely yield a slightly different outcome.

\section{Testing the driving scheme%
  \label{testing-the-driving-scheme}%
}

The mean time between supernovas in the simulation is $\SI{113}{kys}$.
This is slightly higher than the specified SN period of $\SI{102}{kys}$,
which is expected, as a supernova is only injected during the next occupied
timesteps for that it is scheduled. Also only one supernova can occur per
timestep, which might limit the number of SN in time intervals with large
timesteps. This is also visible in Fig. \DUrole{ref}{fig:1u-feedback-400pc-ecdf}, with
most of the ECDF laying below the CDF of the exponential distribution\DUfootnotemark{id6}{id7}{3}.
\DUfootnotetext{id7}{id6}{3}{%
The cumulative distribution function (CDF) $F_X(x)$ gives the
probability, that a continuous random variable of a probability distribution is
smaller than $x$, i.e. $F_X(x) = P(X < x)$
\DUrole{citep}{Wikipediacontributors2015}. The CDF of the exponential distribution is
$F_X(x) = 1 - e^{\lambda x}$ \DUrole{citep}{Wikipediacontributors2015b}. The
empirical cumulative distribution function (ECDF) is the CDF of an actual experimental
sample. It is a step function that increases by $1/n$, at the value of
each of the $n$ samples \DUrole{citet}{Wikipediacontributors2015a}.
}

The standard deviation of the time between supernovas is $\SI{112}{kys}$,
close to the mean and matches the value expected from the exponential
distribution. Also an Anderson-Darling test can not reject the null-hypothesis
that the distribution is drawn from the exponential distribution (with an AD
statistic of 0.207 and critical values of 0.919,  1.074,  1.337,  1.601 and
1.951 for an significance level of 15\%, 10\%, 5\%, 2.5\% and 1\% respectively
\DUrole{citep}{Stephens1977,EricJones2001}; see also \DUrole{ref}{code:plot-sne-py}).
Therefore it seems likely that the implementation shows the expected behaviour.

\begin{figure}
\noindent\makebox[\textwidth][c]{\includegraphics[width=1.000\linewidth]{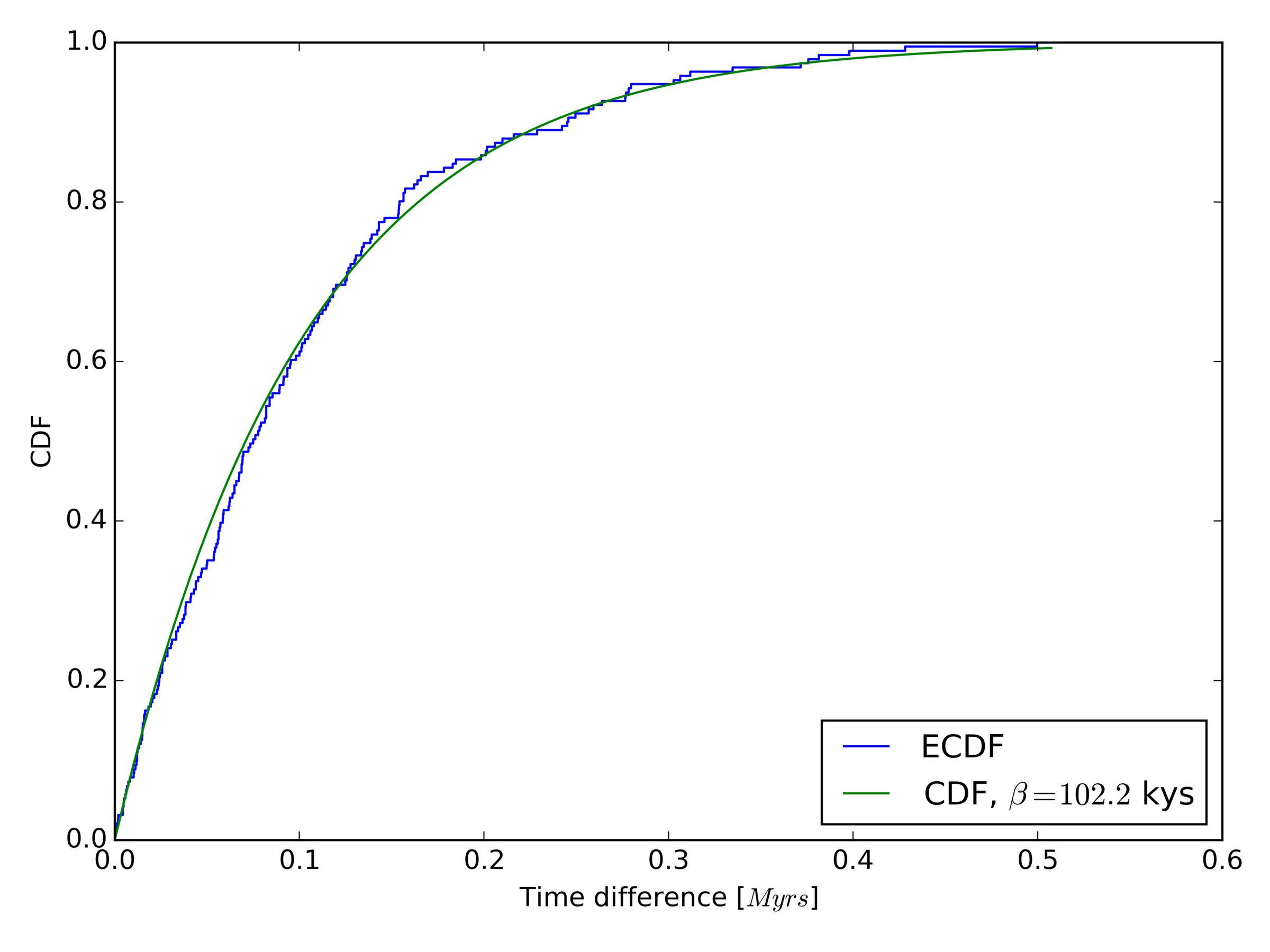}}
\caption{\DUrole{label}{fig:1u-feedback-400pc-ecdf}
Cumulative distribution
function of the time between SN events, simulation with initial density
$\SI{1}{u/cm^3}$.}
\end{figure}

\chapter{Conclusion%
  \label{conclusion}%
}

A simple treatment of supernova feedback for the astrophysical simulation
software \DUrole{sw}{Arepo} in the energy conserving regime only was implemented for
this thesis. A fixed amount of thermal energy is injected into a cells inside a
spherical region containing a certain mass, with the location and time of
supernova events determined stochastically.

The general approach was first tested by simulating a single supernova with
energy injection already done in the initial conditions, with and without
cooling. Large errors in the estimation of the volume of the cells in the SN energy
injection region limit the significance of these simulations. Still overcooling
effects were likely observed for the largest injection radii.

For testing the implementation in Arepo, a simulation of an initially uniform
density box ($\rho = \SI{1}{u/cm^3}$) was run, with supernovas set off at
random locations at a rate derived from the Kennicutt-Schmidt relation
\DUrole{citep}{Kennicutt1998}. These lead to gravitational collapse of the gas, and
formation of molecular gas. Velocity dispersions of the different gas fractions
are similar to values reported in comparable simulations by \DUrole{citet}{Gatto2014}.
The implementation of the stochastic SN generation process is tested by
comparison of the empirical cumulative distribution function (ECDF) with the
expected CDF and shows the expected behaviour of a slightly lower SN rate than
specified. Therefore the simulation seems to work well for fully resolved
simulations, where random driving is desired.

\section{Further work%
  \label{further-work}%
}

The implementation of supernova feedback for this thesis doesn’t have the
capability for momentum injection. However, this is something that is present in
the full cosmological version of the Arepo code and has been addressed more
fully by other authors.

\cleardoublepage
\appendix
\printbibliography

\chapter{Source code%
  \label{source-code}%
}

\label{appendixsource}
\labeledslstinputlistingc{arepo-sne/sne/sne.c}{code:sne-c}
\labeledslstinputlistingc{arepo-sne/sne/sne.h}{code:sne-h}
\labeledslstinputlistingc{arepo-sne/sne/sne_def.c}{code:sne-def-c}
\labeledslstinputlistingc{arepo-sne/sne/sne_def.h}{code:sne-def-h}
\labeledslstinputlisting{pint/constants_en.txt}{code:constants-en-txt}
\labeledslstinputlisting{pint/default_en.txt}{code:default-en-txt}
\labeledslstinputlisting{supernova/configfiles/1000Mys.in}{code:1000Mys-in}
\labeledslstinputlisting{supernova/configfiles/100Mys.in}{code:100Mys-in}
\labeledslstinputlisting{supernova/configfiles/1Mys.in}{code:1Mys-in}
\labeledslstinputlisting{supernova/configfiles/72h.in}{code:72h-in}
\labeledslstinputlisting{supernova/configfiles/box_size_100pc.in}{code:box-size-100pc-in}
\labeledslstinputlisting{supernova/configfiles/box_size_10kpc.in}{code:box-size-10kpc-in}
\labeledslstinputlisting{supernova/configfiles/box_size_1kpc.in}{code:box-size-1kpc-in}
\labeledslstinputlisting{supernova/configfiles/box_size_200pc.in}{code:box-size-200pc-in}
\labeledslstinputlisting{supernova/configfiles/box_size_2kpc.in}{code:box-size-2kpc-in}
\labeledslstinputlisting{supernova/configfiles/box_size_400pc.in}{code:box-size-400pc-in}
\labeledslstinputlisting{supernova/configfiles/defaults.in}{code:defaults-in}
\labeledslstinputlisting{supernova/configfiles/density_100u.in}{code:density-100u-in}
\labeledslstinputlisting{supernova/configfiles/density_10u.in}{code:density-10u-in}
\labeledslstinputlisting{supernova/configfiles/density_10u_400pc.in}{code:density-10u-400pc-in}
\labeledslstinputlisting{supernova/configfiles/density_1u.in}{code:density-1u-in}
\labeledslstinputlisting{supernova/configfiles/feedback.in}{code:feedback-in}
\labeledslstinputlisting{supernova/configfiles/job_radius_0.5.in}{code:job-radius-0-5-in}
\labeledslstinputlisting{supernova/configfiles/job_radius_0.in}{code:job-radius-0-in}
\labeledslstinputlisting{supernova/configfiles/job_radius_1.in}{code:job-radius-1-in}
\labeledslstinputlisting{supernova/configfiles/job_radius_10.in}{code:job-radius-10-in}
\labeledslstinputlisting{supernova/configfiles/job_radius_2.in}{code:job-radius-2-in}
\labeledslstinputlisting{supernova/configfiles/job_radius_5.in}{code:job-radius-5-in}
\labeledslstinputlisting{supernova/configfiles/job_radius_7.5.in}{code:job-radius-7-5-in}
\labeledslstinputlisting{supernova/configfiles/job_run_1M.in}{code:job-run-1M-in}
\labeledslstinputlisting{supernova/configfiles/job_run_2M.in}{code:job-run-2M-in}
\labeledslstinputlisting{supernova/configfiles/job_run_8M.in}{code:job-run-8M-in}
\labeledslstinputlisting{supernova/configfiles/machine_grable.in}{code:machine-grable-in}
\labeledslstinputlisting{supernova/configfiles/machine_milkyway.in}{code:machine-milkyway-in}
\labeledslstinputlisting{supernova/configfiles/mpi_192_processors.in}{code:mpi-192-processors-in}
\labeledslstinputlisting{supernova/configfiles/mpi_96_processors.in}{code:mpi-96-processors-in}
\labeledslstinputlisting{supernova/configfiles/name_single.in}{code:name-single-in}
\labeledslstinputlisting{supernova/configfiles/nochemistry.in}{code:nochemistry-in}
\labeledslstinputlisting{supernova/configfiles/noenergy.in}{code:noenergy-in}
\labeledslstinputlisting{supernova/configfiles/norefinement.in}{code:norefinement-in}
\labeledslstinputlisting{supernova/configfiles/noselfgravity.in}{code:noselfgravity-in}
\labeledslstinputlisting{supernova/configfiles/optical_thin_chemistry.in}{code:optical-thin-chemistry-in}
\labeledslstinputlisting{supernova/configfiles/supernova_test.in}{code:supernova-test-in}
\labeledslstinputlisting{supernova/env.py}{code:env-py}
\labeledslstinputlistingsh{supernova/make_jobs_grable.sh}{code:make-jobs-grable-sh}
\labeledslstinputlistingsh{supernova/make_jobs_milkyway.sh}{code:make-jobs-milkyway-sh}
\labeledslstinputlisting{supernova/mktmpl.py}{code:mktmpl-py}
\labeledslstinputlisting{supernova/plot_max.py}{code:plot-max-py}
\labeledslstinputlistingsh{supernova/postprocess_all.sh}{code:postprocess-all-sh}
\labeledslstinputlisting{supernova/py/cache_quantities.py}{code:cache-quantities-py}
\labeledslstinputlisting{supernova/py/plot.py}{code:plot-py}
\labeledslstinputlisting{supernova/py/plot_energy.py}{code:plot-energy-py}
\labeledslstinputlisting{supernova/py/plot_radial.py}{code:plot-radial-py}
\labeledslstinputlisting{supernova/py/plot_sne.py}{code:plot-sne-py}
\labeledslstinputlisting{supernova/py/sim.py}{code:sim-py}
\labeledslstinputlisting{supernova/py/util.py}{code:util-py}
\labeledslstinputlistingsh{supernova/templates/Config.sh}{code:Config-sh}
\labeledslstinputlistingsh{supernova/templates/clean.sh}{code:clean-sh}
\labeledslstinputlistingsh{supernova/templates/jellyrc}{code:jellyrc}
\labeledslstinputlistingsh{supernova/templates/movies.sh}{code:movies-sh}
\labeledslstinputlisting{supernova/templates/params.txt}{code:params-txt}
\labeledslstinputlisting{supernova/templates/plot_radial.yaml}{code:plot-radial-yaml}
\labeledslstinputlistingsh{supernova/templates/plots.sh}{code:plots-sh}
\labeledslstinputlistingsh{supernova/templates/postprocess.sh}{code:postprocess-sh}
\labeledslstinputlistingsh{supernova/templates/prepare.sh}{code:prepare-sh}
\labeledslstinputlistingsh{supernova/templates/print_env.sh}{code:print-env-sh}
\labeledslstinputlistingsh{supernova/templates/redo_slices.sh}{code:redo-slices-sh}
\labeledslstinputlistingsh{supernova/templates/run_clean.sh}{code:run-clean-sh}
\labeledslstinputlistingsh{supernova/templates/run_mpi.sh}{code:run-mpi-sh}
\labeledslstinputlistingsh{supernova/templates/run_mpi_plots.sh}{code:run-mpi-plots-sh}
\labeledslstinputlistingsh{supernova/templates/run_mpi_restart.sh}{code:run-mpi-restart-sh}
\labeledslstinputlisting{supernova/templates/units.txt}{code:units-txt}
\labeledslstinputlisting{supernova/templates/plot.yaml}{code:plot-yaml}
\labeledslstinputlistingsh{supernova/tests.sh}{code:tests-sh}

\chapter{Licenses%
  \label{licenses}%
}

\section{Pint%
  \label{pint}%
}

The \DUrole{sw}{Python} package \DUrole{sw}{Pint} is available under the following license:

\begin{verbatim}
Copyright (c) 2012 by Hernan E. Grecco and contributors.  See AUTHORS
for more details.

Some rights reserved.

Redistribution and use in source and binary forms of the software as well
as documentation, with or without modification, are permitted provided
that the following conditions are met:

* Redistributions of source code must retain the above copyright
  notice, this list of conditions and the following disclaimer.

* Redistributions in binary form must reproduce the above
  copyright notice, this list of conditions and the following
  disclaimer in the documentation and/or other materials provided
  with the distribution.

* The names of the contributors may not be used to endorse or
  promote products derived from this software without specific
  prior written permission.

THIS SOFTWARE AND DOCUMENTATION IS PROVIDED BY THE COPYRIGHT HOLDERS AND
CONTRIBUTORS "AS IS" AND ANY EXPRESS OR IMPLIED WARRANTIES, INCLUDING, BUT
NOT LIMITED TO, THE IMPLIED WARRANTIES OF MERCHANTABILITY AND FITNESS FOR
A PARTICULAR PURPOSE ARE DISCLAIMED. IN NO EVENT SHALL THE COPYRIGHT OWNER
OR CONTRIBUTORS BE LIABLE FOR ANY DIRECT, INDIRECT, INCIDENTAL, SPECIAL,
EXEMPLARY, OR CONSEQUENTIAL DAMAGES (INCLUDING, BUT NOT LIMITED TO,
PROCUREMENT OF SUBSTITUTE GOODS OR SERVICES; LOSS OF USE, DATA, OR
PROFITS; OR BUSINESS INTERRUPTION) HOWEVER CAUSED AND ON ANY THEORY OF
LIABILITY, WHETHER IN CONTRACT, STRICT LIABILITY, OR TORT (INCLUDING
NEGLIGENCE OR OTHERWISE) ARISING IN ANY WAY OUT OF THE USE OF THIS
SOFTWARE AND DOCUMENTATION, EVEN IF ADVISED OF THE POSSIBILITY OF SUCH
DAMAGE.
\end{verbatim}

\begin{verbatim}
Pint is written and maintained by Hernan E. Grecco <hernan.grecco@gmail.com>.

Other contributors, listed alphabetically, are:

* Alexander Böhn <fish2000@gmail.com>
* Brend Wanders <b.wanders@utwente.nl>
* choloepus
* Daniel Sokolowski <daniel.sokolowski@danols.com>
* Dave Brooks <dave@bcs.co.nz>
* David Linke
* Eduard Bopp <eduard.bopp@aepsil0n.de>
* Felix Hummel <felix@felixhummel.de>
* Giel van Schijndel <me@mortis.eu>
* James Rowe <jnrowe@gmail.com>
* Jim Turner <jturner314@gmail.com>
* Joel B. Mohler <joel@kiwistrawberry.us>
* John David Reaver <jdreaver@adlerhorst.com>
* Jonas Olson <jolson@kth.se>
* Kenneth D. Mankoff <mankoff@gmail.com>
* Luke Campbell <luke.s.campbell@gmail.com>
* Matthieu Dartiailh <marul@laposte.net>
* Nate Bogdanowicz <natezb@gmail.com>
* Peter Grayson <jpgrayson@gmail.com>
* Richard Barnes <rbarnes@umn.edu>
* Ryan Kingsbury <RyanSKingsbury@alumni.unc.edu>
* Sundar Raman <cybertoast@gmail.com>
* Tiago Coutinho <coutinho@esrf.fr>
* Tom Ritchford <tom@swirly.com>
* Virgil Dupras <virgil.dupras@savoirfairelinux.com>
* Ryan Dwyer <ryanpdwyer@gmail.com>

(If you think that your name belongs here, please let the maintainer know)
\end{verbatim}

\section{GPL%
  \label{gpl}%
}

The source code in \DUrole{ref}{appendixsource} is available under the GNU General Public License version 3.0:

\begin{verbatim}
                    GNU GENERAL PUBLIC LICENSE
                       Version 3, 29 June 2007

 Copyright (C) 2007 Free Software Foundation, Inc. <http://fsf.org/>
 Everyone is permitted to copy and distribute verbatim copies
 of this license document, but changing it is not allowed.

                            Preamble

  The GNU General Public License is a free, copyleft license for
software and other kinds of works.

  The licenses for most software and other practical works are designed
to take away your freedom to share and change the works.  By contrast,
the GNU General Public License is intended to guarantee your freedom to
share and change all versions of a program--to make sure it remains free
software for all its users.  We, the Free Software Foundation, use the
GNU General Public License for most of our software; it applies also to
any other work released this way by its authors.  You can apply it to
your programs, too.

  When we speak of free software, we are referring to freedom, not
price.  Our General Public Licenses are designed to make sure that you
have the freedom to distribute copies of free software (and charge for
them if you wish), that you receive source code or can get it if you
want it, that you can change the software or use pieces of it in new
free programs, and that you know you can do these things.

  To protect your rights, we need to prevent others from denying you
these rights or asking you to surrender the rights.  Therefore, you have
certain responsibilities if you distribute copies of the software, or if
you modify it: responsibilities to respect the freedom of others.

  For example, if you distribute copies of such a program, whether
gratis or for a fee, you must pass on to the recipients the same
freedoms that you received.  You must make sure that they, too, receive
or can get the source code.  And you must show them these terms so they
know their rights.

  Developers that use the GNU GPL protect your rights with two steps:
(1) assert copyright on the software, and (2) offer you this License
giving you legal permission to copy, distribute and/or modify it.

  For the developers' and authors' protection, the GPL clearly explains
that there is no warranty for this free software.  For both users' and
authors' sake, the GPL requires that modified versions be marked as
changed, so that their problems will not be attributed erroneously to
authors of previous versions.

  Some devices are designed to deny users access to install or run
modified versions of the software inside them, although the manufacturer
can do so.  This is fundamentally incompatible with the aim of
protecting users' freedom to change the software.  The systematic
pattern of such abuse occurs in the area of products for individuals to
use, which is precisely where it is most unacceptable.  Therefore, we
have designed this version of the GPL to prohibit the practice for those
products.  If such problems arise substantially in other domains, we
stand ready to extend this provision to those domains in future versions
of the GPL, as needed to protect the freedom of users.

  Finally, every program is threatened constantly by software patents.
States should not allow patents to restrict development and use of
software on general-purpose computers, but in those that do, we wish to
avoid the special danger that patents applied to a free program could
make it effectively proprietary.  To prevent this, the GPL assures that
patents cannot be used to render the program non-free.

  The precise terms and conditions for copying, distribution and
modification follow.

                       TERMS AND CONDITIONS

  0. Definitions.

  "This License" refers to version 3 of the GNU General Public License.

  "Copyright" also means copyright-like laws that apply to other kinds of
works, such as semiconductor masks.

  "The Program" refers to any copyrightable work licensed under this
License.  Each licensee is addressed as "you".  "Licensees" and
"recipients" may be individuals or organizations.

  To "modify" a work means to copy from or adapt all or part of the work
in a fashion requiring copyright permission, other than the making of an
exact copy.  The resulting work is called a "modified version" of the
earlier work or a work "based on" the earlier work.

  A "covered work" means either the unmodified Program or a work based
on the Program.

  To "propagate" a work means to do anything with it that, without
permission, would make you directly or secondarily liable for
infringement under applicable copyright law, except executing it on a
computer or modifying a private copy.  Propagation includes copying,
distribution (with or without modification), making available to the
public, and in some countries other activities as well.

  To "convey" a work means any kind of propagation that enables other
parties to make or receive copies.  Mere interaction with a user through
a computer network, with no transfer of a copy, is not conveying.

  An interactive user interface displays "Appropriate Legal Notices"
to the extent that it includes a convenient and prominently visible
feature that (1) displays an appropriate copyright notice, and (2)
tells the user that there is no warranty for the work (except to the
extent that warranties are provided), that licensees may convey the
work under this License, and how to view a copy of this License.  If
the interface presents a list of user commands or options, such as a
menu, a prominent item in the list meets this criterion.

  1. Source Code.

  The "source code" for a work means the preferred form of the work
for making modifications to it.  "Object code" means any non-source
form of a work.

  A "Standard Interface" means an interface that either is an official
standard defined by a recognized standards body, or, in the case of
interfaces specified for a particular programming language, one that
is widely used among developers working in that language.

  The "System Libraries" of an executable work include anything, other
than the work as a whole, that (a) is included in the normal form of
packaging a Major Component, but which is not part of that Major
Component, and (b) serves only to enable use of the work with that
Major Component, or to implement a Standard Interface for which an
implementation is available to the public in source code form.  A
"Major Component", in this context, means a major essential component
(kernel, window system, and so on) of the specific operating system
(if any) on which the executable work runs, or a compiler used to
produce the work, or an object code interpreter used to run it.

  The "Corresponding Source" for a work in object code form means all
the source code needed to generate, install, and (for an executable
work) run the object code and to modify the work, including scripts to
control those activities.  However, it does not include the work's
System Libraries, or general-purpose tools or generally available free
programs which are used unmodified in performing those activities but
which are not part of the work.  For example, Corresponding Source
includes interface definition files associated with source files for
the work, and the source code for shared libraries and dynamically
linked subprograms that the work is specifically designed to require,
such as by intimate data communication or control flow between those
subprograms and other parts of the work.

  The Corresponding Source need not include anything that users
can regenerate automatically from other parts of the Corresponding
Source.

  The Corresponding Source for a work in source code form is that
same work.

  2. Basic Permissions.

  All rights granted under this License are granted for the term of
copyright on the Program, and are irrevocable provided the stated
conditions are met.  This License explicitly affirms your unlimited
permission to run the unmodified Program.  The output from running a
covered work is covered by this License only if the output, given its
content, constitutes a covered work.  This License acknowledges your
rights of fair use or other equivalent, as provided by copyright law.

  You may make, run and propagate covered works that you do not
convey, without conditions so long as your license otherwise remains
in force.  You may convey covered works to others for the sole purpose
of having them make modifications exclusively for you, or provide you
with facilities for running those works, provided that you comply with
the terms of this License in conveying all material for which you do
not control copyright.  Those thus making or running the covered works
for you must do so exclusively on your behalf, under your direction
and control, on terms that prohibit them from making any copies of
your copyrighted material outside their relationship with you.

  Conveying under any other circumstances is permitted solely under
the conditions stated below.  Sublicensing is not allowed; section 10
makes it unnecessary.

  3. Protecting Users' Legal Rights From Anti-Circumvention Law.

  No covered work shall be deemed part of an effective technological
measure under any applicable law fulfilling obligations under article
11 of the WIPO copyright treaty adopted on 20 December 1996, or
similar laws prohibiting or restricting circumvention of such
measures.

  When you convey a covered work, you waive any legal power to forbid
circumvention of technological measures to the extent such circumvention
is effected by exercising rights under this License with respect to
the covered work, and you disclaim any intention to limit operation or
modification of the work as a means of enforcing, against the work's
users, your or third parties' legal rights to forbid circumvention of
technological measures.

  4. Conveying Verbatim Copies.

  You may convey verbatim copies of the Program's source code as you
receive it, in any medium, provided that you conspicuously and
appropriately publish on each copy an appropriate copyright notice;
keep intact all notices stating that this License and any
non-permissive terms added in accord with section 7 apply to the code;
keep intact all notices of the absence of any warranty; and give all
recipients a copy of this License along with the Program.

  You may charge any price or no price for each copy that you convey,
and you may offer support or warranty protection for a fee.

  5. Conveying Modified Source Versions.

  You may convey a work based on the Program, or the modifications to
produce it from the Program, in the form of source code under the
terms of section 4, provided that you also meet all of these conditions:

    a) The work must carry prominent notices stating that you modified
    it, and giving a relevant date.

    b) The work must carry prominent notices stating that it is
    released under this License and any conditions added under section
    7.  This requirement modifies the requirement in section 4 to
    "keep intact all notices".

    c) You must license the entire work, as a whole, under this
    License to anyone who comes into possession of a copy.  This
    License will therefore apply, along with any applicable section 7
    additional terms, to the whole of the work, and all its parts,
    regardless of how they are packaged.  This License gives no
    permission to license the work in any other way, but it does not
    invalidate such permission if you have separately received it.

    d) If the work has interactive user interfaces, each must display
    Appropriate Legal Notices; however, if the Program has interactive
    interfaces that do not display Appropriate Legal Notices, your
    work need not make them do so.

  A compilation of a covered work with other separate and independent
works, which are not by their nature extensions of the covered work,
and which are not combined with it such as to form a larger program,
in or on a volume of a storage or distribution medium, is called an
"aggregate" if the compilation and its resulting copyright are not
used to limit the access or legal rights of the compilation's users
beyond what the individual works permit.  Inclusion of a covered work
in an aggregate does not cause this License to apply to the other
parts of the aggregate.

  6. Conveying Non-Source Forms.

  You may convey a covered work in object code form under the terms
of sections 4 and 5, provided that you also convey the
machine-readable Corresponding Source under the terms of this License,
in one of these ways:

    a) Convey the object code in, or embodied in, a physical product
    (including a physical distribution medium), accompanied by the
    Corresponding Source fixed on a durable physical medium
    customarily used for software interchange.

    b) Convey the object code in, or embodied in, a physical product
    (including a physical distribution medium), accompanied by a
    written offer, valid for at least three years and valid for as
    long as you offer spare parts or customer support for that product
    model, to give anyone who possesses the object code either (1) a
    copy of the Corresponding Source for all the software in the
    product that is covered by this License, on a durable physical
    medium customarily used for software interchange, for a price no
    more than your reasonable cost of physically performing this
    conveying of source, or (2) access to copy the
    Corresponding Source from a network server at no charge.

    c) Convey individual copies of the object code with a copy of the
    written offer to provide the Corresponding Source.  This
    alternative is allowed only occasionally and noncommercially, and
    only if you received the object code with such an offer, in accord
    with subsection 6b.

    d) Convey the object code by offering access from a designated
    place (gratis or for a charge), and offer equivalent access to the
    Corresponding Source in the same way through the same place at no
    further charge.  You need not require recipients to copy the
    Corresponding Source along with the object code.  If the place to
    copy the object code is a network server, the Corresponding Source
    may be on a different server (operated by you or a third party)
    that supports equivalent copying facilities, provided you maintain
    clear directions next to the object code saying where to find the
    Corresponding Source.  Regardless of what server hosts the
    Corresponding Source, you remain obligated to ensure that it is
    available for as long as needed to satisfy these requirements.

    e) Convey the object code using peer-to-peer transmission, provided
    you inform other peers where the object code and Corresponding
    Source of the work are being offered to the general public at no
    charge under subsection 6d.

  A separable portion of the object code, whose source code is excluded
from the Corresponding Source as a System Library, need not be
included in conveying the object code work.

  A "User Product" is either (1) a "consumer product", which means any
tangible personal property which is normally used for personal, family,
or household purposes, or (2) anything designed or sold for incorporation
into a dwelling.  In determining whether a product is a consumer product,
doubtful cases shall be resolved in favor of coverage.  For a particular
product received by a particular user, "normally used" refers to a
typical or common use of that class of product, regardless of the status
of the particular user or of the way in which the particular user
actually uses, or expects or is expected to use, the product.  A product
is a consumer product regardless of whether the product has substantial
commercial, industrial or non-consumer uses, unless such uses represent
the only significant mode of use of the product.

  "Installation Information" for a User Product means any methods,
procedures, authorization keys, or other information required to install
and execute modified versions of a covered work in that User Product from
a modified version of its Corresponding Source.  The information must
suffice to ensure that the continued functioning of the modified object
code is in no case prevented or interfered with solely because
modification has been made.

  If you convey an object code work under this section in, or with, or
specifically for use in, a User Product, and the conveying occurs as
part of a transaction in which the right of possession and use of the
User Product is transferred to the recipient in perpetuity or for a
fixed term (regardless of how the transaction is characterized), the
Corresponding Source conveyed under this section must be accompanied
by the Installation Information.  But this requirement does not apply
if neither you nor any third party retains the ability to install
modified object code on the User Product (for example, the work has
been installed in ROM).

  The requirement to provide Installation Information does not include a
requirement to continue to provide support service, warranty, or updates
for a work that has been modified or installed by the recipient, or for
the User Product in which it has been modified or installed.  Access to a
network may be denied when the modification itself materially and
adversely affects the operation of the network or violates the rules and
protocols for communication across the network.

  Corresponding Source conveyed, and Installation Information provided,
in accord with this section must be in a format that is publicly
documented (and with an implementation available to the public in
source code form), and must require no special password or key for
unpacking, reading or copying.

  7. Additional Terms.

  "Additional permissions" are terms that supplement the terms of this
License by making exceptions from one or more of its conditions.
Additional permissions that are applicable to the entire Program shall
be treated as though they were included in this License, to the extent
that they are valid under applicable law.  If additional permissions
apply only to part of the Program, that part may be used separately
under those permissions, but the entire Program remains governed by
this License without regard to the additional permissions.

  When you convey a copy of a covered work, you may at your option
remove any additional permissions from that copy, or from any part of
it.  (Additional permissions may be written to require their own
removal in certain cases when you modify the work.)  You may place
additional permissions on material, added by you to a covered work,
for which you have or can give appropriate copyright permission.

  Notwithstanding any other provision of this License, for material you
add to a covered work, you may (if authorized by the copyright holders of
that material) supplement the terms of this License with terms:

    a) Disclaiming warranty or limiting liability differently from the
    terms of sections 15 and 16 of this License; or

    b) Requiring preservation of specified reasonable legal notices or
    author attributions in that material or in the Appropriate Legal
    Notices displayed by works containing it; or

    c) Prohibiting misrepresentation of the origin of that material, or
    requiring that modified versions of such material be marked in
    reasonable ways as different from the original version; or

    d) Limiting the use for publicity purposes of names of licensors or
    authors of the material; or

    e) Declining to grant rights under trademark law for use of some
    trade names, trademarks, or service marks; or

    f) Requiring indemnification of licensors and authors of that
    material by anyone who conveys the material (or modified versions of
    it) with contractual assumptions of liability to the recipient, for
    any liability that these contractual assumptions directly impose on
    those licensors and authors.

  All other non-permissive additional terms are considered "further
restrictions" within the meaning of section 10.  If the Program as you
received it, or any part of it, contains a notice stating that it is
governed by this License along with a term that is a further
restriction, you may remove that term.  If a license document contains
a further restriction but permits relicensing or conveying under this
License, you may add to a covered work material governed by the terms
of that license document, provided that the further restriction does
not survive such relicensing or conveying.

  If you add terms to a covered work in accord with this section, you
must place, in the relevant source files, a statement of the
additional terms that apply to those files, or a notice indicating
where to find the applicable terms.

  Additional terms, permissive or non-permissive, may be stated in the
form of a separately written license, or stated as exceptions;
the above requirements apply either way.

  8. Termination.

  You may not propagate or modify a covered work except as expressly
provided under this License.  Any attempt otherwise to propagate or
modify it is void, and will automatically terminate your rights under
this License (including any patent licenses granted under the third
paragraph of section 11).

  However, if you cease all violation of this License, then your
license from a particular copyright holder is reinstated (a)
provisionally, unless and until the copyright holder explicitly and
finally terminates your license, and (b) permanently, if the copyright
holder fails to notify you of the violation by some reasonable means
prior to 60 days after the cessation.

  Moreover, your license from a particular copyright holder is
reinstated permanently if the copyright holder notifies you of the
violation by some reasonable means, this is the first time you have
received notice of violation of this License (for any work) from that
copyright holder, and you cure the violation prior to 30 days after
your receipt of the notice.

  Termination of your rights under this section does not terminate the
licenses of parties who have received copies or rights from you under
this License.  If your rights have been terminated and not permanently
reinstated, you do not qualify to receive new licenses for the same
material under section 10.

  9. Acceptance Not Required for Having Copies.

  You are not required to accept this License in order to receive or
run a copy of the Program.  Ancillary propagation of a covered work
occurring solely as a consequence of using peer-to-peer transmission
to receive a copy likewise does not require acceptance.  However,
nothing other than this License grants you permission to propagate or
modify any covered work.  These actions infringe copyright if you do
not accept this License.  Therefore, by modifying or propagating a
covered work, you indicate your acceptance of this License to do so.

  10. Automatic Licensing of Downstream Recipients.

  Each time you convey a covered work, the recipient automatically
receives a license from the original licensors, to run, modify and
propagate that work, subject to this License.  You are not responsible
for enforcing compliance by third parties with this License.

  An "entity transaction" is a transaction transferring control of an
organization, or substantially all assets of one, or subdividing an
organization, or merging organizations.  If propagation of a covered
work results from an entity transaction, each party to that
transaction who receives a copy of the work also receives whatever
licenses to the work the party's predecessor in interest had or could
give under the previous paragraph, plus a right to possession of the
Corresponding Source of the work from the predecessor in interest, if
the predecessor has it or can get it with reasonable efforts.

  You may not impose any further restrictions on the exercise of the
rights granted or affirmed under this License.  For example, you may
not impose a license fee, royalty, or other charge for exercise of
rights granted under this License, and you may not initiate litigation
(including a cross-claim or counterclaim in a lawsuit) alleging that
any patent claim is infringed by making, using, selling, offering for
sale, or importing the Program or any portion of it.

  11. Patents.

  A "contributor" is a copyright holder who authorizes use under this
License of the Program or a work on which the Program is based.  The
work thus licensed is called the contributor's "contributor version".

  A contributor's "essential patent claims" are all patent claims
owned or controlled by the contributor, whether already acquired or
hereafter acquired, that would be infringed by some manner, permitted
by this License, of making, using, or selling its contributor version,
but do not include claims that would be infringed only as a
consequence of further modification of the contributor version.  For
purposes of this definition, "control" includes the right to grant
patent sublicenses in a manner consistent with the requirements of
this License.

  Each contributor grants you a non-exclusive, worldwide, royalty-free
patent license under the contributor's essential patent claims, to
make, use, sell, offer for sale, import and otherwise run, modify and
propagate the contents of its contributor version.

  In the following three paragraphs, a "patent license" is any express
agreement or commitment, however denominated, not to enforce a patent
(such as an express permission to practice a patent or covenant not to
sue for patent infringement).  To "grant" such a patent license to a
party means to make such an agreement or commitment not to enforce a
patent against the party.

  If you convey a covered work, knowingly relying on a patent license,
and the Corresponding Source of the work is not available for anyone
to copy, free of charge and under the terms of this License, through a
publicly available network server or other readily accessible means,
then you must either (1) cause the Corresponding Source to be so
available, or (2) arrange to deprive yourself of the benefit of the
patent license for this particular work, or (3) arrange, in a manner
consistent with the requirements of this License, to extend the patent
license to downstream recipients.  "Knowingly relying" means you have
actual knowledge that, but for the patent license, your conveying the
covered work in a country, or your recipient's use of the covered work
in a country, would infringe one or more identifiable patents in that
country that you have reason to believe are valid.

  If, pursuant to or in connection with a single transaction or
arrangement, you convey, or propagate by procuring conveyance of, a
covered work, and grant a patent license to some of the parties
receiving the covered work authorizing them to use, propagate, modify
or convey a specific copy of the covered work, then the patent license
you grant is automatically extended to all recipients of the covered
work and works based on it.

  A patent license is "discriminatory" if it does not include within
the scope of its coverage, prohibits the exercise of, or is
conditioned on the non-exercise of one or more of the rights that are
specifically granted under this License.  You may not convey a covered
work if you are a party to an arrangement with a third party that is
in the business of distributing software, under which you make payment
to the third party based on the extent of your activity of conveying
the work, and under which the third party grants, to any of the
parties who would receive the covered work from you, a discriminatory
patent license (a) in connection with copies of the covered work
conveyed by you (or copies made from those copies), or (b) primarily
for and in connection with specific products or compilations that
contain the covered work, unless you entered into that arrangement,
or that patent license was granted, prior to 28 March 2007.

  Nothing in this License shall be construed as excluding or limiting
any implied license or other defenses to infringement that may
otherwise be available to you under applicable patent law.

  12. No Surrender of Others' Freedom.

  If conditions are imposed on you (whether by court order, agreement or
otherwise) that contradict the conditions of this License, they do not
excuse you from the conditions of this License.  If you cannot convey a
covered work so as to satisfy simultaneously your obligations under this
License and any other pertinent obligations, then as a consequence you may
not convey it at all.  For example, if you agree to terms that obligate you
to collect a royalty for further conveying from those to whom you convey
the Program, the only way you could satisfy both those terms and this
License would be to refrain entirely from conveying the Program.

  13. Use with the GNU Affero General Public License.

  Notwithstanding any other provision of this License, you have
permission to link or combine any covered work with a work licensed
under version 3 of the GNU Affero General Public License into a single
combined work, and to convey the resulting work.  The terms of this
License will continue to apply to the part which is the covered work,
but the special requirements of the GNU Affero General Public License,
section 13, concerning interaction through a network will apply to the
combination as such.

  14. Revised Versions of this License.

  The Free Software Foundation may publish revised and/or new versions of
the GNU General Public License from time to time.  Such new versions will
be similar in spirit to the present version, but may differ in detail to
address new problems or concerns.

  Each version is given a distinguishing version number.  If the
Program specifies that a certain numbered version of the GNU General
Public License "or any later version" applies to it, you have the
option of following the terms and conditions either of that numbered
version or of any later version published by the Free Software
Foundation.  If the Program does not specify a version number of the
GNU General Public License, you may choose any version ever published
by the Free Software Foundation.

  If the Program specifies that a proxy can decide which future
versions of the GNU General Public License can be used, that proxy's
public statement of acceptance of a version permanently authorizes you
to choose that version for the Program.

  Later license versions may give you additional or different
permissions.  However, no additional obligations are imposed on any
author or copyright holder as a result of your choosing to follow a
later version.

  15. Disclaimer of Warranty.

  THERE IS NO WARRANTY FOR THE PROGRAM, TO THE EXTENT PERMITTED BY
APPLICABLE LAW.  EXCEPT WHEN OTHERWISE STATED IN WRITING THE COPYRIGHT
HOLDERS AND/OR OTHER PARTIES PROVIDE THE PROGRAM "AS IS" WITHOUT WARRANTY
OF ANY KIND, EITHER EXPRESSED OR IMPLIED, INCLUDING, BUT NOT LIMITED TO,
THE IMPLIED WARRANTIES OF MERCHANTABILITY AND FITNESS FOR A PARTICULAR
PURPOSE.  THE ENTIRE RISK AS TO THE QUALITY AND PERFORMANCE OF THE PROGRAM
IS WITH YOU.  SHOULD THE PROGRAM PROVE DEFECTIVE, YOU ASSUME THE COST OF
ALL NECESSARY SERVICING, REPAIR OR CORRECTION.

  16. Limitation of Liability.

  IN NO EVENT UNLESS REQUIRED BY APPLICABLE LAW OR AGREED TO IN WRITING
WILL ANY COPYRIGHT HOLDER, OR ANY OTHER PARTY WHO MODIFIES AND/OR CONVEYS
THE PROGRAM AS PERMITTED ABOVE, BE LIABLE TO YOU FOR DAMAGES, INCLUDING ANY
GENERAL, SPECIAL, INCIDENTAL OR CONSEQUENTIAL DAMAGES ARISING OUT OF THE
USE OR INABILITY TO USE THE PROGRAM (INCLUDING BUT NOT LIMITED TO LOSS OF
DATA OR DATA BEING RENDERED INACCURATE OR LOSSES SUSTAINED BY YOU OR THIRD
PARTIES OR A FAILURE OF THE PROGRAM TO OPERATE WITH ANY OTHER PROGRAMS),
EVEN IF SUCH HOLDER OR OTHER PARTY HAS BEEN ADVISED OF THE POSSIBILITY OF
SUCH DAMAGES.

  17. Interpretation of Sections 15 and 16.

  If the disclaimer of warranty and limitation of liability provided
above cannot be given local legal effect according to their terms,
reviewing courts shall apply local law that most closely approximates
an absolute waiver of all civil liability in connection with the
Program, unless a warranty or assumption of liability accompanies a
copy of the Program in return for a fee.

                     END OF TERMS AND CONDITIONS

            How to Apply These Terms to Your New Programs

  If you develop a new program, and you want it to be of the greatest
possible use to the public, the best way to achieve this is to make it
free software which everyone can redistribute and change under these terms.

  To do so, attach the following notices to the program.  It is safest
to attach them to the start of each source file to most effectively
state the exclusion of warranty; and each file should have at least
the "copyright" line and a pointer to where the full notice is found.

    <one line to give the program's name and a brief idea of what it does.>
    Copyright (C) <year>  <name of author>

    This program is free software: you can redistribute it and/or modify
    it under the terms of the GNU General Public License as published by
    the Free Software Foundation, either version 3 of the License, or
    (at your option) any later version.

    This program is distributed in the hope that it will be useful,
    but WITHOUT ANY WARRANTY; without even the implied warranty of
    MERCHANTABILITY or FITNESS FOR A PARTICULAR PURPOSE.  See the
    GNU General Public License for more details.

    You should have received a copy of the GNU General Public License
    along with this program.  If not, see <http://www.gnu.org/licenses/>.

Also add information on how to contact you by electronic and paper mail.

  If the program does terminal interaction, make it output a short
notice like this when it starts in an interactive mode:

    <program>  Copyright (C) <year>  <name of author>
    This program comes with ABSOLUTELY NO WARRANTY; for details type `show w'.
    This is free software, and you are welcome to redistribute it
    under certain conditions; type `show c' for details.

The hypothetical commands `show w' and `show c' should show the appropriate
parts of the General Public License.  Of course, your program's commands
might be different; for a GUI interface, you would use an "about box".

  You should also get your employer (if you work as a programmer) or school,
if any, to sign a "copyright disclaimer" for the program, if necessary.
For more information on this, and how to apply and follow the GNU GPL, see
<http://www.gnu.org/licenses/>.

  The GNU General Public License does not permit incorporating your program
into proprietary programs.  If your program is a subroutine library, you
may consider it more useful to permit linking proprietary applications with
the library.  If this is what you want to do, use the GNU Lesser General
Public License instead of this License.  But first, please read
<http://www.gnu.org/philosophy/why-not-lgpl.html>.
\end{verbatim}

\chapter{Pint unit definitions}
\label{appendixpint}
Pint is available under the License in \ref{pint}.
\slstinputlisting{pint/constants_en.txt}
\slstinputlisting{pint/default_en.txt}

\KOMAoptions{cleardoublepage=empty}

\cleardoublepage

 \chapter*{Erklärung}

 \thispagestyle{empty}
 Ich versichere, dass ich diese Arbeit selbstständig verfasst habe und keine
 anderen als die angegebenen Quellen und Hilfsmittel benutzt habe.

 \vspace{10mm}

 Winnen, den 31. Juli 2015

 \vspace{15mm}

 (André-Patrick Bubel)

\end{document}